\documentclass[twocolumn,floatfix]{revtex4}
\pdfoutput=1
\usepackage{graphicx}
\usepackage{dcolumn}
\usepackage{longtable}
\usepackage{amsmath}
\usepackage{amssymb}
\usepackage{bm}

\begin{document}

\title{Islands of shape coexistence from single-particle spectra in covariant density functional theory}

\author
{Dennis Bonatsos$^1$, K.E. Karakatsanis$^{1,2,3}$, Andriana Martinou$^1$, T.J. Mertzimekis$^4$, and N. Minkov$^5$ }

\affiliation
{$^1$Institute of Nuclear and Particle Physics, National Centre for Scientific Research ``Demokritos'', GR-15310 Aghia Paraskevi, Attiki, Greece}

\affiliation
{$^2$  Department of Physics, Faculty of Science, University of Zagreb, HR-10000 Zagreb, Croatia}

\affiliation
{$^3$ Physics Department, Aristotle University of Thessaloniki, Thessaloniki GR-54124, Greece}

\affiliation
{$^4$  Department of Physics, National and Kapodistrian University of Athens, Zografou Campus, GR-15784 Athens, Greece}

\affiliation
{$^5$Institute of Nuclear Research and Nuclear Energy, Bulgarian Academy of Sciences, 72 Tzarigrad Road, 1784 Sofia, Bulgaria}

\begin{abstract}

Using covariant density functional theory with the DDME2 functional and labeling single-particle energy orbitals by Nilsson quantum numbers, a search for particle-hole (p-h) excitations connected to the appearance of shape coexistence is performed for $Z=38$ to 84. Islands of shape coexistence are found near the magic numbers $Z=82$ and $Z=50$, restricted in regions around the relevant neutron midshells $N=104$ and $N=66$ respectively, in accordance to the well accepted  p-h interpretation of shape coexistence in these regions, which we call neutron-induced shape coexistence, since the neutrons act as elevators creating holes in the proton orbitals. Similar but smaller islands of shape coexistence are found near $N=90$ and $N=60$, restricted in regions around the relevant proton midshells $Z=66$ and $Z=39$ respectively, related to p-h excitations across the 3-dimensional isotropic harmonic  oscillator (3D-HO) magic numbers $N=112$ and $N=70$, which correspond to the beginning of the participation of the opposite parity orbitals $1i_{13/2}$ and $1h_{11/2}$ respectively to the onset of deformation. 
We call this case proton-induced shape coexistence, since the protons act as elevators creating holes in the neutron orbitals, thus offering a possible microscopic mechanism for the appearance of shape coexistence in these regions. In the region around $N=40$, $Z=40$, an island is located on which both neutron p-h excitations and proton p-h excitations are present.   

 \end{abstract}

\maketitle
\section{Introduction}

Shape coexistence \cite{Meyer,Wood,Heyde,Garrett} in atomic nuclei is receiving recently intense experimental and theoretical attention, since it is a subtle effect depending on the details of nuclear structure, as shaped up by specific features of the nucleon-nucleon interaction \cite{Otsuka}. Shape coexistence is said to occur in a nucleus if the ground state band is accompanied by another similar band possessing the same angular momenta at similar energies, but with the two bands having radically different structure, for example with one of them being spherical and the other one deformed. First proposed in 1956 by Morinaga in $^{16}$O \cite{Morinaga}, it has led to numerous investigations, reviewed in \cite{Meyer} for odd nuclei and in \cite{Wood,Heyde,Garrett} for even-even nuclei. The availability of radioactive ion beams in several laboratories around the world has fostered extended experimental searches for shape coexistence, recently reviewed in Ref. \cite{Garrett}, in which the experimental fingerprints of shape coexistence are also discussed in detail.  

The regions in which shape coexistence had been observed experimentally in even-even nuclei up to that time (2011) have been schematically summarized in Fig.~8 of Ref.~\cite{Heyde}, showing the nuclear chart with $Z$ as the vertical axis and $N$ as the horizontal axis. In medium mass and heavy nuclei they include regions elongated in the $N$ direction around the magic numbers $Z=82$ and $Z=50$, shorter regions slightly elongated along the $Z$ axis in the regions of $N=90$, $Z=64$ and $N=60$, $Z=40$, as well as a region around $N=40$, $Z=34$. In addition, in lighter nuclei an elongated region along the diagonal is seen for $N=Z$ nuclei, along with a few other regions. Over the years the suspicion has been growing that shape coexistence is an effect which can appear anywhere on the nuclear chart. 

An opposite view has been presented in Ref. \cite{EPJASC}, in which a mechanism for shape coexistence based on the Elliott \cite{Elliott1,Elliott2,Elliott3} and proxy-SU(3) \cite{proxy1,proxy2} models has been presented. Using algebraic arguments, it has been found that shape coexistence can only occur within certain horizontal and vertical stripes of the nuclear chart, bordered by 7-8, 17-20, 34-40, 59-70, 96-112, 146-184 neutrons or protons. The predicted stripes of shape coexistence contain all regions shown in Fig.~8 of Ref.~\cite{Heyde}, described above, but they are considerably wider. Therefore being within the limits of these stripes appears to be a necessary condition for shape coexistence, but not a sufficient one. It would be desirable to find additional arguments narrowing down these stripes into specific regions, in order to provide more specific guidance to the experimental effort. 

Concerning the microscopic mechanism underlying the appearance of shape coexistence, particle-hole excitations across the proton magic numbers $Z=82$ and $Z=50$ have been proposed \cite{Wood,Heyde}. The microscopic mechanism underlying shape coexistenece in the $N=90$, $Z=64$ region, as well as in the very similar region with $N=60$, $Z=40$, as pointed out in Ref.~\cite{Garrett}, has remained an open question up to date. It should be mentioned at this point that the possible connection between shape coexistence and shape phase transition, an old standing question \cite{Jolie},  has been considered recently \cite{Ramos} in the region $N=60$, $Z=40$ in the framework of the interacting boson model with configuration mixing \cite{IAbook}. In parallel, the region of $N=90$, $Z=64$ is known as being the best example of the occurrence of the critical point symmetry X(5) \cite{IacX5}, related to the shape phase transition \cite{McCutchan,Cejnar} from spherical to deformed nuclei. 

It is clear that the proton-neutron interaction is expected to be playing a leading role in the creation of shape coexistence, as it does for the onset and further development of deformation, already realized by Federman and Pittel in Refs.~\cite{FP1,FP2,FP3} and demonstrated through the success of the $N_pN_n$ scheme \cite{Casten1,Casten2}  and the P-factor $P=N_p N_n/(N_p+N_n)$ \cite{Brenner}, in estimating the degree of collectivity of nuclei throughout the nuclear chart, where $N_p$, $N_n$ are the numbers of valence proton and valence neutron pairs measured from the nearest closed shell, also used in the framework of the interacting boson model \cite{IAbook}.

Measuring valence protons and neutrons from the nearest closed shell brings shell gaps and magic numbers into the present discussion. The shell model magic numbers 2, 8, 20, 28, 50, 82, 126, 184, \dots are known to be derived from the 3-dimensional (3D) isotropic harmonic oscillator (HO) magic numbers 2, 8, 20, 40, 70, 112, 168, \dots through the influence of the spin-orbit interaction \cite{Mayer}  and they are valid at zero deformation. When deformation sets in, as seen in the standard Nilsson diagrams \cite{NR}, corresponding to the single-particle spectrum of a deformed 3D-HO with cylindrical symmetry, the shell model magic numbers rapidly get blurred and disappear. Magic numbers are also modified as one is moving away from the valley of stability, as reviewed in Ref. \cite{Sorlin}. 

It is the purpose of the present work to attempt to shed light on the microscopic mechanism underlying shape coexistence in the $N=90$, $Z=64$ and $N=60$, $Z=40$ regions, which remains an open question. The tool to be used is standard covariant density functional theory (CDFT), using the DDME2 functional of Ref.~\cite{Lalazissis}, which is an improved relativistic mean-field effective interaction with explicit density dependence of the meson-nucleon couplings. The relativistic self-consistent mean field code DIRHB \cite{Niksic} is implemented. Single-particle energy levels \cite{NR} are derived using the approach described in detail in Refs. \cite{Prassa,Karakatsanis1,Karakatsanis2}. 
     
Our paper is organized as follows. In Section II there is a short summary of the different ways CDFT has been used for the study of shape coexistence along with a more detailed description of the present theoretical method. In Section III the regions of shape coexistence in which particle-hole excitations across proton shell model magic numbers have been proposed, are considered, in order to demonstrate the applicability of the present approach. Then in Section IV the regions $N=90$, $Z=64$ and $N=60$, $Z=40$ are considered, demonstrating that particle-hole excitations can also be proposed as a possible mechanism behind shape coexistence there, with the important difference that the particle-hole excitations take place across neutron magic numbers of the 3D-HO, thus involving pairs of nucleons of opposite parity. It should be remembered that proton-neutron pairs of opposite parity have been known to play a major role in the onset and development of nuclear deformation \cite{FP1,FP2,FP3,Cakirli,EPJASM,SDANCA21,HINP21SM}. A preliminary account of some of the present results has been given in \cite{PLB}. 
 
\section{Theoretical framework}

Nuclear energy density functionals are one of the most successful tools for the description of nuclear structure phenomena all over the chart of isotopes. Their success lies on the fact that the complex nuclear many-body problem can be reduced to essentially a single-particle problem with the energy of the system expressed as a functional of the local density. This is typically achieved at the mean-field level where the nucleons are approximated as independent particles, moving under a potential produced effectively by a microscopic two-body interaction \cite{RingBook,Bender2003}. The most widely used interactions are phenomenological, with a relatively small number of parameters adjusted on experimental data from stable nuclei such as binding energies and radii.
In their relativistic version they have been used for several decades for the description of ground and excited nuclear states from light nuclei with only a few nucleons up to the super-heavy region \cite{Bender2003,Ring1996,Vretenar2005,Niksic2011,Meng2006,Meng2015,Liang2015}. 

Specifically for the study of shape coexistence or shape phase transitions, the simplest approach is to solve the relativistic self-consistent mean-field (SCMF) equations and calculate the expectation values of observables related to the shape of the ground state of the nucleus. These observables are usually connected with the multipole moment operators $\hat{Q}_{\lambda\mu}(\bm{r}) = r^2Y_{\lambda\mu}(\theta,\phi)$, when the nuclear shape is parametrised by spherical harmonics $Y_{\lambda\mu}(\theta,\phi)$. A more elaborate method in the same context, is using one or more constraints on the value of the multipole moments while solving the SCMF equations for a certain isotope. The objective is the construction of a potential energy curve or multidimensional surface, depending on the number of constraints, that can reveal the existence of one or more minima in the potential energy. Doing this for an isotopic chain shows the evolution of the nuclear shape with the nucleon number as well as the coexistence of more than one shapes as additional equilibrium points with similar energy. Examples of this approach can be found in \cite{Lalazissis1999,Niksic2002,Vretenar2005,Niksic2007,Niksic2010,Agbemava2014,Agbemava2016,Cao2020,Kumar2022}.

One of the limitations of the mean-field is that it neglects the dynamical collective correlations of the system. However signatures of shape coexistence are directly imprinted in the spectroscopic properties of the 
low-lying excitation states. Therefore, a theoretical description in the framework of energy density functionals that goes beyond the mean-field level and includes many-body collective correlations is extremely important, in order to overcome this limitation. 

A straightforward extension consists of the generator coordinate method (GCM). A collective wavefunction is built by configuration mixing of a set of wavefunctions determined by one or more collective variables which act as the generator coordinates. In practical applications these generating wavefunctions are solutions of the constrained mean-field equations used for the creation of the potential energy surface. Correlations are included through the restoration of symmetries that are broken in the mean-field level by projecting the generating wavefunctions onto good quantum numbers, related to those symmetries. 
The application of this method extending the relativistic mean-field has been applied in \cite{Niksic2006-1,Niksic2006-2,Yao2009,Yao2010,Hei2016,Yao2011,Marevic2018}. However, extensive calculations with the GCM are computationally extremely demanding especially for heavier nuclei and/or for large number of generating coordinates. 

Alternative methods to the GCM have been developed that are computationally less intensive and can be applied in a broader scale. One approach is the construction of a Bohr-like collective Hamiltonian (CH). The same constrained solutions used in a GCM application, determine now the moments of inertia, mass parameters and collective potential which are the parameters of the Hamiltonian. The subsequent diagonalization results in the excitation spectra and the corresponding collective wavefunctions of the excited states, and allows the calculation of several spectroscopic properties such as energy ratios and transition probabilities. There have been numerous studies of shape coexistence or shape transitions involving the CH within relativistic functionals, that cover specific nuclei, isotopic chains or provide a global analysis of the nuclear 
chart \cite{Niksic2009,ZPLi2009,ZPLi2010,ZPLi2011,ZPLi2012,Xiang2012,Prassa2012, Prassa,Niksic2014,KQLu2015,Quan2017,Shi2019,Yang2021,Prassa2021}.
Finally, a similar approach that uses as a starting point the same set of constrained calculations has been produced by the combination of the mean-field and the interacting boson model (IBM). The microscopic input, like single-particle energies and occupation probabilities, are mapped onto the parameters of an IBM Hamiltonian describing the collective dynamics of the system
\cite{Nomura2008,Nomura2010,Nomura2011}. 
 
Our approach in the present study remains at the mean-field level, but it deviates from the standard way it has been applied for the analysis of shape coexistence. 
In particular, instead of looking for the expectation values of certain quantities, related to macroscopic properties such as the quadrupole moment, we focus on the single-particle energies that come out as solutions of the SCMF. 
More specifically, we begin by solving the relativistic Hartree--Bogoliubov (RHB) equations, which is a framework that unifies the long-range particle-hole (p-h) correlations and the short-range particle-particle (p-p) pairing correlations. This is crucial since the majority of nuclei examined here are open shell nuclei where pairing is important. So the DDME2 relativistic functional \cite{Lalazissis} is employed for the long-range part of the nucleon-nucleon interaction and the Gogny like, finite-range force known as TMR \cite{Tian2009_PLB676-44} is used for the pairing. The RHB equations are solved with the axial version of the DIRHB code \cite{Niksic}, where the single nucleon wavefunctions are expanded in an axially symmetric harmonic oscillator basis. We thus end up with the static ground state which corresponds to the global minimum point of the projected energy surface and is represented as a product state of single-(quasi)particle states. For a more detailed explanation see \cite{Niksic}. 

The resulting single-particle energies and respective occupation probabilities are the basic quantities employed in an effort to provide a microscopic background for the results obtained recently \cite{EPJASC} within the proxy-SU(3) model. Our objective is to demonstrate the appearance of particle-hole excitations through the evolution of the single-particle structure as the number of nucleons change. More specifically, we examine how the relative position of orbitals, that belong to certain spherical shells, changes with respect to the Fermi surface for a series of isotopes or isotones. Orbitals located below the Fermi surface correspond to particle states while orbitals that sit above it correspond to holes. Following a series of isotopes, we observe if there are proton states that change from particles to holes and vice versa. When this happens for a certain range of neutrons we propose that in that section neutron-induced shape coexistence occurs. By analogy, following a series of isotones, we observe if there are neutron states that change from particles to holes and vice versa. When this happens for a certain range of protons we propose that in that section proton-induced shape coexistence occurs.

\section{Neutron-induced shape coexistence}

\subsection{$Z=82$ region}

In Figs.~1-5 CDFT results obtained with the DDME2 functional are shown for the evolution, along the Po, Pb, Hg, Pt ($Z=84$, 82, 80, 78) series of isotopes, of the single-particle energies of the orbitals $3s_{1/2}$, $2d_{3/2}$, $2d_{5/2}$, $1g_{7/2}$, $1h_{11/2}$, forming the 50-82 proton shell, as well as of the orbital $1h_{9/2}$, belonging to the next shell, lying above $Z=82$. 

In Fig.~1 we see that the orbital 11/2[505] of $1h_{11/2}$, which should have lied below the Fermi energy, surfaces above it in the region $N=98$-110, related to the region $N=96$-112 determined in \cite{EPJASC}, thus leaving two holes in the 50-82 shell. 

In. Fig.~2 we see that similar holes in the 50-82 shell are created in the $N=98$-110 region by the orbital 1/2[400] of $3s_{1/2}$, as well as by the orbital 3/2[402] of $2d_{3/2}$ (the latter except in Po).  

In contrast, in Fig.~3 we see that the orbitals  1/2[541] and 3/2[532] of $1h_{9/2}$ are occupied in the $N=98$-110 region, although they should have lied above $Z=82$ (approaching the Fermi surface in the Pt isotopes).

As a consequence, particle-hole excitations are observed in the $N=98$-110 region, formed by particles appearing in Fig.~3 and holes appearing in Figs.~1, 2. 

For the sake of comparison, the $2d_{5/2}$ orbitals are shown in Fig.~4. We see that they stay below the Fermi surface, except in the case of the Pt isotopes, where 5/2[402] starts getting above it, a reasonable result since Pt has fewer protons than Po, Pb, Hg.  

A similar trend is seen in Fig.~5, in which the $1g_{7/2}$ orbitals stay below the Fermi surface, with 7/2[404] gradually approaching the Fermi surface as we move from Po to Pt. 

In conclusion, we see that p-h excitations occur, in which particles come from the 1/2[541] and 3/2[532] orbitals of $1h_{9/2}$, while holes come from the 11/2[505] orbital of $1h_{11/2}$, as well as from the 1/2[400] orbital of $3s_{1/2}$ and  the 3/2[402] orbital of $2d_{3/2}$. Notice in Table~1 that the orbitals $1h_{9/2}$ and $1h_{11/2}$ form a Federman-Pittel pair related to the onset of deformation.  

Results for the Os isotopes indicate that  only the orbital 1/2[541] of $1h_{9/2}$ is occupied, while the orbital 5/2[402] of $2d_{5/2}$ becomes clearly empty, thus producing a weaker effect, which is expected to die out as one moves deeper down into the 50-82 proton shell. 

In the standard Nilsson diagrams, the neutron $1i_{13/2}$ orbital lies at $N=100$-114. As seen in Table~1, the $1i_{13/2}$ orbital forms a Federman-Pittel pair related to strong deformation with the proton $1h_{11/2}$ orbital, which should be full below $Z=82$. Thus strong p-n interaction starts, in which the $1i_{13/2}$ neutron orbital acts as an elevator for the $1h_{11/2}$ proton orbital, managing to elevate the 11/2[505] suborbital of $1h_{11/2}$ above the Fermi surface for $N=98$-110.     
      
The findings of this subsection are corroborated by the results provided for the Hg isotopes within the interacting boson model, based on a Gogny energy density functional \cite{Nomura2013}. Indeed, shape coexistence starts at $N=96$, when a second minimum appears in the potential energy surface (PES), and stops at $N=112$ with the disappearance of the second minimum of the PES. In between, at $98\leq N \leq 110$, a prolate minimum and an oblate one are seen, signaling the occurrence of shape coexistence. A similar calculation has been carried out in the Pb isotopes \cite{Nomura2012}, limited to $100 \leq N \leq 110$. In this case shape coexistence is seen throughout the considered region, in agreement with the present findings. Furthermore,  a similar calculation has been carried out in the Pt isotopes \cite{Nomura2011,Nomura2011b,Nomura2011c}, limited to $112 \leq N \leq 120$. In this case no signs of shape coexistence are seen throughout the considered region, again in agreement with the present findings.
      
Additional evidence corroborating the present results is found in Ref. \cite{Ramos2014}, in which the $^{172-194}_{78}$Pt$_{94-116}$ isotopes have been considered both within the interacting boson model with configuration mixing (IBM-CM) \cite{IAbook}, as well as within a self-consistent Hartree--Fock--Bogoliubov calculation using the Gogny-D1S interaction, with a comparison of the potential energy surfaces (PES) produced by the two different approaches being carried out. Two different coexisting configurations have been found by both approaches in $^{176-186}_{78}$Pt$_{98-108}$, in agreement with the present findings. Similar IBM-CM calculations have been performed in $^{172-200}_{80}$Hg$_{92-120}$, with clear evidence of two types of coexisting configurations found in $^{180-188}_{80}$Hg$_{100-108}$ \cite{Ramos2014b}, in agreement with the present findings. Similar IBM-CM calculations have also been carried out in $^{190-208}_{84}$Po$_{106-124}$ \cite{Ramos2015}, with some signs for shape coexistence seen in the PES of $^{190-194}_{84}$Po$_{106-110}$, while no shape coexistence is seen for $112\leq N \leq 124$, in agreement with the present findings. 

Since the above mentioned calculations  \cite{Ramos2014b} have been produced using the IBM-CM formalism, while in the present work the DDME2 functional is used, the PES for $^{182}_{80}$Hg$_{102}$ resulting from calculations using the DDME2 functional is shown in Fig.~6. We remark that the present PES is very similar to Fig. 18(a) of Ref. \cite{Ramos2014b}, exhibiting a clear minimum along the $\gamma=0$ axis and another clear minimum along the $\gamma=\pi /3$ axis. In the present figure the first minimum occurs at $\beta\approx 0.30$, while the second one is located at $\beta\approx 0.15$, thus offering a clear example of shape coexistence of two bands drastically different in both the $\beta$ and $\gamma$ collective degrees of freedom. Further comparisons of this type can be pursued in subsequent work.  

Further corroboration of the present results comes from Fig. 3 of Ref. \cite{Nomura2013}, in which the proton and neutron single particle levels of  $^{182}_{80}$Hg$_{102}$ resulting from a self-consistent mean-field calculation employing the  Gogny energy density functional are shown. Since $^{182}_{80}$Hg$_{102}$ lies in the middle of the parabolic region characteristic of shape coexistence in the Hg isotopes, as seen in Fig. 10 of Ref. \cite{Heyde}, it makes a textbook example of a nucleus exhibiting shape coexistence in the $Z=82$ region. The deformation of $^{182}_{80}$Hg$_{102}$ within the D1S Gogny interaction is estimated to be 0.322 \cite{Delaroche}. At this deformation we see that the proton 1/2[541] and 3/2[532] orbitals (the lowest two orbitals of $1h_{9/2}$) do dive below the Fermi energy in Fig. 3 of Ref. \cite{Nomura2013}, while the 1/2[400] orbital (the $3s_{1/2}$ orbital) and the 3/2[402] orbital (the highest $2d_{3/2}$ orbital) do jump above the Fermi energy, and the 11/2[505] orbital (the highest $1h_{11/2}$ orbital) lies in the close vicinity of the Fermi energy, thus corroborating the proton particle-hole picture summarized in Table II. In relation to the proton-neutron interaction in the framework of the Federman--Pittel mechanism, we see in Fig. 3 of Ref. \cite{Nomura2013} that the proton suborbitals of $1h_{11/2}$ (1/2[550], 3/2[541], 5/2[532], 7/2[523], 9/2[514]) lie below the Fermi surface (except the highest one, 11/2[505], which lies at its vicinity, as already mentioned), while in parallel the lowest four neutron suborbitals (1/2[660], 3/2[651], 5/2[642], 7/2[633]) of $1i_{13/2}$ lie below the Fermi surface, thus allowing for strong proton-neutron interaction to develop, which acts as elevator of the $1h_{11/2}$ proton suborbitals. These observations are also in full agreement with the fact that proton-neutron interaction is maximized \cite{Cakirli,Karampagia} between proton and neutron orbitals differing by $\Delta K [\Delta N \Delta n_z \Delta \Lambda] = 0[110]$ in the usual Nilsson notation \cite{NR}. Indeed, the occupied proton orbitals 1/2[550], 3/2[541], 5/2[532], 7/2[523] form 0[110] pairs with the occupied neutron orbitals 1/2[660], 3/2[651], 5/2[642], 7/2[633] respectively. As emphasized in Refs. \cite{proxy1,EPJASM},  the similarity between orbitals differing by 0[110] is the key feature lying behind the validity of the proxy-SU(3) scheme. 

\subsection{$Z=50$ region}

In Fig.~7 the Te ($Z=52$) isotopes are considered. 2p-2h proton excitations are seen for $N=64$-68. The orbital 9/2[404] of $1g_{9/2}$ (a) (normally lying below $Z=50$) is vacant, while the orbital 3/2[422] of $1g_{7/2}$ (b) (normally lying above $Z=50$) is occupied. Notice in Table~1 that the orbitals $1g_{7/2}$ and $1g_{9/2}$ form a Federman-Pittel pair related to the onset of deformation.  

In view of the above findings a small island of shape coexistence is expected to be formed by $^{116,118,120}$Te. 
 
In the standard Nilsson diagrams, the neutron $1h_{11/2}$ orbital lies at $N=66$-78. As seen in Table~1, the $1h_{11/2}$ orbital forms a Federman-Pittel pair related to strong deformation with the proton $1g_{9/2}$ orbital, which should be full below $Z=50$. Thus strong p-n interaction starts, in which the $1h_{11/2}$ neutron orbital acts as an elevator for the $1g_{9/2}$ proton orbital, managing to elevate the 9/2[404] suborbital of $1g_{9/2}$ above the Fermi surface for $N=64$-68.   

The findings of this subsection are corroborated by the results provided for $^{104-144}_{52}$Te$_{52-92}$ by density dependent point coupling and meson exchange effective interactions within the relativistic 
Hartree--Bogoliubov approach \cite{Sharma2019}. Potential energy surfaces provided by this method exhibit shape coexistence in $^{116-120}_{52}$Te$_{64-68}$ \cite{Sharma2019}, in agreement with the present findings. 
 
\subsection{$Z=40$ region}
 
In Fig.~8 the Zr ($Z=40$) isotopes are shown. 6p-6h proton excitations are seen for $N=38$-40. The orbitals 1/2[301] of $2p_{1/2}$, as well as 3/2[301], 5/2[303] of $1f_{5/2}$ (a) (normally lying below $Z=40$) are vacant, while the orbitals 1/2[440], 3/2[431], 5/2[422] of $1g_{9/2}$   (b) (normally lying above $Z=40$) are occupied.
 
In Fig.~9 the Sr ($Z=38$) isotopes are shown. 4p-4h proton excitations are seen for $N=38$-40. The orbitals 3/2[301], 5/2[303] of $1f_{5/2}$ (a) (normally lying below $Z=40$) are vacant, while the orbitals 1/2[440], 3/2[431] of $1g_{9/2}$   (b) (normally lying above $Z=40$) are occupied.

The net outcome of these figures is that a small island of shape coexistence formed by $^{78}$Sr, $^{78}$Zr and $^{80}$Zr  is expected.

These results are corroborated by several mean field calculations. Potential energy surfaces obtained by covariant density functional theory calculations using the PC-PK1 parameter set for $^{88-106}_{38}$Sr$_{50-68}$ and $^{90-108}_{40}$Zr$_{50-68}$ have singled out $^{98}_{38}$Sr$_{60}$ and $^{100}_{40}$Zr$_{60}$ as examples of coexistence of prolate and oblate shapes \cite{Xiang2012}. Shape coexistence in $^{80}_{40}$Zr$_{40}$ has been found through beyond mean field methods using the Gogny D1S interaction \cite{Rodriguez2011}. Total-Ruthian-surface calculations have found shape coexistence in $^{76-80}_{38}$Sr$_{38-42}$ and $^{78-82}_{40}$Zr$_{38-42}$ \cite{Zheng2014}, pointing out $^{80}_{38}$Sr$_{42}$ and $^{82}_{40}$Zr$_{42}$ as the best examples, in rough agreement with the present results.  

Further discussion on this region will appear in subsec.~IV.C. 

The results of the present section are summarized in Table~II. 

\section{Proton-induced shape coexistence}

\subsection{$N=90$ region}

In the $N=90$ region a shape phase transition \cite{Cejnar} from spherical to deformed shapes is known to occur, characterized by the X(5) critical point symmetry \cite{IacX5}, of which the $N=90$ isotones $^{150}$Nd, $^{152}$Sm and $^{154}$Gd ($Z=60$-64) are known to be the best experimental manifestations \cite{McCutchan}. We are going to show that the orbitals responsible for the rapid increase of the proton-neutron interaction, causing the shape phase transition, exhibit particle-hole excitations which provide a possible mechanism for the appearance of shape coexistence in this region. 

The particle-hole excitations seen in this region are taking place not across a major shell gap, as in Sec.~II, but across the $N=112$ magic number of the 3-dimensional isotropic harmonic oscillator (3D-HO). The following aspects should be stressed.

1) The usual shell model magic numbers 28, 50, 82, 126, 184 are valid at zero deformation. As deformation sets in, they are destroyed, as seen in the standard Nilsson diagrams \cite{NR}, since the various orbitals move up or down, depending on their individual values of quantum numbers.  

2) What is not destroyed by the deformation, is the distinction between positive parity and negative parity orbitals. In the case of the 82-126 neutron shell, the orbitals present are $3p_{1/2}$, $3p_{3/2}$, $2f_{5/2}$, $2f_{7/2}$, $1h_{9/2}$, possessing negative parity, and the orbital $1i_{13/2}$, possessing positive parity. The former lie within the pfh shell of the 3D-HO, extending over 70-112 particles, while the latter lies in the sdgi shell of the 3D-HO, extending over 112-168 particles. 

In Figs.~10-12 the relevant orbitals of the 82-126 neutron shell are shown. 

In Fig.~10(a), the neutron orbital 1/2[660] of $1i_{13/2}$ is seen to fall below the Fermi surface for $N=90$ at $Z=60$-64, thus becoming a candidate for being the particle partner in p-h excitations.  In Fig.~10(b) the orbital 1/2[660] is sunk lower in energy for $N=92$, since two more neutrons have been added, but in its turn the 3/2[651] orbital is seen to fall below the Fermi surface at $Z=60$-64, thus becoming a candidate for being the particle partner in p-h excitations. In Figs.~10(c), (d) it is seen that at $N=94$, 96 both orbitals are always occupied, because of the additional neutrons added. In the standard Nilsson diagrams \cite{NR} one sees that both orbitals are rapidly falling with increasing prolate deformation. They both belong to the $1i_{13/2}$ orbital, which plays a crucial role in the creation of deformation through the Federman-Pittel mechanism, as seen in Table~I. From $N=94$ onwards, a large number of $1i_{13/2}$ suborbitals lies below the Fermi surface, thus being able to participate in building up deformation, therefore no isolated p-h excitations can be seen anymore. The 1/2[660] and 3/2[651] orbitals contain  the particles which are candidates for participating in the formation of p-h excitations across the $N=112$ 3D-HO magic number. Their possible hole partners will be searched for in the next paragraph. 

In Fig.~11 we see that the neutron orbital 5/2[523] of $2f_{7/2}$ is empty at $N=90$, 92 for $Z=60$-64, thus providing hole candidates for participation in p-h excitations. At $N=94$ the orbital 5/2[523] has already reached the Fermi surface, while it sinks below the Fermi energy at $N=96$, because of the additional neutrons added. The orbital 5/2[523] therefore provides hole candidates for the formation of p-h excitations for $N=90$, 92. In the standard Nilsson diagrams \cite{NR} one sees that the 5/2[523] orbital is keeping its energy, falling only slowly with increasing prolate deformation. 

The net result so far is that in the $N=90$ isotones $^{150}$Nd, $^{152}$Sm, and $^{154}$Gd, p-h excitations across the $N=112$ magic number of the 3D-HO are formed by particles in the 1/2[660] orbital and holes in the 5/2[523] orbital, while in the $N=92$ isotones $^{152}$Nd, $^{154}$Sm, and $^{156}$Gd, p-h excitations across the $N=112$ magic number of the 3D-HO are formed by particles in the 3/2[651] orbital and holes in the 5/2[523] orbital. These 6 nuclei are expected to form an island of shape coexistence, which we characterize as proton-induced, since it is due to neutron p-h excitations, caused by the protons which act as elevators of the neutrons through the proton-neutron interaction.

In order to clarify the origin of the development of proton-neutron interaction, one can consider Table~I. The protons responsible for p-n interaction in the rare earths region are the ones occupying the $1h_{11/2}$ orbital, lying in the proton Nilsson diagram in the region $Z=64$-76. These could interact with neutrons in the $1i_{13/2}$ orbital, lying at $N=100$-114, thus expected to be empty in the nuclei with $N=90$-96 considered here, or with neutrons in the $1h_{9/2}$ orbital, lying at $N=90$-100, which is expected to start getting filled in the $N=90$-96 region considered here.  In order to check this, the orbital $1h_{9/2}$ is shown in Fig.~12. We see that the neutron orbital   3/2[521] of $1h_{9/2}$, moderately falling in the Nilsson diagrams, is mostly empty for $N=90$, sinking lower and lower in relation to the Fermi surface with N increasing to 92 and 94, and finally sinking entirely below the Fermi surface at $N=96$. It is therefore corroborated, as expected from Table~I, that $1h_{11/2}$ protons together with $1h_{9/2}$ neutrons are responsible for the onset of deformation in this region.  

This conclusion is  corroborated by  the detailed results provided by relativistic energy-density functionals reported in Ref. \cite{ZPLi2009}. Indeed, 
 in Fig. 9 of Ref. \cite{ZPLi2009}, one can see that in $^{150}_{60}$Nd$_{90}$ both the neutron $1h_{9/2}$ orbital and the proton $1h_{11/2}$ orbital dive below the Fermi surface at moderate deformations, thus being able to trigger the onset of deformation according to the Federman--Pittel mechanism reported in Table I.  

The findings of this subsection are further corroborated by the detailed results provided by relativistic energy-density functionals in Ref. \cite{ZPLi2009}, in Fig. 8 of which neutron single-particle levels are reported as a function of the quadrupole deformation $\beta$ for $^{146-152}_{60}$Nd$_{86-92}$. It is seen there that the orbitals 1/2[660] and 3/2[651] (the lowest two orbitals of $1i_{13/2}$) do dive below the Fermi surface for $N=90$, 92 within a wide range of $\beta$, in agreement to the present Fig. 10, while they do not do so for $N=86$, 88. Furthermore, the 5/2[523] and 7/2[514] orbitals (the top two orbitals of $2f_{7/2}$) do remain above the Fermi surface for any $\beta$ for $N=90$, 92, in agreement to the present Fig. 11. 

The rapid increase of the quadrupole deformation with increasing neutron number in the Nd, Sm, and Gd series of isotopes in the $N=90$ region  within the proxy-SU(3) approach can be seen in Fig. 6 of Ref. \cite{BJP44}, as well as in Figs. 5 and 6 of Ref. \cite{HINP2017}, in which the parameter-free proxy-SU(3) predictions are compared to the experimental values, as well as to alternative theoretical predictions, with very good  agreement observed. Further investigation of the possible connection between the present particle-hole mechanism and the appearance of shape/phase transitions is in place.     

\subsection{$N=60$ region} 

The $Z=40$, $N=60$ region is known \cite{Garrett} to be very similar to the $Z=64$, $N=90$ region considered in the previous subsection. Therefore a similar study is attempted in this region, in which the relevant 3D-HO magic number across which p-h excitations are to be expected is $N=70$. 

The $N=60$ isotones are considered in Fig.~13. 4p-4h proton excitations are seen for $Z=40$-42. The orbitals 1/2[411] of $3s_{1/2}$ and 5/2[413] of $2d_{5/2}$ (a) (normally lying below $N=70$) are vacant for $Z=40$, 42, while the orbitals 1/2[550] and 3/2[541] of $1h_{11/2}$ (b) (normally lying above $N=70$) are occupied. Therefore shape coexistence is expected to occur in $^{100}$Zr ($Z=40$) and $^{102}$Mo ($Z=42$).  

Furthermore, the $N=58$ isotones are considered in Fig.~14. 4p-4h proton excitations are seen for $Z=40$-42. The orbitals 1/2[411] of $3s_{1/2}$ and 5/2[413] of $2d_{5/2}$ (a) (normally lying below $N=70$) are vacant everywhere, while the orbital 1/2[550] of $1h_{11/2}$ (b) (normally lying above $N=70$) is occupied. Therefore 2p-2h excitations are seen for $N=58$, $Z=40$, i.e. in $^{98}$Zr. 

In view of the above findings a small island of shape coexistence is expected to be formed by $^{98}$Zr, $^{100}$Zr and $^{102}$Mo. 

In order to clarify the origin of the development of proton-neutron interaction, one can consider Table~I. The protons responsible for p-n interaction in the relevant intermediate region are the ones occupying the $1g_{9/2}$ orbital, lying in the proton Nilsson diagram in the region $Z=40$-50. These could interact with neutrons in the $1h_{11/2}$ orbital, lying at $N=66$-78, thus expected to be empty in the nuclei with $N=60$ considered here, or with neutrons in the $1g_{7/2}$ orbital, lying at $N=56$-64, which is expected to be partially filled in the $N=60$ region considered here. Figs.~13, 14 suggest that the pairs of $1g_{9/2}$ protons with $1g_{7/2}$ neutrons are the ones responsible for the onset of deformation in this region.  

The findings of this subsection are in qualitative agreement with the results provided in the $A\approx 100$ region within the interacting boson model, mapped on the Gogny energy density functional \cite{Nomura2016}. Indeed, pronounced competition between prolate and oblate minima, which is a fingerprint of shape coexistence, is found in the deformation-energy surfaces (see Figs. 2 and 3 of \cite{Nomura2016}) for the Zr isotopes with $60\leq N \leq 64$. Shape coexistence has also  been seen experimentally in $^{102}_{42}$Mo$_{60}$ \cite{Esmaylzadeh}, in agreement to the present findings, although the PES produced by the previous IBM plus Gogny method does not show shape coexistence \cite{Esmaylzadeh} in this particular case. Furthermore, potential energy surfaces for $^{94-110}_{40}$Zr$_{54-70}$ have been calculated recently \cite{Ramos} within the interacting boson model with configuration mixing (IBM-CM) \cite{IAbook}, showing in $^{100}_{40}$Zr$_{60}$ two coexisting minima with almost equal energies, i.e. a clear case of shape coexistence, in agreement with the present findings. In addition, shape coexistence in $^{98}_{40}$Zr$_{58}$ has been predicted within the VAMPIR approach \cite{Petrovici}, while shape coexistence in $^{100}_{40}$Zr$_{60}$ has been suggested through configuration mixing \cite{Wu2003}. 

\subsection{$N=40$ region}

The  $N=40$ isotones are considered in Fig.~15. 6p-6h proton excitations are seen for $Z=38$-40. The orbitals 5/2[303], 3/2[301] of $1f_{5/2}$ and 1/2[301] of $2p_{1/2}$ (a) (normally lying below $N=40$) are vacant for $Z=40$, while the orbitals 1/2[440], 3/2[431], 5/2[422] of $1g_{9/2}$ (b) (normally lying above $N=40$) are occupied.

Furthermore, the $N=38$ isotones are shown in Fig.~16. 4p-4h proton excitations are seen for $Z=40$. The orbitals 5/2[303], 3/2[301] of $1f_{5/2}$ (a) (normally lying below $N=40$) are vacant for $Z=40$, while the orbitals 1/2[440], 3/2[431] of $1g_{9/2}$ (b) (normally lying above $N=40$) are occupied.

The net outcome of these figures is that a small island of shape coexistence formed by $^{78}$Sr, $^{78}$Zr and $^{80}$Zr  is expected. In other words, the result of subsec.~III.C is obtained again. 

These results are corroborated by several mean field calculations, as already described in detail in subsec. III.C. 

The results of the present section are summarized in Table~III. The islands of shape coexistence found in both Sections III and IV are shown in Fig.~17. 

\section{Conclusions}

In this paper we have used standard covariant density functional theory (using the DDME2 functional) calculations of the single-particle energy levels of a wide range of even-even nuclei throughout the nuclear chart between $N=Z=38$ up to the beginning of the actinides region. Our main findings are

1) In the $Z=82$ and $Z=50$ regions we have found proton particle-hole (p-h) excitations across the proton magic numbers 82 and 50 respectively, in nuclei in which shape coexistence is known to occur. We have attributed these p-h excitations to the influence of neutrons, through a relatively strong proton-neutron (p-n) interaction, known to be related to the onset of deformation, calling this type of shape coexistence a neutron-induced one. As a consequence, neutron-induced shape coexistence does not occur along the whole isotopic chains, but is limited in regions around the neutron midshell. The p-h interpretation of shape coexistence (SC) in these regions is well accepted in the litearure. The existence of SC borders on the isotopic chains is a novel result, in agreement with Ref.~\cite{EPJASC}, based on algebraic arguments. 

2) In the $N=90$ and $N=60$ regions we have found neutron particle-hole (p-h) excitations across the harmonic oscillator neutron magic numbers 70 and 40 respectively, in nuclei in which shape coexistence is known to occur, with the mechanism causing it being still unclear. In this case we have attributed the p-h excitations to the influence of protons, through a relatively strong proton-neutron (p-n) interaction, known to be related to the onset of deformation, calling this type of shape coexistence a proton-induced one. Proton-induced shape coexistence does not occur along the whole isotonic chains, but is limited in regions around the proton midshell. The p-h interpretation of shape coexistence (SC) in these regions is a novel one, attributed to the beginning of participation of the intruder levels of opposite parity to the onset of deformation within the 50-82 and 28-50 proton shells respectively. The existence of SC borders on the isotonic chains is a novel result, in agreement with Ref.~\cite{EPJASC}, based on algebraic arguments. 

3) In the $Z=40$, $N=40$ region both proton p-h and neutron p-h excitations are observed. The fact that in this case protons and neutrons occupy the same shells and the possible influnce of the Wigner SU(4) supermultiplet should be taken into account \cite{Sahu}.  

4) Neutron-induced SC, in which neutrons are acting as elevators for the protons, is more robust, exhibited over wide regions along isotopic chains, while proton-induced SC, in which the protons are acting as elevators of the neutrons, is weaker, limited in rather narrow regions along isotonic chains. Since both are attributed to the p-n interaction, and the number of neutrons is usually much larger than the number of protons, this difference appears plausible. 

Shape coexistence and/or p-h excitations might occur in other regions of the nuclear chart, based on different mechanisms. The present method is confined within regions in which the deformation created by the p-n interaction  prevails. 

It would be interesting to perform calculations using different relativistic or non-relativistic interactions, in order to check the robustness of the results presented in this article.

\section*{Acknowledgements} 

Support by the Tenure Track Pilot Programme of the Croatian Science Foundation and the Ecole Polytechnique F\'{e}d\'{e}rale de Lausanne, the Project TTP-2018-07-3554 Exotic Nuclear Structure and Dynamics with funds of the Croatian-Swiss Research Programme, 
as well as by the Bulgarian National Science Fund (BNSF) under Contract No. KP-06-N48/1  is gratefully acknowledged.
Results presented in this work have been produced using the Aristotle University of Thessaloniki (AUTH) computational infrastructure and resources. The authors would like to acknowledge the support provided by the IT Center of AUTH throughout the progress of this research work. 


\begin{table}[htb]
\centering
\caption{Proton-neutron pairs of orbitals playing a leading role in the development of deformation in different mass regions of the nuclear chart according to Federman and Pittel \cite{FP1,FP2,FP3}. The pairs on the left part of the table contribute in the beginning of the relevant shell, while the pairs on the right become important further within the shell. Adapted from Ref.~\cite{HINP21SM}.} \label{FPpairs}
\begin{tabular}{ r r r r r   }
\hline\noalign{\smallskip}
 &  protons  & neutrons &      protons  &     neutrons  \\ 
\noalign{\smallskip}\hline\noalign{\smallskip}
light       & $1d_{5/2}$     & $1d_{3/2}$    &    $1d_{5/2}$      &  $1f_{7/2}$        \\
intermediate& $1g_{9/2}$     & $1g_{7/2}$    &    $1g_{9/2}$      &  $1h_{11/2}$       \\
rare earths & $1h_{11/2}$    & $1h_{9/2}$    &    $1h_{11/2}$     &  $1i_{13/2}$       \\
actinides   & $1i_{13/2}$    & $1i_{11/2}$   &    $1i_{13/2}$     &  $1j_{15/2}$       \\

\noalign{\smallskip}\hline
\end{tabular}
\end{table}

\begin{table*}

\caption{Proton single-particle energy levels participating in proton particle-hole formation in various isotopes across different regions of the nuclear chart.  Since the proton particle-hole excitations are caused by the neutrons, we say that neutron-induced shape coexistence is expected in these isotopes. See Section III for further discussion. 
}

\bigskip

\begin{tabular}{ l  l  l  }

\hline
nuclei                           & occupied $Z>40$      & vacant $Z<40$ \\ 
\hline 
\medskip
$^{78}_{38}$Sr$_{40}$          &  1/2[440] 3/2[431]          &   3/2[301] 5/2[303]     \\ \medskip
$^{78,80}_{40}$Zr$_{38,40}$    &  1/2[440] 3/2[431] 5/2[422] &   1/2[301]  3/2[301] 5/2[303]     \\ 

\hline
nuclei                          & occupied $Z>50$      & vacant $Z<50$ \\ 
\hline 
\medskip
$^{116-120}_{52}$Te$_{64-68}$   &      3/2[422]     &    9/2[404]   \\ 

\hline
nuclei                           & occupied $Z>82$      & vacant $Z<82$ \\ 
\hline 
\medskip

$^{176,188}_{78}$Pt$_{98,110}$   & 1/2[541]             &                1/2[400]  3/2[402] \\ \medskip
$^{178-186}_{78}$Pt$_{100-108}$  & 1/2[541] 3/2[532]   &                 1/2[400]  3/2[402]    \\ \medskip
$^{176}_{80}$Hg$_{96}$           & 1/2[541]             &  11/2[505]     1/2[400]  3/2[402]  \\ \medskip
$^{178-190}_{80}$Hg$_{98-110}$   & 1/2[541] 3/2[532]   &   11/2[505]     1/2[400]  3/2[402] \\ \medskip
$^{180-190}_{82}$Pb$_{98-108}$   & 1/2[541] 3/2[532]   &   11/2[505]     1/2[400]  3/2[402]  \\ \medskip 
$^{182-192}_{84}$Po$_{98-108}$   & 1/2[541] 3/2[532]   &   11/2[505]     1/2[400]              \\ 
\hline

\end{tabular}
\end{table*} 

\begin{table*}

\caption{Neutron single-particle energy levels participating in neutron particle--hole formation in various isotones across different regions of the nuclear chart.  Since the neutron particle-hole excitations are caused by the protons, we say that proton-induced shape coexistence is expected in these isotones. See Section IV for further discussion. 
}

\bigskip

\begin{tabular}{ l  l  l  }

\hline
nuclei                           & occupied $N>40$      & vacant $N<40$ \\ 
\hline 
\medskip
$^{78}_{40}$Zr$_{38}$                            &  1/2[440] 3/2[431]          &    3/2[301]  5/2[303]            \\ \medskip
$^{78}_{38}$Sr$_{40}$, $^{80}_{40}$Zr$_{40}$     &  1/2[440] 3/2[431] 5/2[422] &  1/2[301]  3/2[301] 5/2[303]     \\

\hline
nuclei                           & occupied $N>70$      & vacant $N<70$ \\ 
\hline 
\medskip
$^{98}_{40}$Zr$_{58}$                              &  1/2[550]           & 1/2[411]  5/2[413]       \\ \medskip
$^{100}_{40}$Zr$_{60}$, $^{102}_{42}$Mo$_{60}$     &  1/2[550] 3/2[541]  & 1/2[411]  5/2[413]       \\

\hline
nuclei                           & occupied $N>112$      & vacant $N<112$ \\ 
\hline 
\medskip

$^{150}_{60}$Nd$_{90}$,  $^{152}_{62}$Sm$_{90}$, $^{154}_{64}$Gd$_{90}$ & 1/2[660] &   5/2[523]    \\ \medskip
$^{152}_{60}$Nd$_{92}$,  $^{154}_{62}$Sm$_{92}$, $^{156}_{64}$Gd$_{92}$ & 3/2[651] &   5/2[523]    \\ \medskip

\end{tabular}
\end{table*} 



\begin{figure*}[htb]

{\includegraphics[width=75mm]{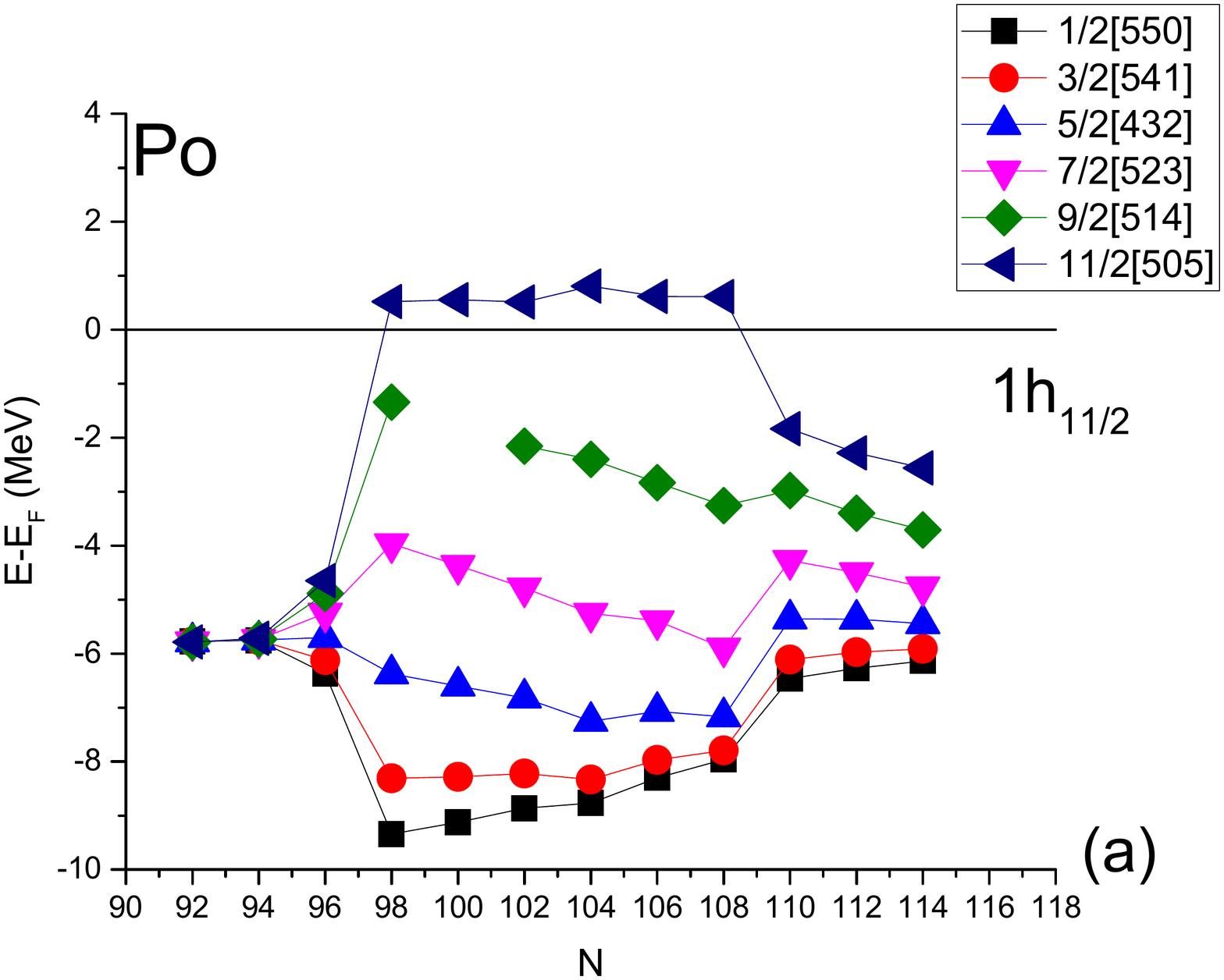}\hspace{5mm}
\includegraphics[width=75mm]{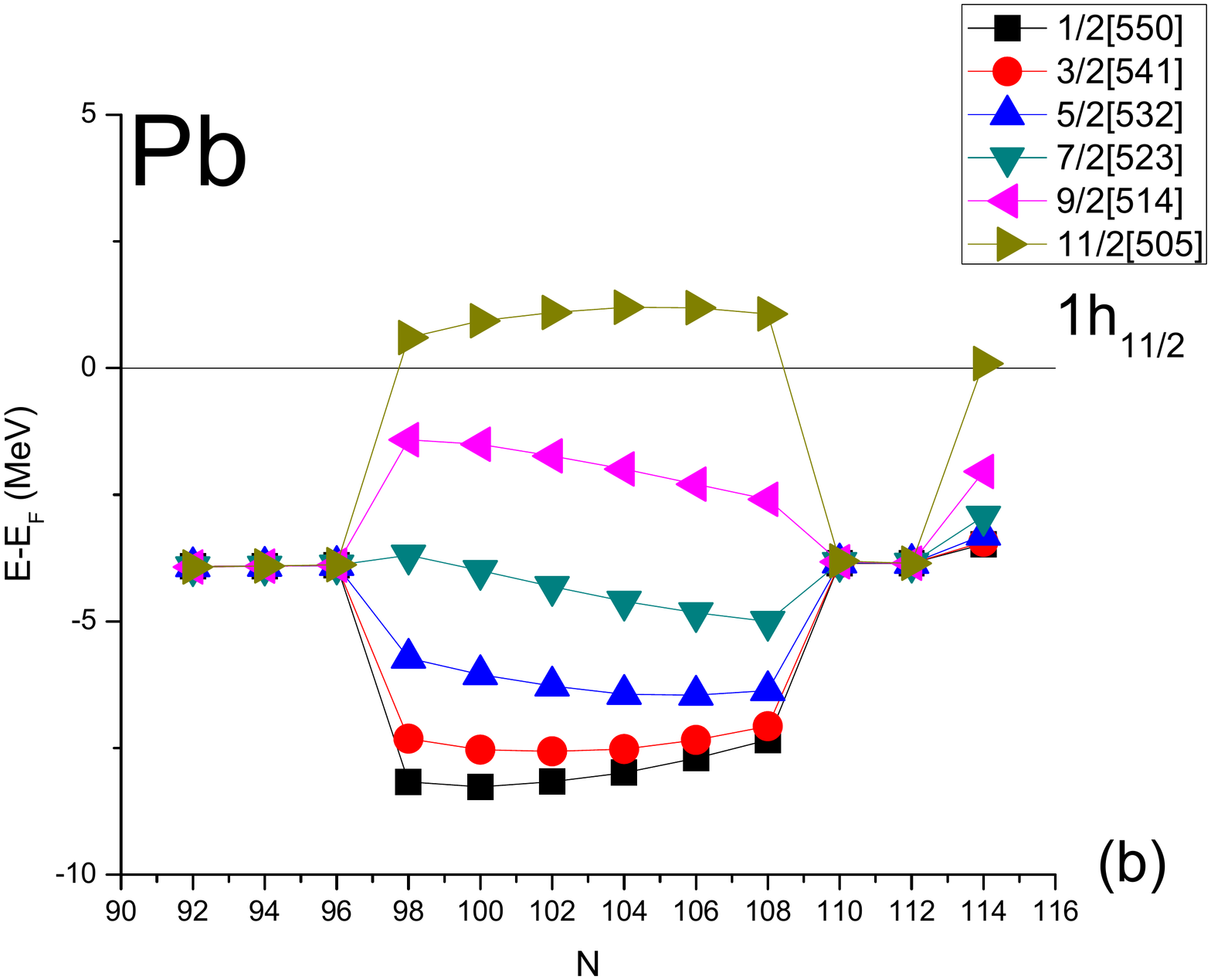}\hspace{5mm}}
{\includegraphics[width=75mm]{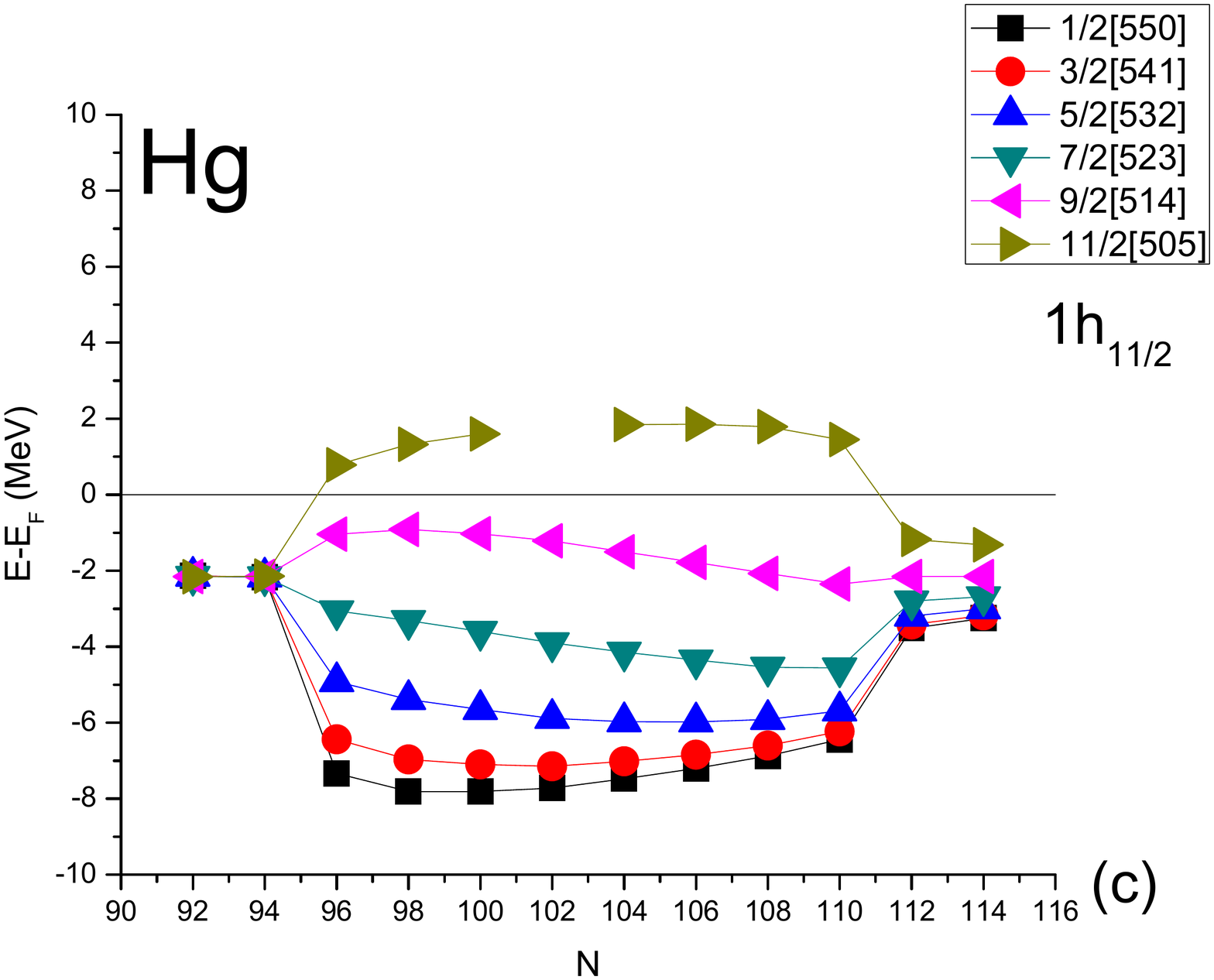}\hspace{5mm}
 \includegraphics[width=75mm]{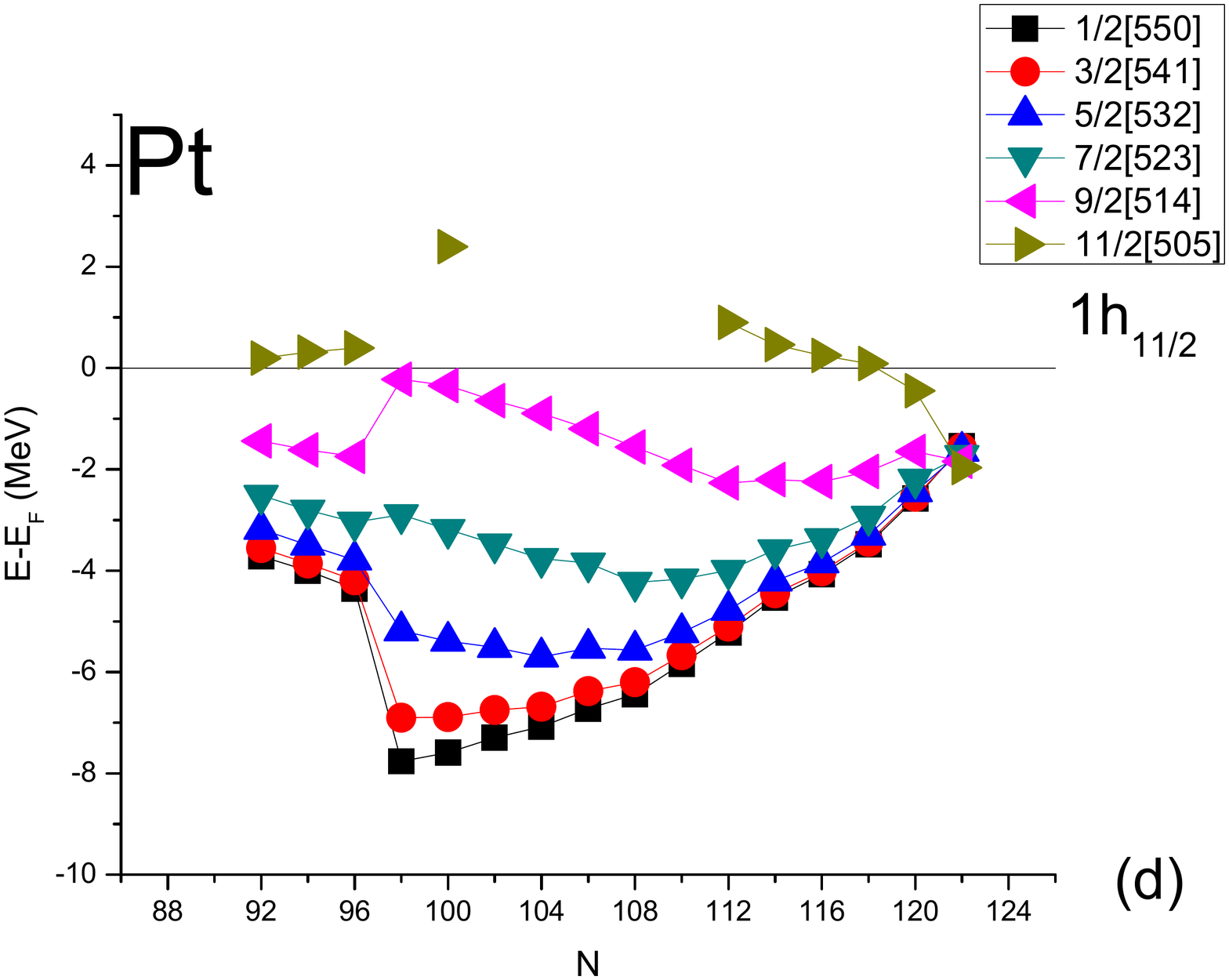}\hspace{5mm}}

\caption{Energies (in MeV) of $1h_{11/2}$ single-particle proton orbitals relative to the Fermi energy obtained by CDFT for $Z=78$-84 isotopes. For $N\approx 98$-110 the orbital 11/2[505] (normally lying below $Z=82$) is vacant, thus creating two holes. Panel (a) adapted from Ref.~\cite{PLB}. In this figure, as well as in all subsequent ones, the spherical shell model quantum labels $nl_j$ \cite{Mayer} are shown in order to indicate the levels from which the reported deformed Nilsson orbitals $K [N n_z \Lambda]$ \cite{NR} come from. 
See Section III for further discussion. 
} 

\end{figure*}


\begin{figure*}[htb]

{\includegraphics[width=75mm]{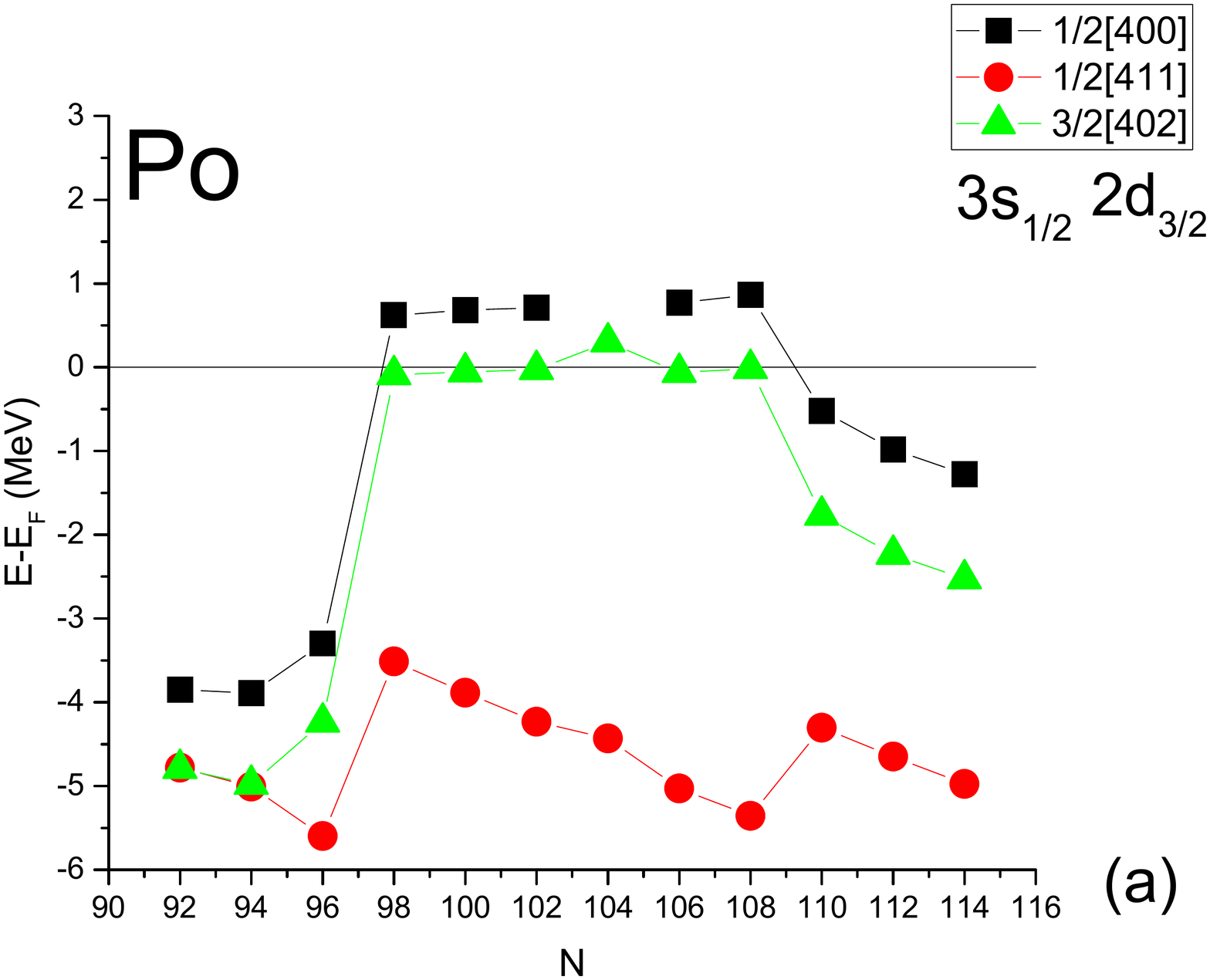}\hspace{5mm}
\includegraphics[width=75mm]{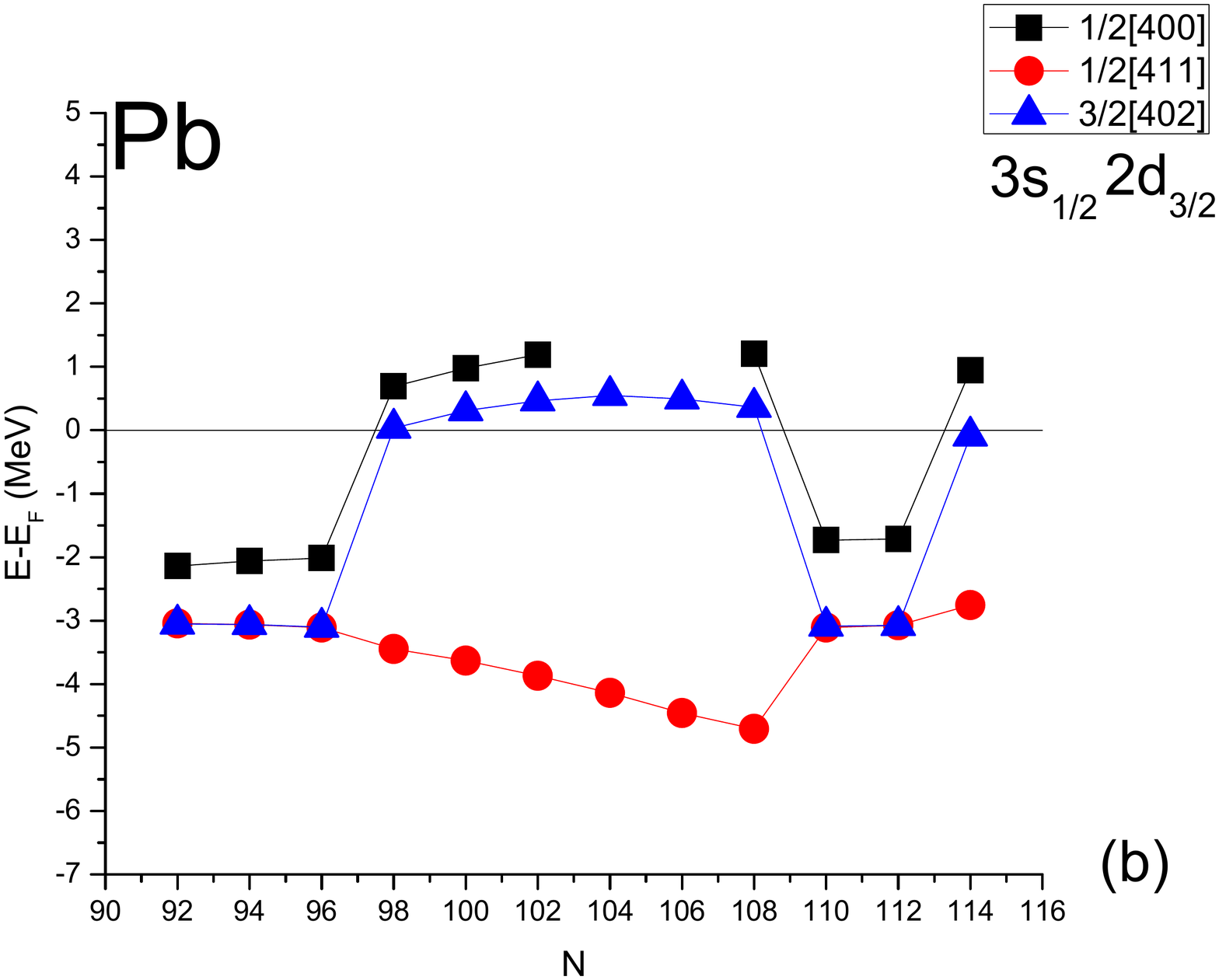}\hspace{5mm}}
{\includegraphics[width=75mm]{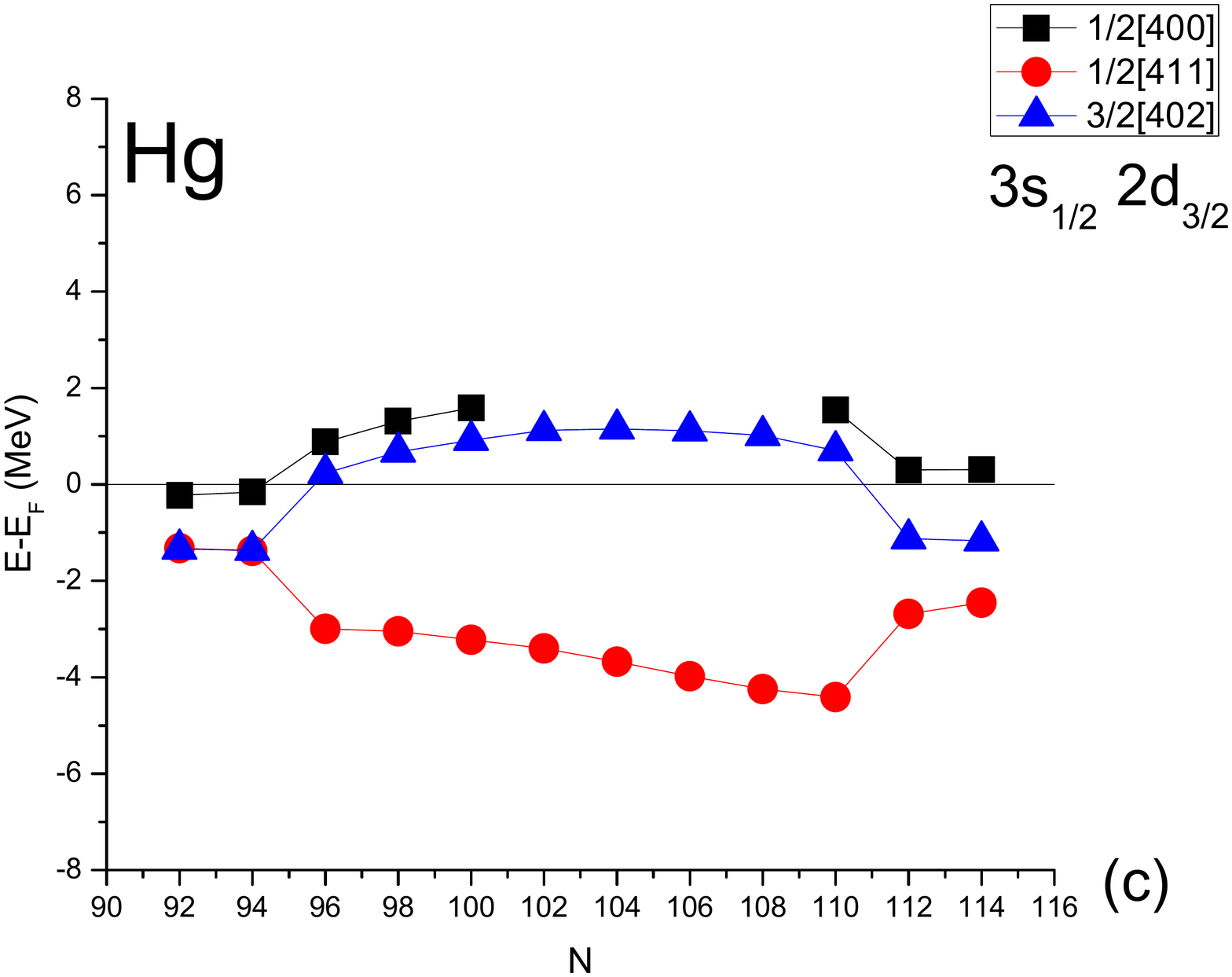}\hspace{5mm}
 \includegraphics[width=75mm]{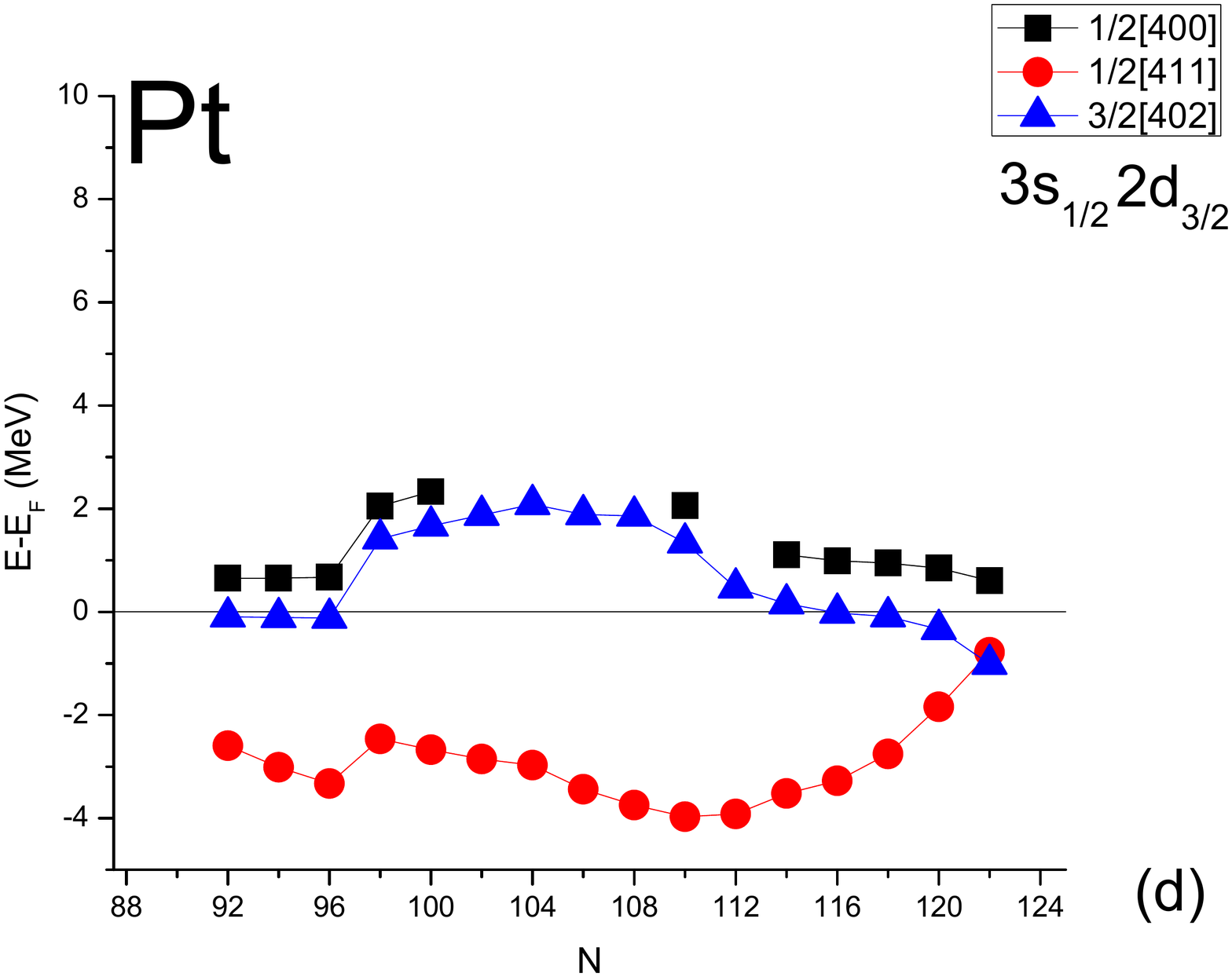}\hspace{5mm}}

\caption{Energies (in MeV) of $3s_{1/2}$ and $2d_{3/2}$ single-particle proton orbitals relative to the Fermi energy obtained by CDFT for $Z=78$-84 isotopes. For $N\approx 98$-110 the orbitals 1/2[400] and 3/2[402] (normally lying below $Z=82$) are vacant, thus creating four holes. Panel (a) adapted from Ref.~\cite{PLB}. See Section III for further discussion. 
} 

\end{figure*}


\begin{figure*}[htb]

{\includegraphics[width=75mm]{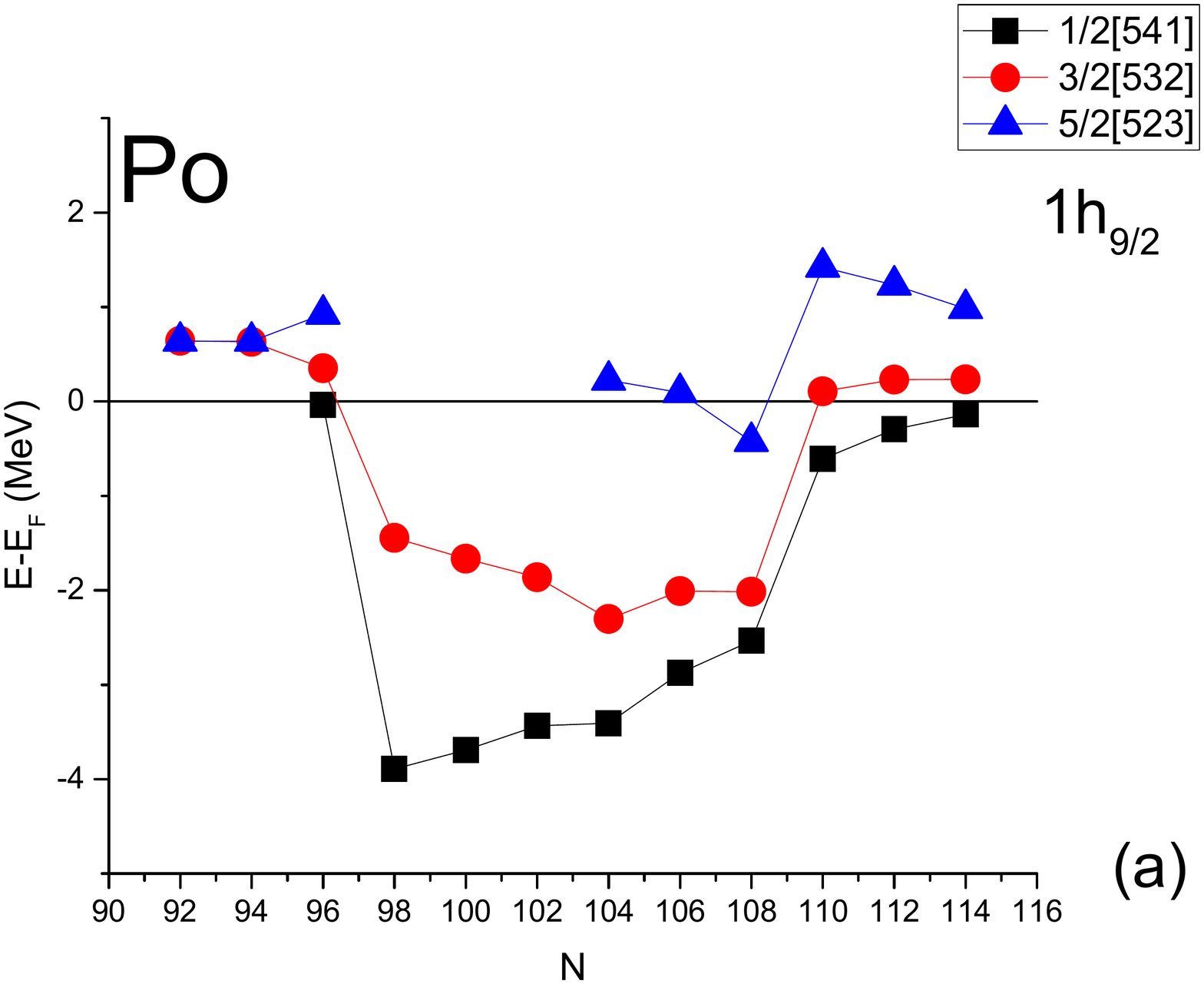}\hspace{5mm}
\includegraphics[width=75mm]{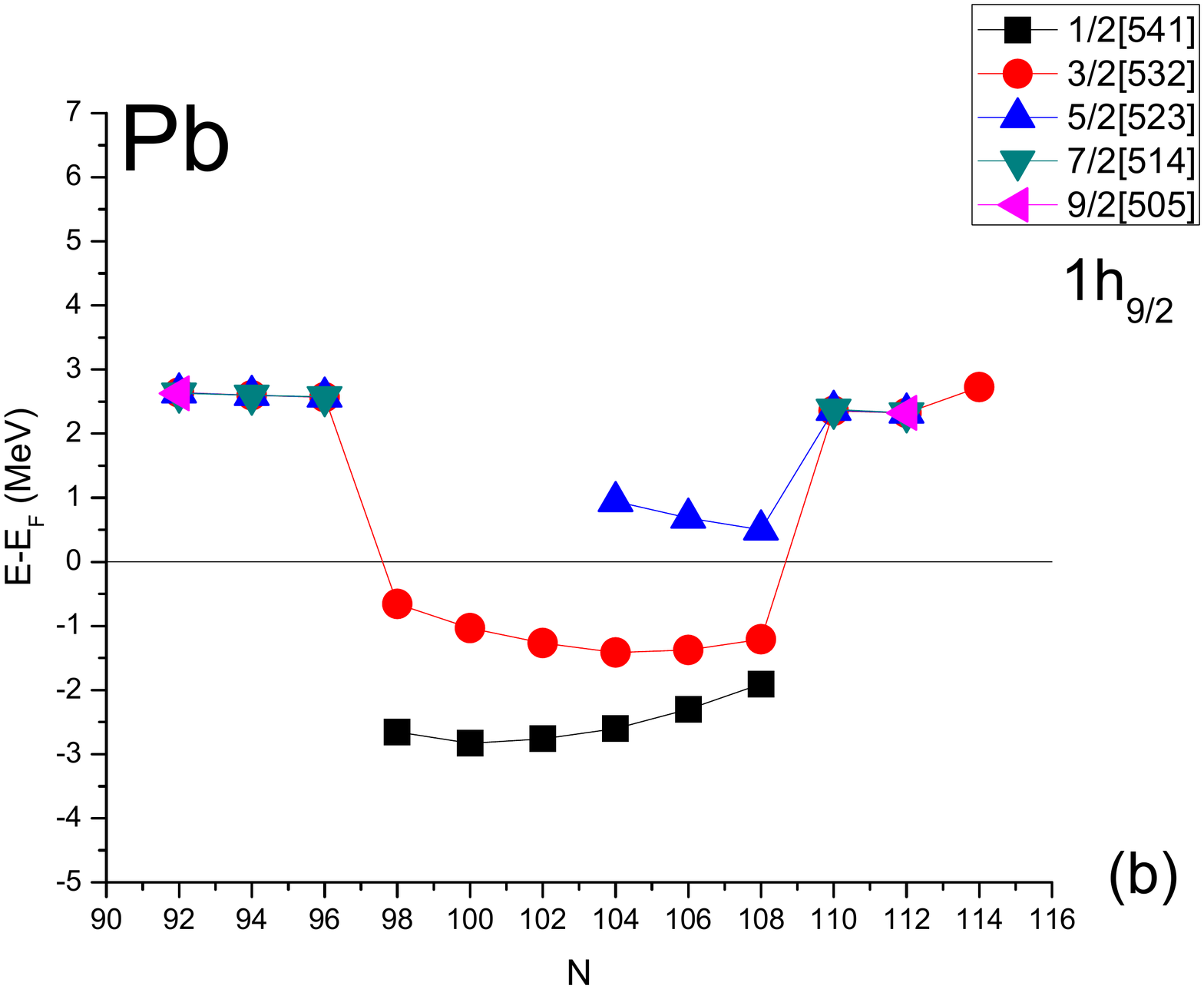}\hspace{5mm}}
{\includegraphics[width=75mm]{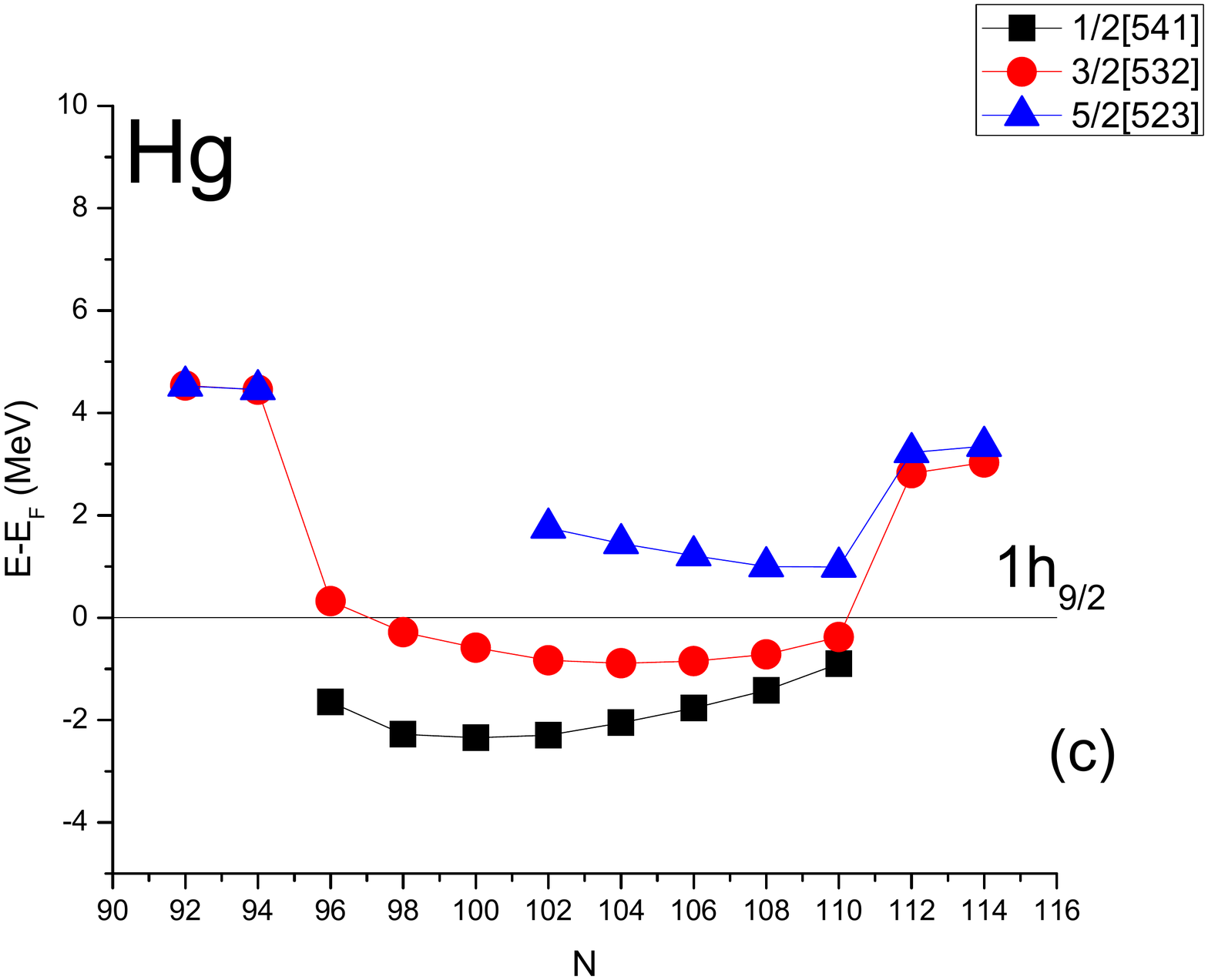}\hspace{5mm}
 \includegraphics[width=75mm]{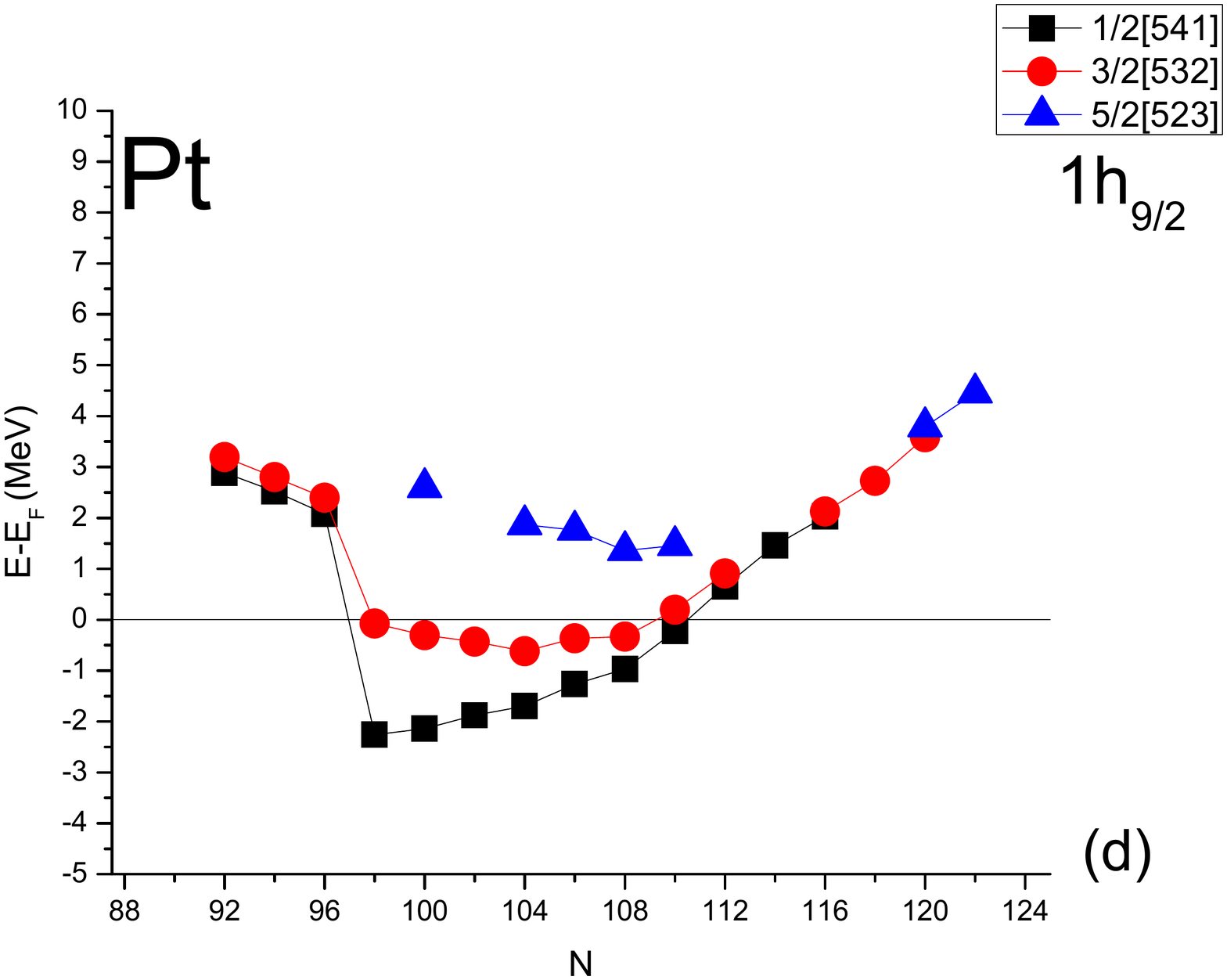}\hspace{5mm}}

\caption{Energies (in MeV) of $1h_{9/2}$ single-particle proton orbitals relative to the Fermi energy obtained by CDFT for $Z=78$-84 isotopes. For $N\approx 98$-110 the orbitals 1/2[541] and 3/2[532] (normally lying above $Z=82$) are occupied, thus creating four particle excitations. Panel (a) adapted from Ref.~\cite{PLB}. See Section III for further discussion. 
} 

\end{figure*}


\begin{figure*}[htb]

{\includegraphics[width=75mm]{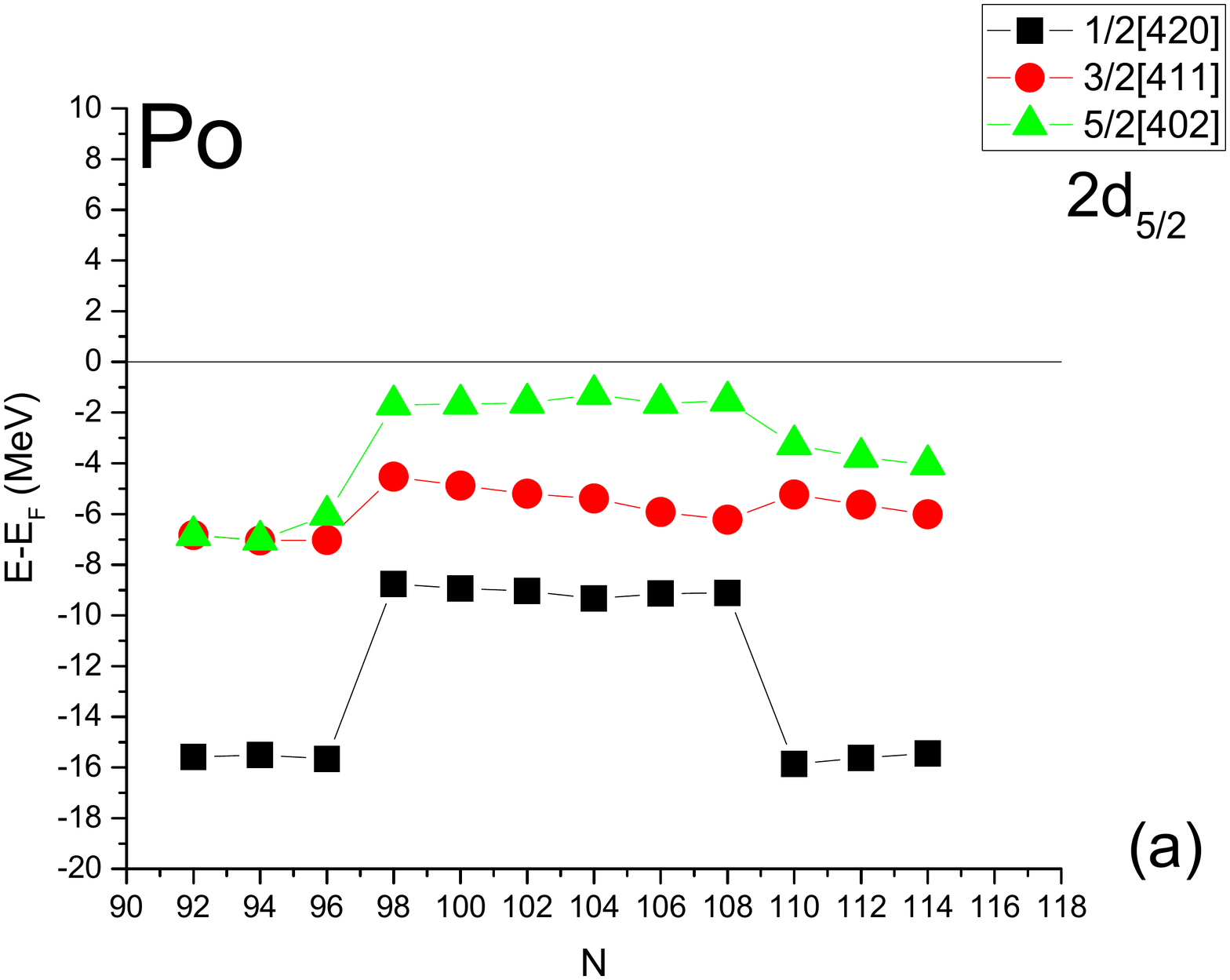}\hspace{5mm}
\includegraphics[width=75mm]{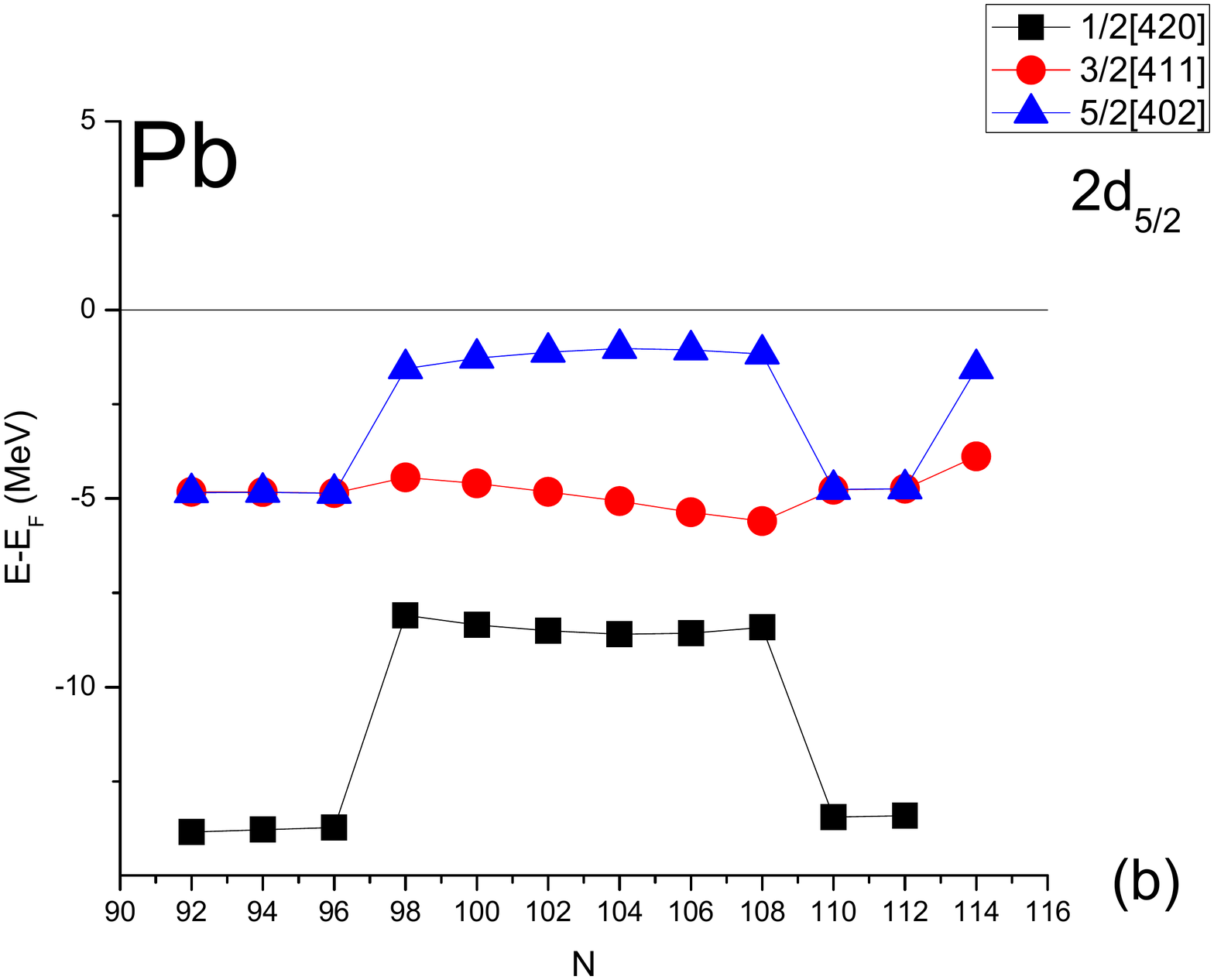}\hspace{5mm}}
{\includegraphics[width=75mm]{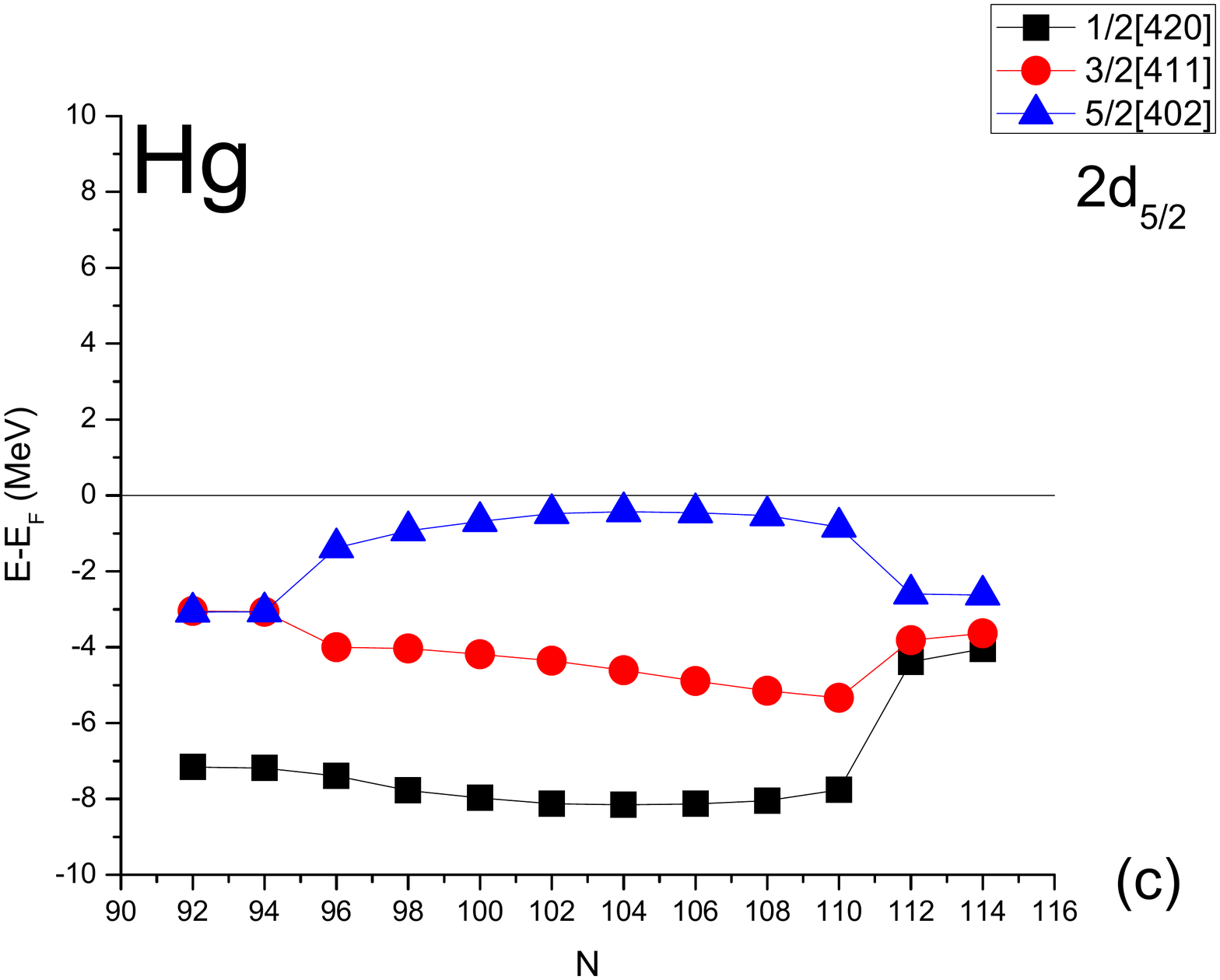}\hspace{5mm}
 \includegraphics[width=75mm]{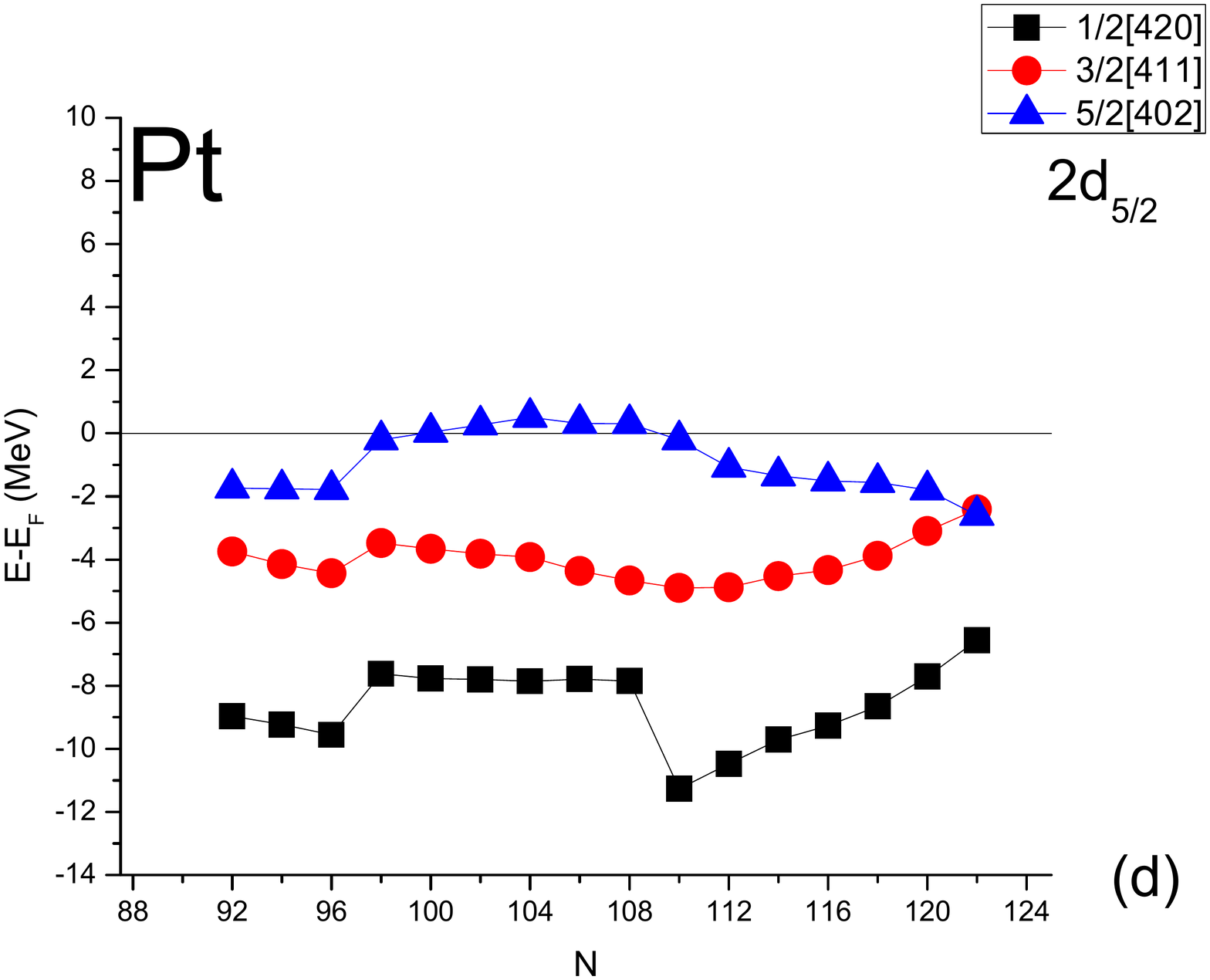}\hspace{5mm}}

\caption{Energies (in MeV) of $2d_{5/2}$ single-particle proton orbitals relative to the Fermi energy obtained by CDFT for $Z=78$-84 isotopes. All orbitals (normally lying below $Z=82$) are occupied, except 5/2[402] , which gets empty for the Pt isotopes with $N=100$-108. See Section III for further discussion. 
} 

\end{figure*}


\begin{figure*}[htb]

{\includegraphics[width=75mm]{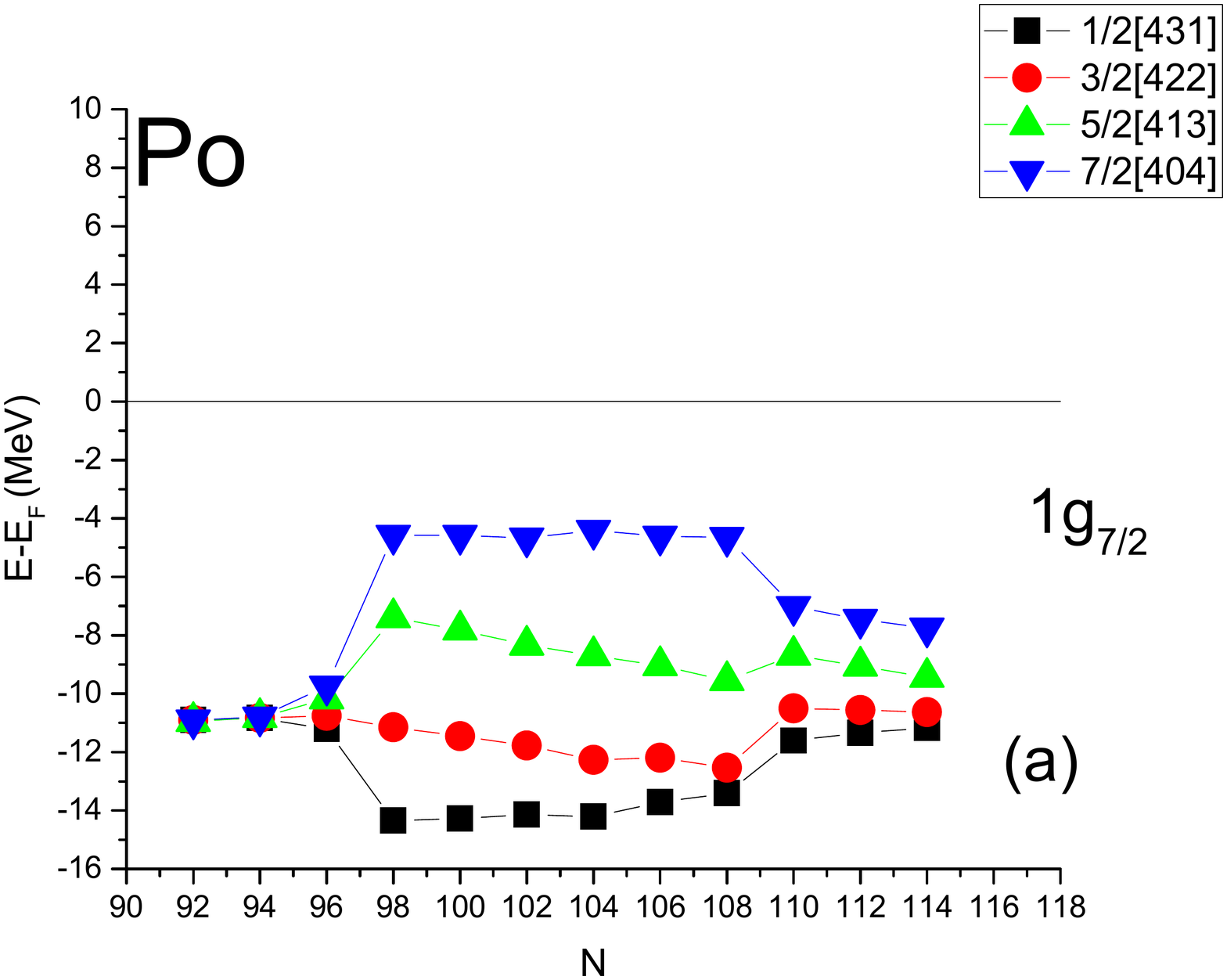}\hspace{5mm}
\includegraphics[width=75mm]{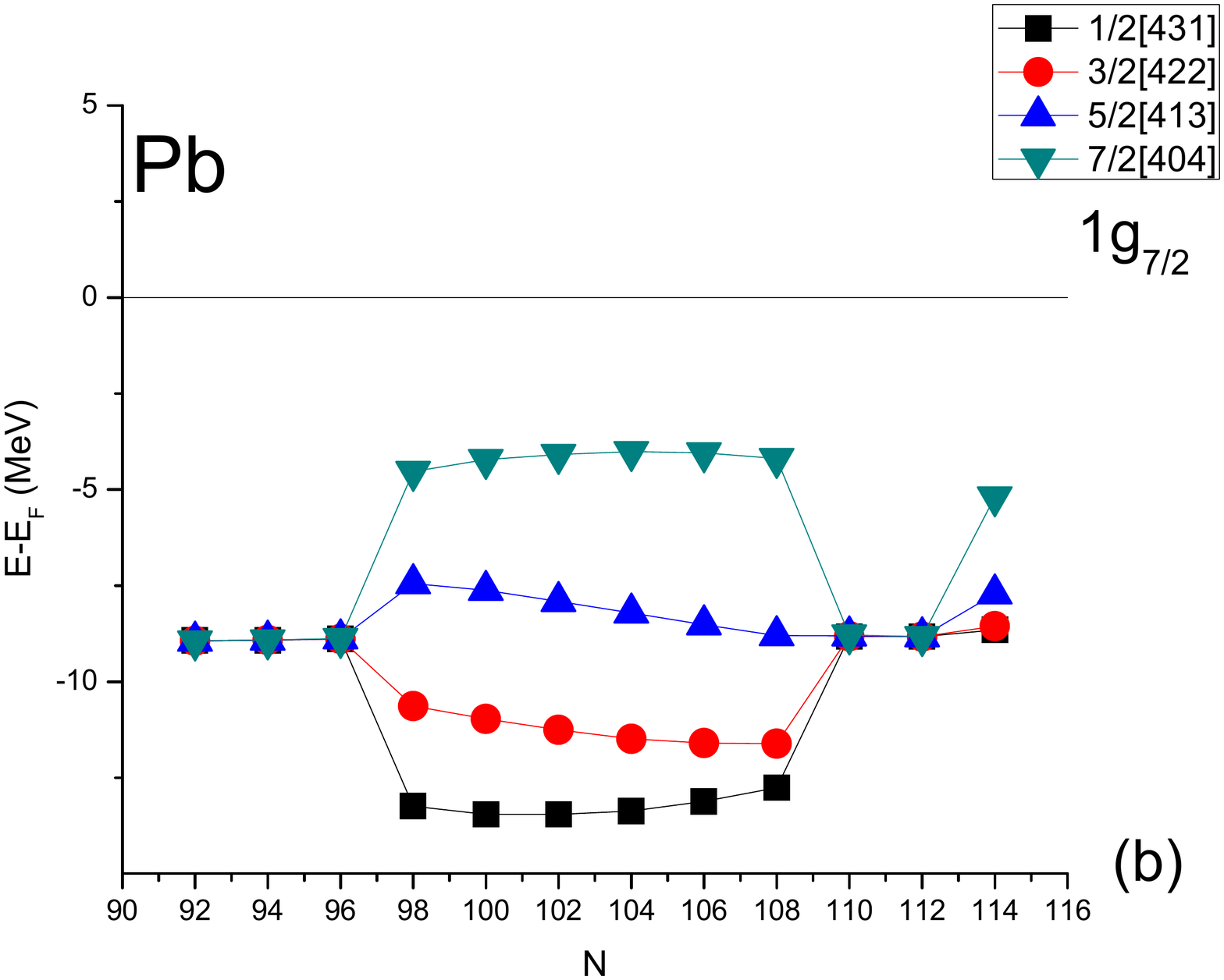}\hspace{5mm}}
{\includegraphics[width=75mm]{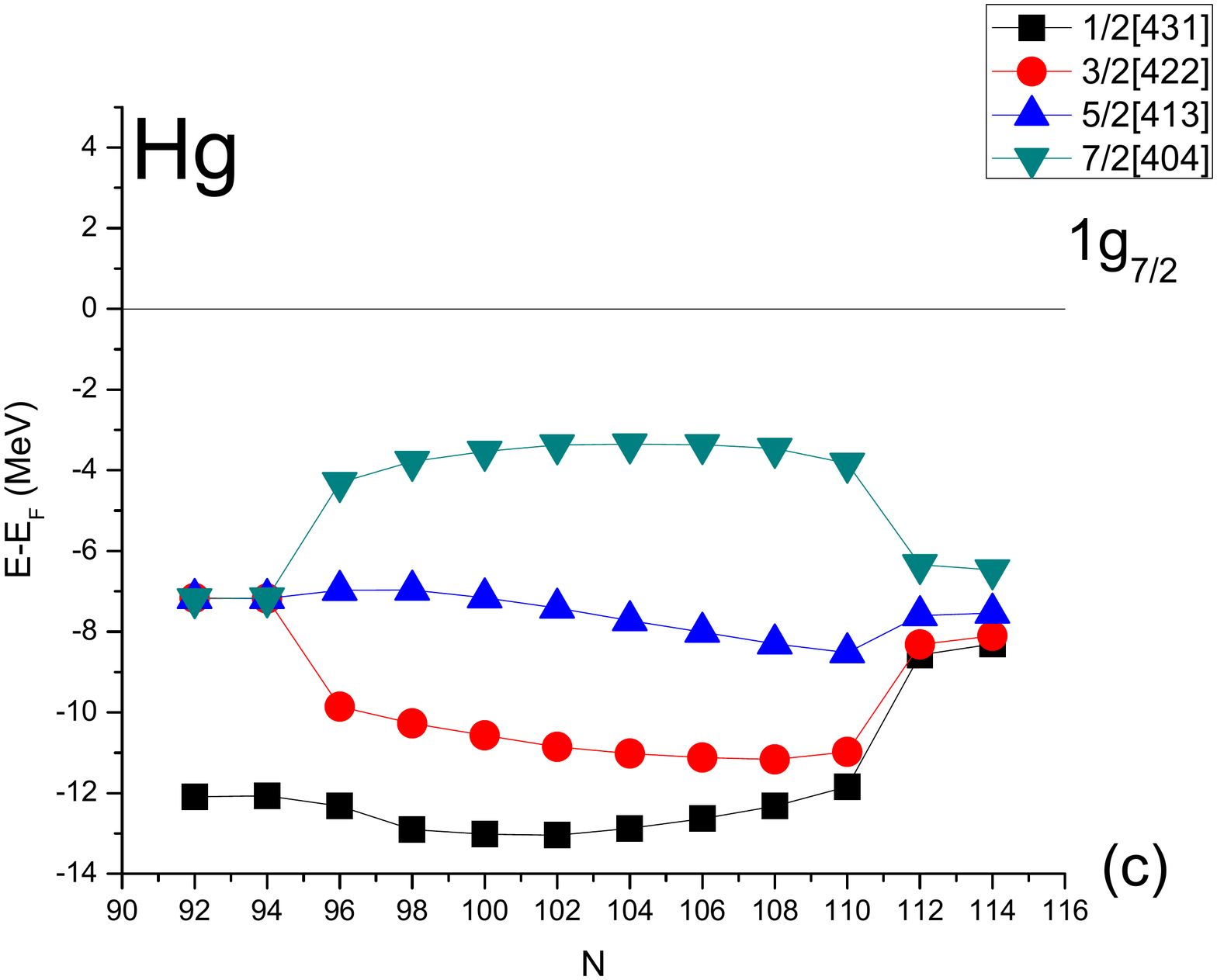}\hspace{5mm}
 \includegraphics[width=75mm]{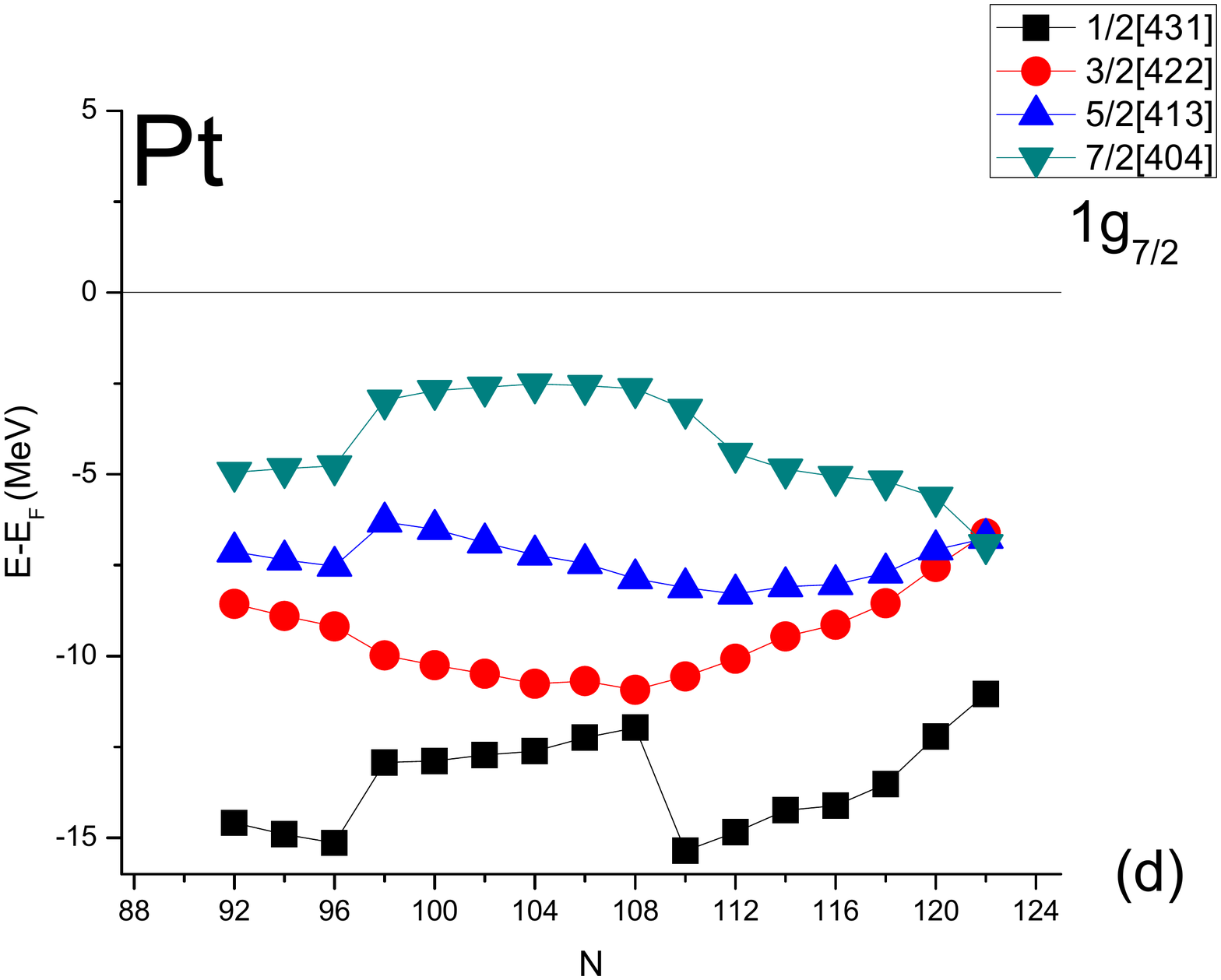}\hspace{5mm}}

\caption{Energies (in MeV) of $1g_{7/2}$ single-particle proton orbitals relative to the Fermi energy obtained by CDFT for $Z=78$-84 isotopes. All orbitals (normally lying below $Z=82$) are occupied. 
See Section III for further discussion. 
} 

\end{figure*}


\begin{figure}[htb]

\includegraphics[width=75mm]{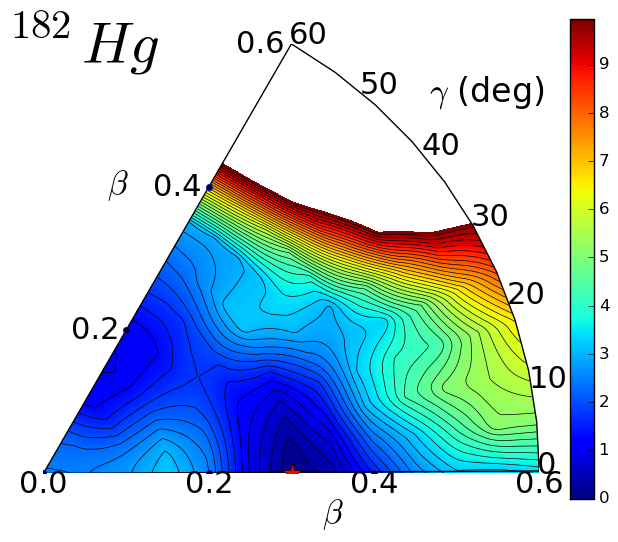}\hspace{5mm}

\caption{Potential energy surface (PES) for the nucleus $^{182}_{80}$Hg$_{102}$, obtained by CDFT using the DDME2 functional used throughout the present work. All energies are normalized with respect to the binding energy of the corresponding ground state. The PES looks very similar to Fig. 18(a) of Ref. \cite{Ramos2014b}, produced for the same nucleus within the IBM-CM framework. See Section III for further discussion. 
} 

\end{figure}



\begin{figure*}[htb]

{\includegraphics[width=75mm]{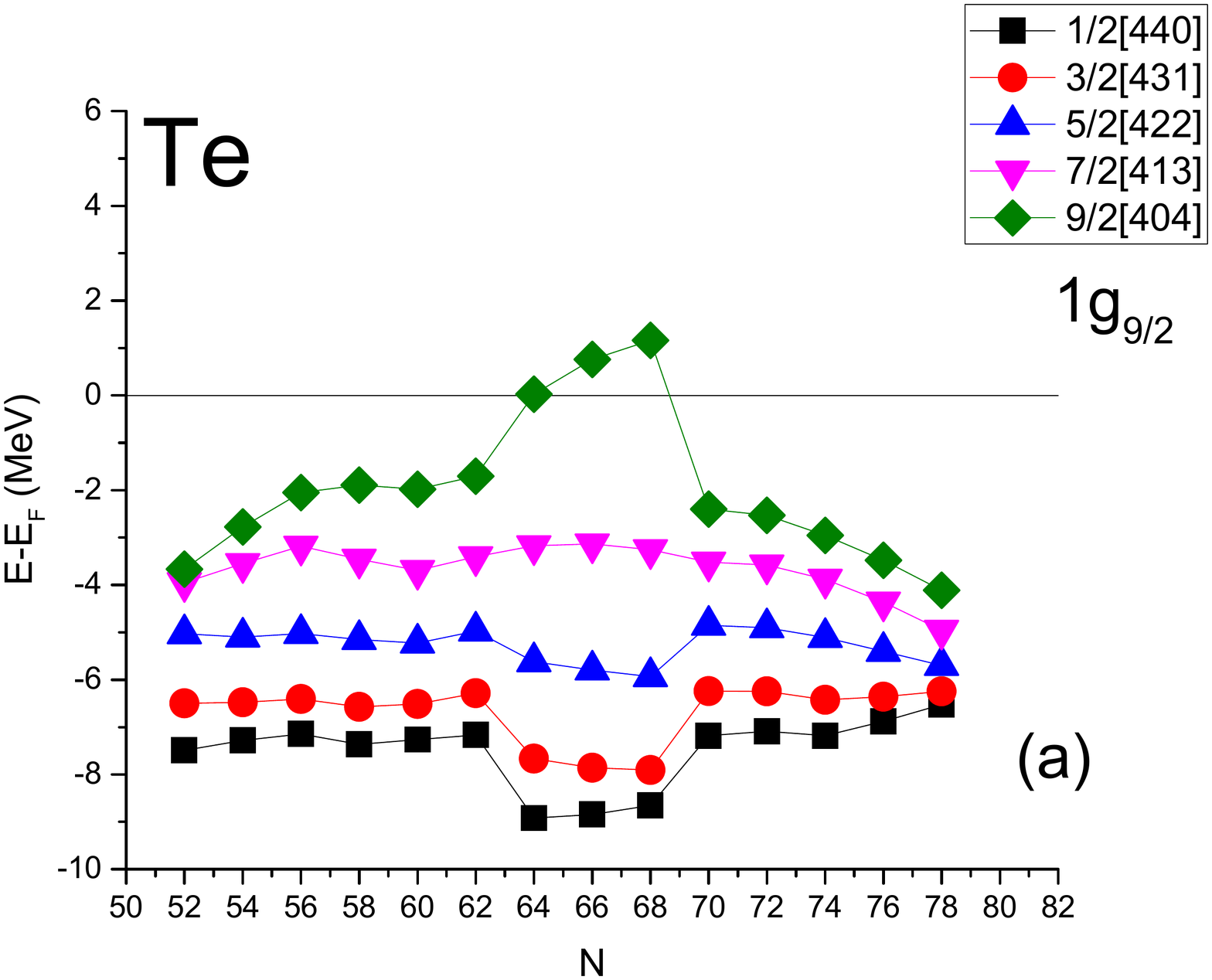}\hspace{5mm}
\includegraphics[width=75mm]{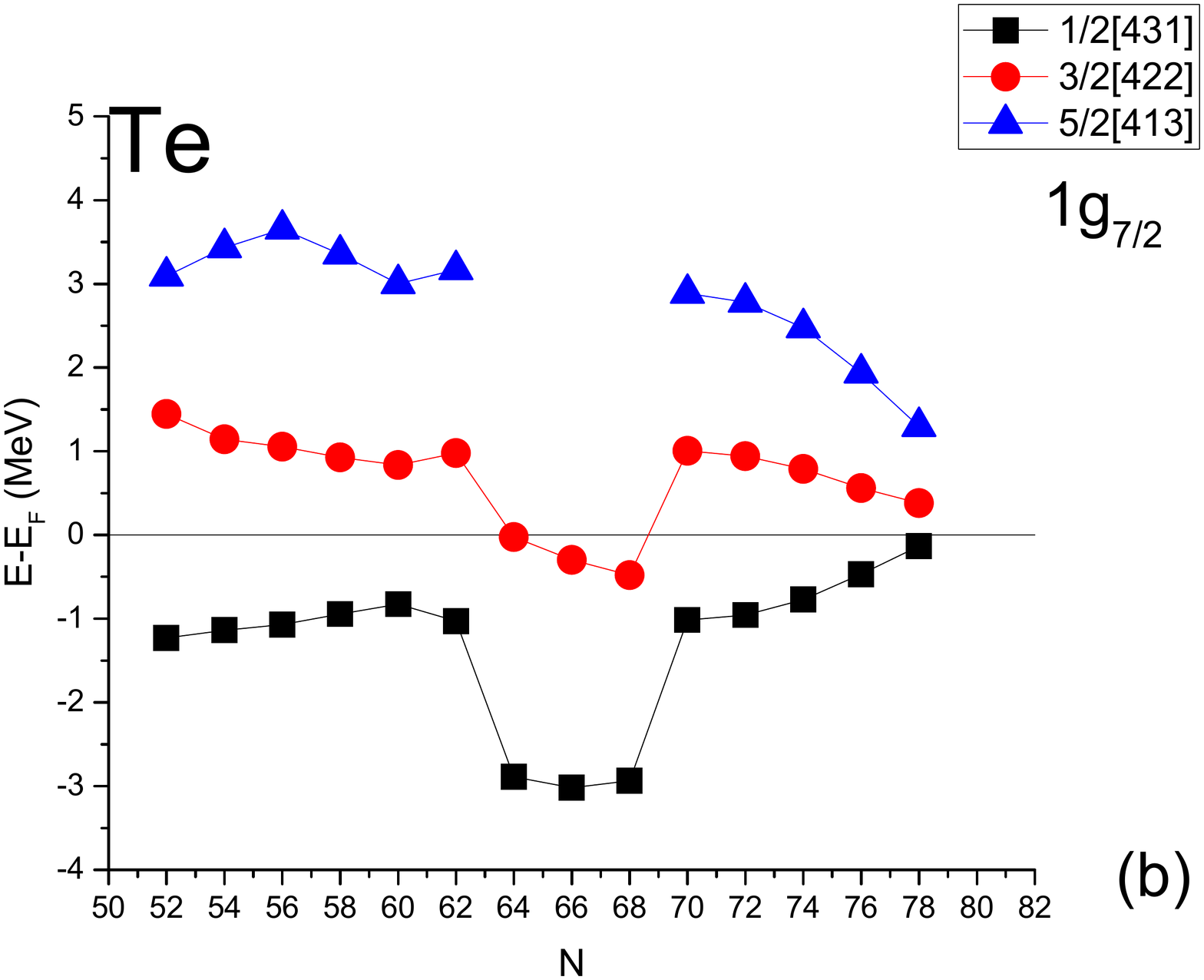}}

\caption{Energies (in MeV) of single-particle proton orbitals relative to the Fermi energy obtained by CDFT for Te ($Z=52$) isotopes. 2p-2h proton excitations are seen for $N=64$-68. The orbital 9/2[404] of $1g_{9/2}$ (a) (normally lying below $Z=50$) is vacant, while the orbital 3/2[422] of $1g_{7/2}$ (b) (normally lying above $Z=50$) is occupied. Adapted from Ref.~\cite{PLB}. See Section III for further discussion. 
} 

\end{figure*}



\begin{figure*}[htb]

{\includegraphics[width=75mm]{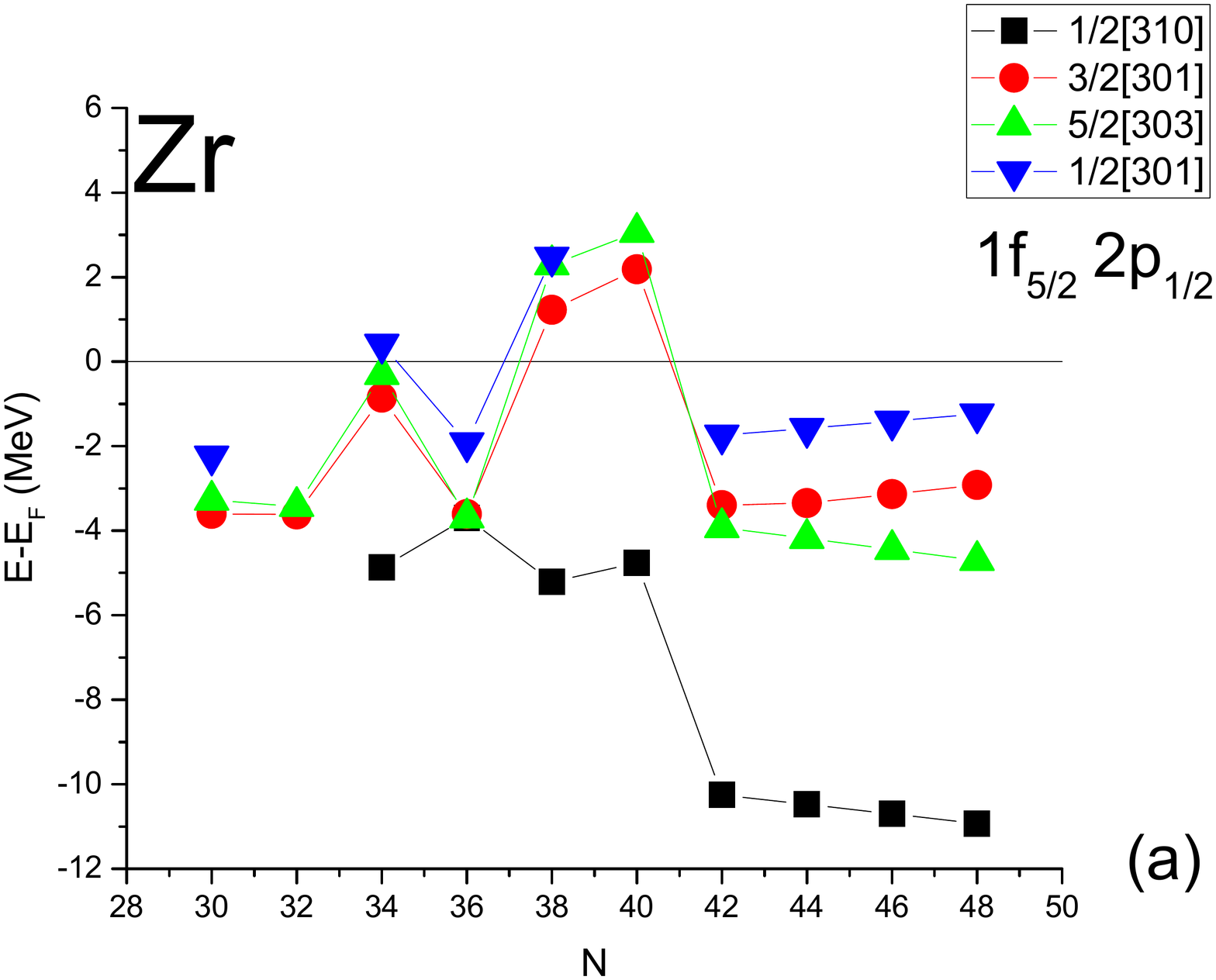}\hspace{5mm}
\includegraphics[width=75mm]{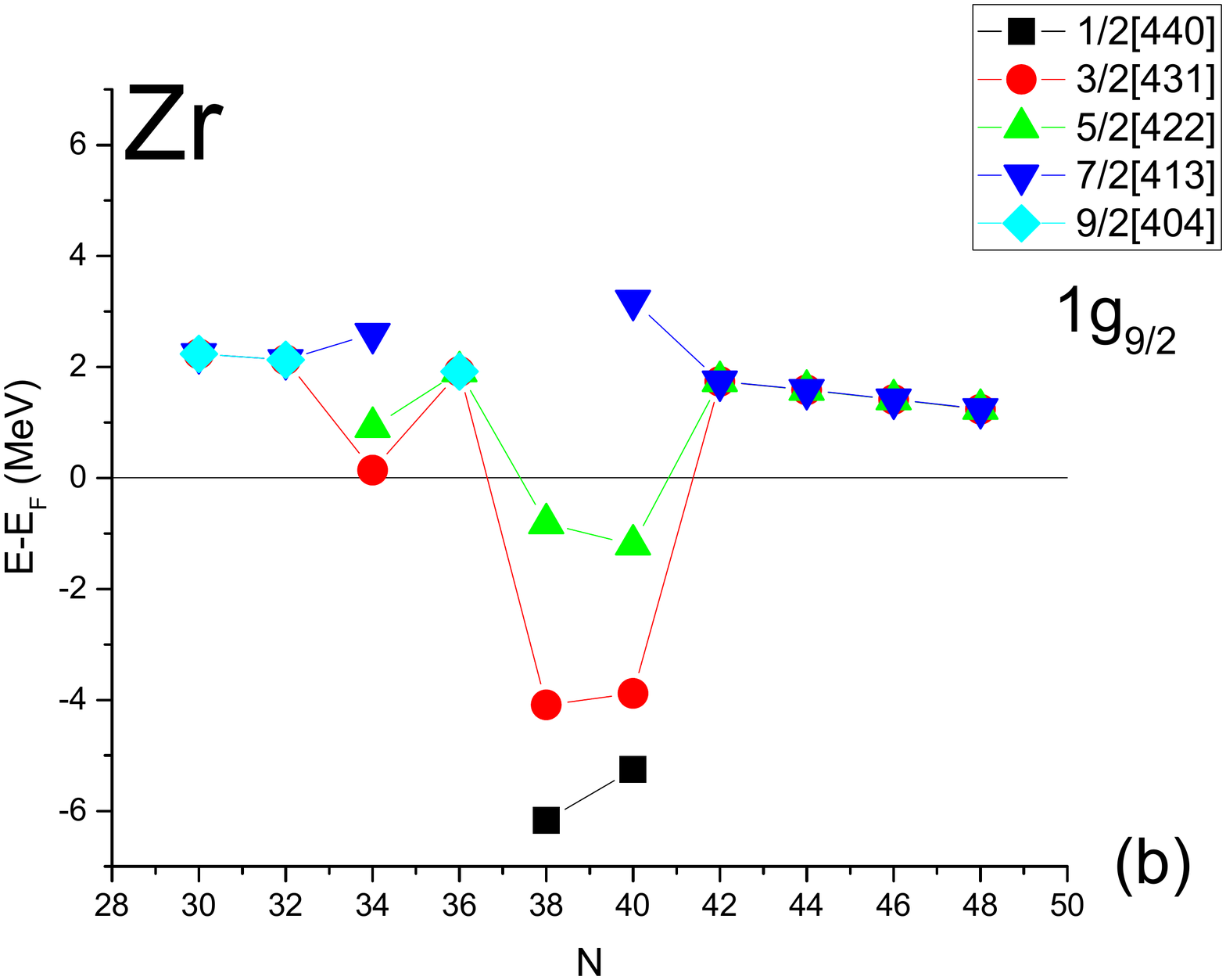}}

\caption{Energies (in MeV) of single-particle proton orbitals relative to the Fermi energy obtained by CDFT for Zr ($Z=40$) isotopes. 6p-6h proton excitations are seen for $N=38$-40. The orbitals 1/2[301] of $2p_{1/2}$, as well as 3/2[301], 5/2[303] of $1f_{5/2}$ (a) (normally lying below $Z=40$) are vacant, while the orbitals 1/2[440], 3/2[431], 5/2[422] of $1g_{9/2}$   (b) (normally lying above $Z=40$) are occupied. Adapted from Ref.~\cite{PLB}. See Section III for further discussion. 
} 

\end{figure*}


\begin{figure*}[htb]

{\includegraphics[width=75mm]{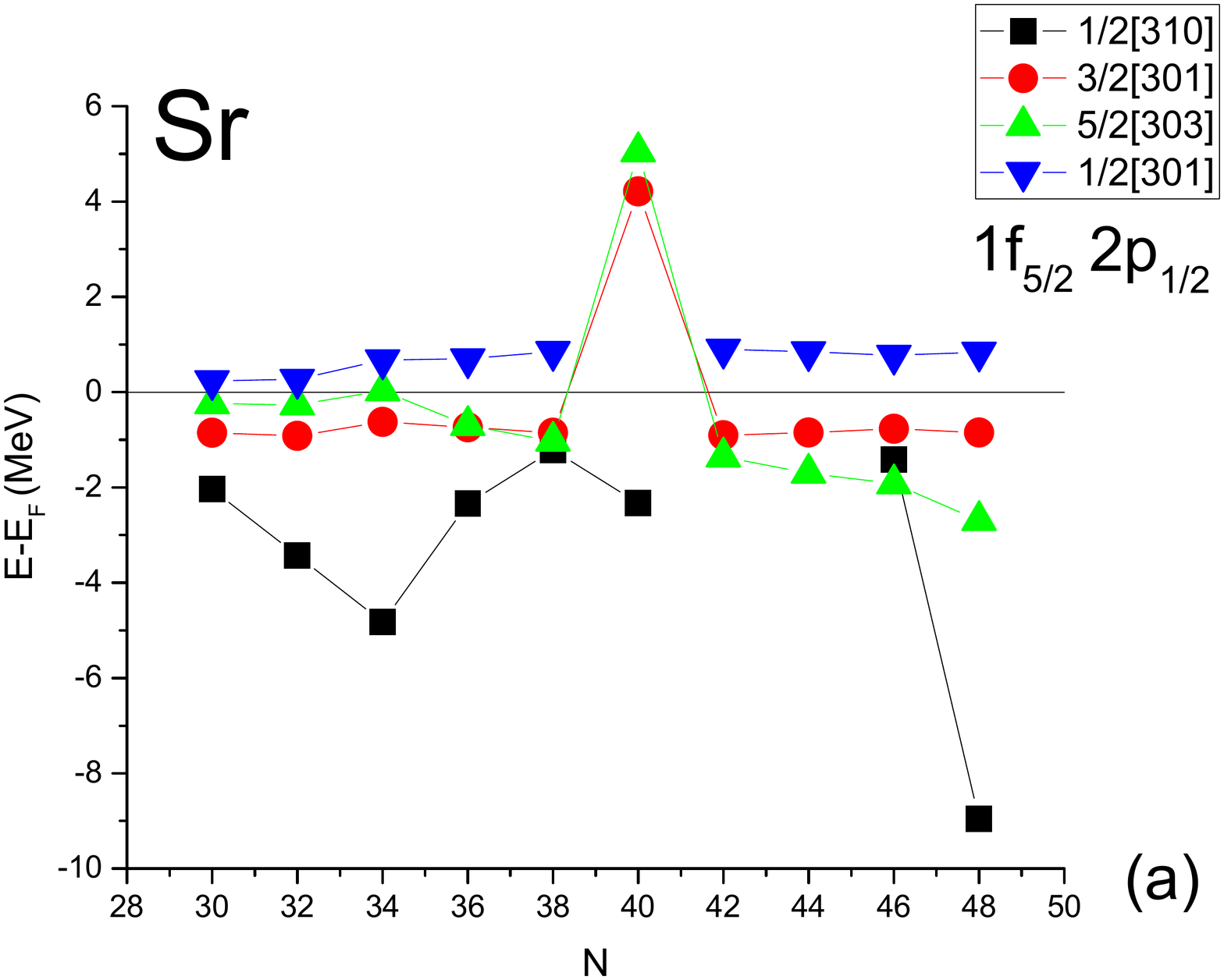}\hspace{5mm}
\includegraphics[width=75mm]{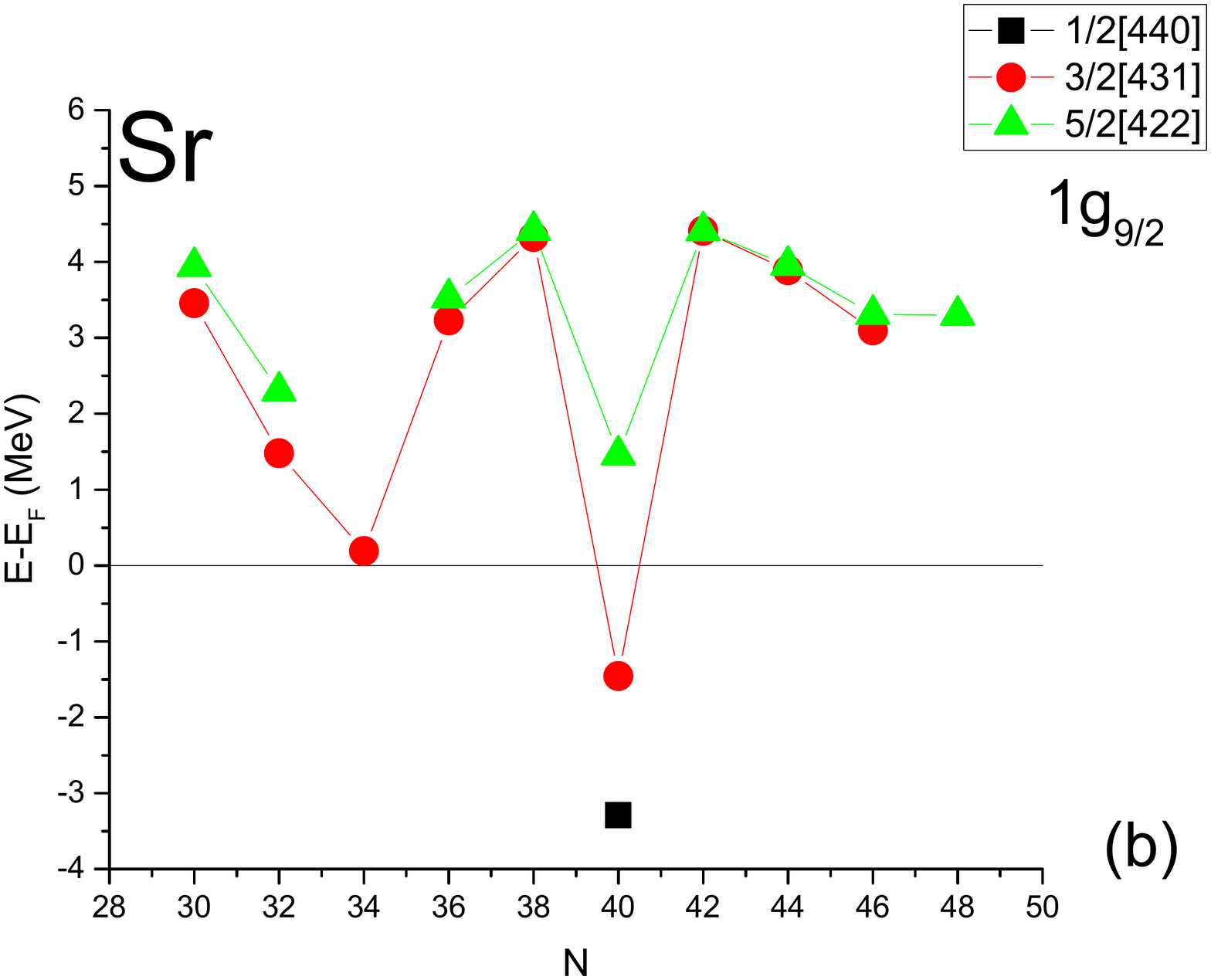}}

\caption{Energies (in MeV) of single-particle proton orbitals relative to the Fermi energy obtained by CDFT for Sr ($Z=38$) isotopes. 4p-4h proton excitations are seen for $N=38$-40. The orbitals 3/2[301], 5/2[303] of $1f_{5/2}$ (a) (normally lying below $Z=40$) are vacant, while the orbitals 1/2[440], 3/2[431] of $1g_{9/2}$   (b) (normally lying above $Z=40$) are occupied. See Section III for further discussion. 
} 

\end{figure*}



\begin{figure*}[htb]

{\includegraphics[width=75mm]{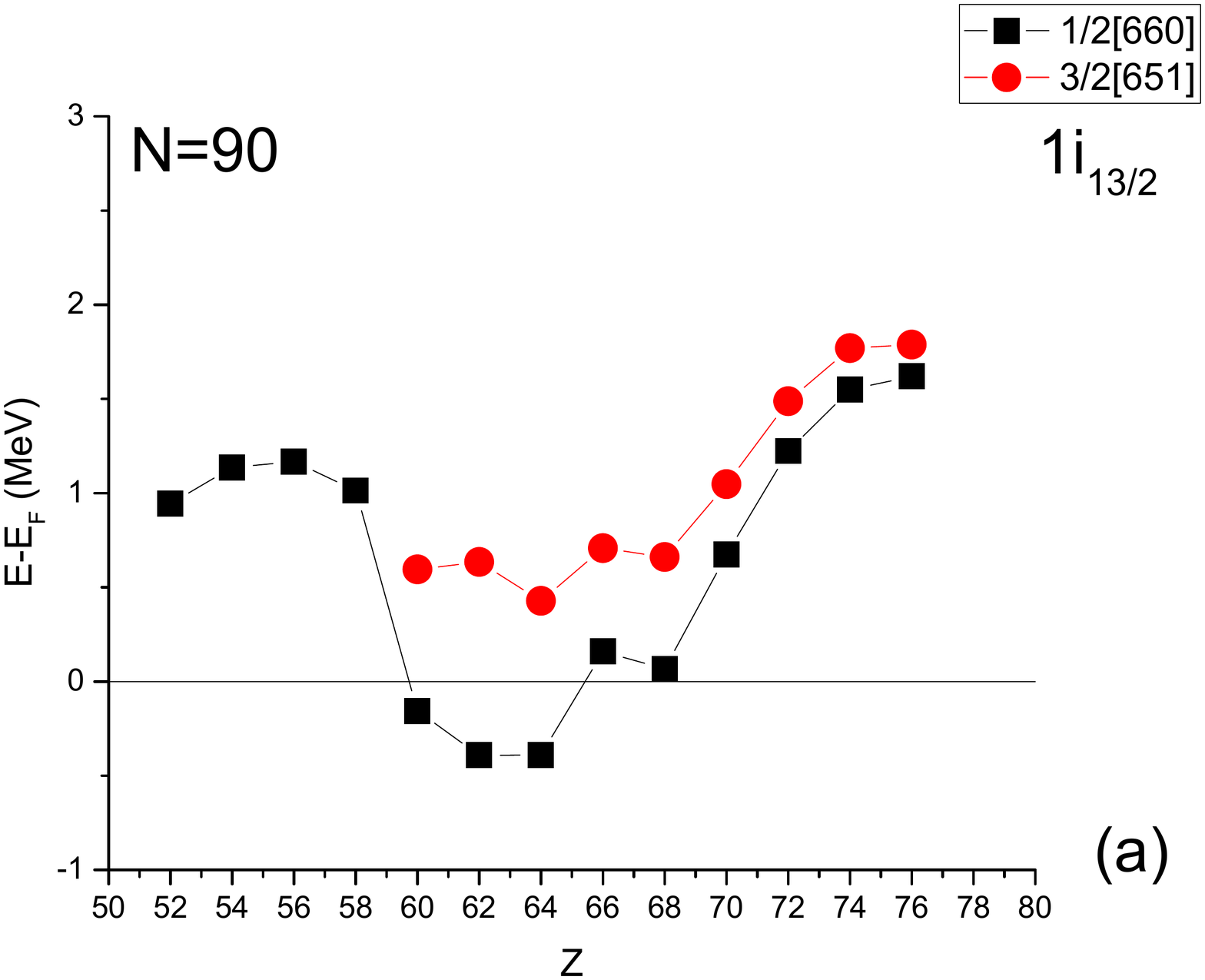}\hspace{5mm}
\includegraphics[width=75mm]{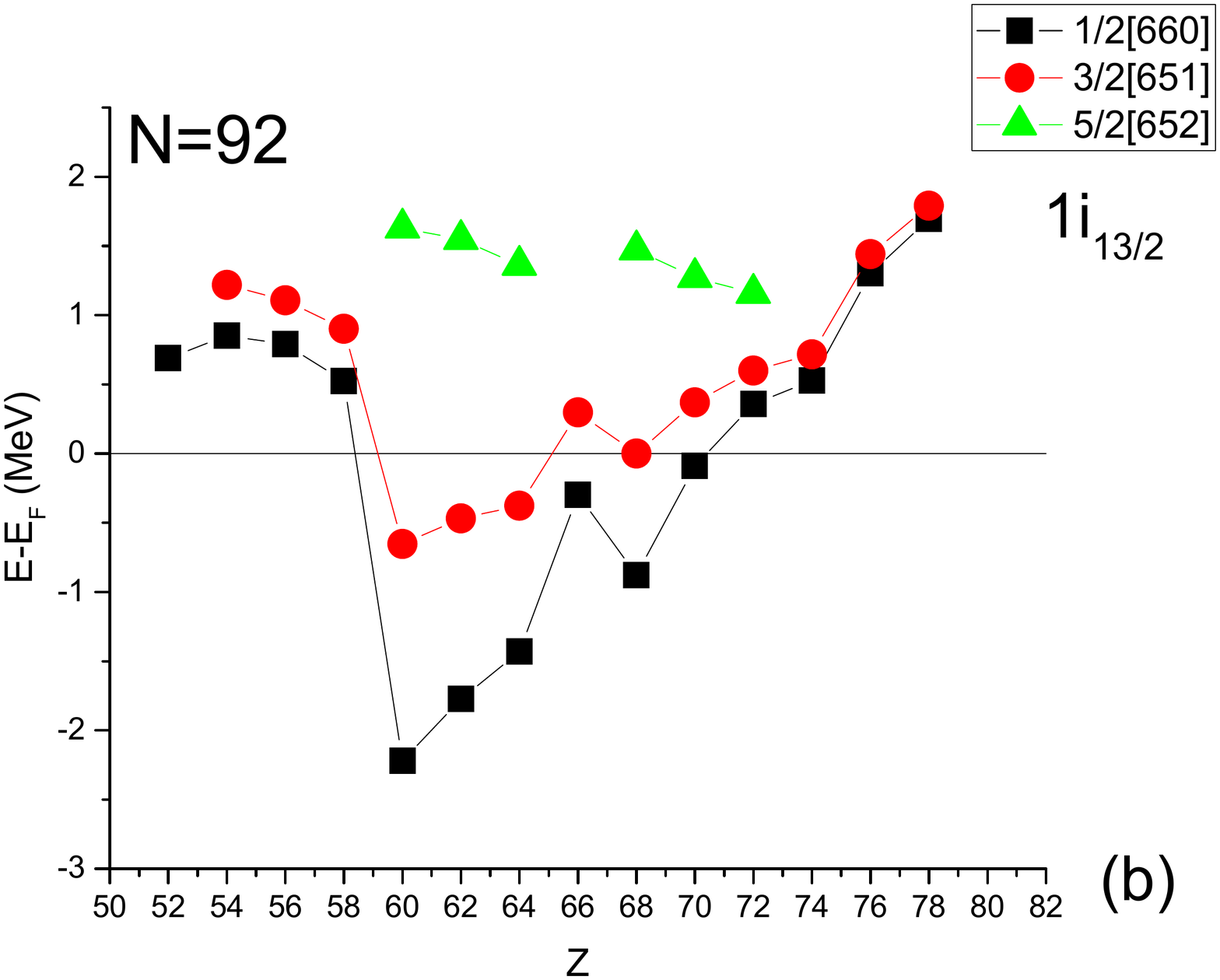}\hspace{5mm}}
{\includegraphics[width=75mm]{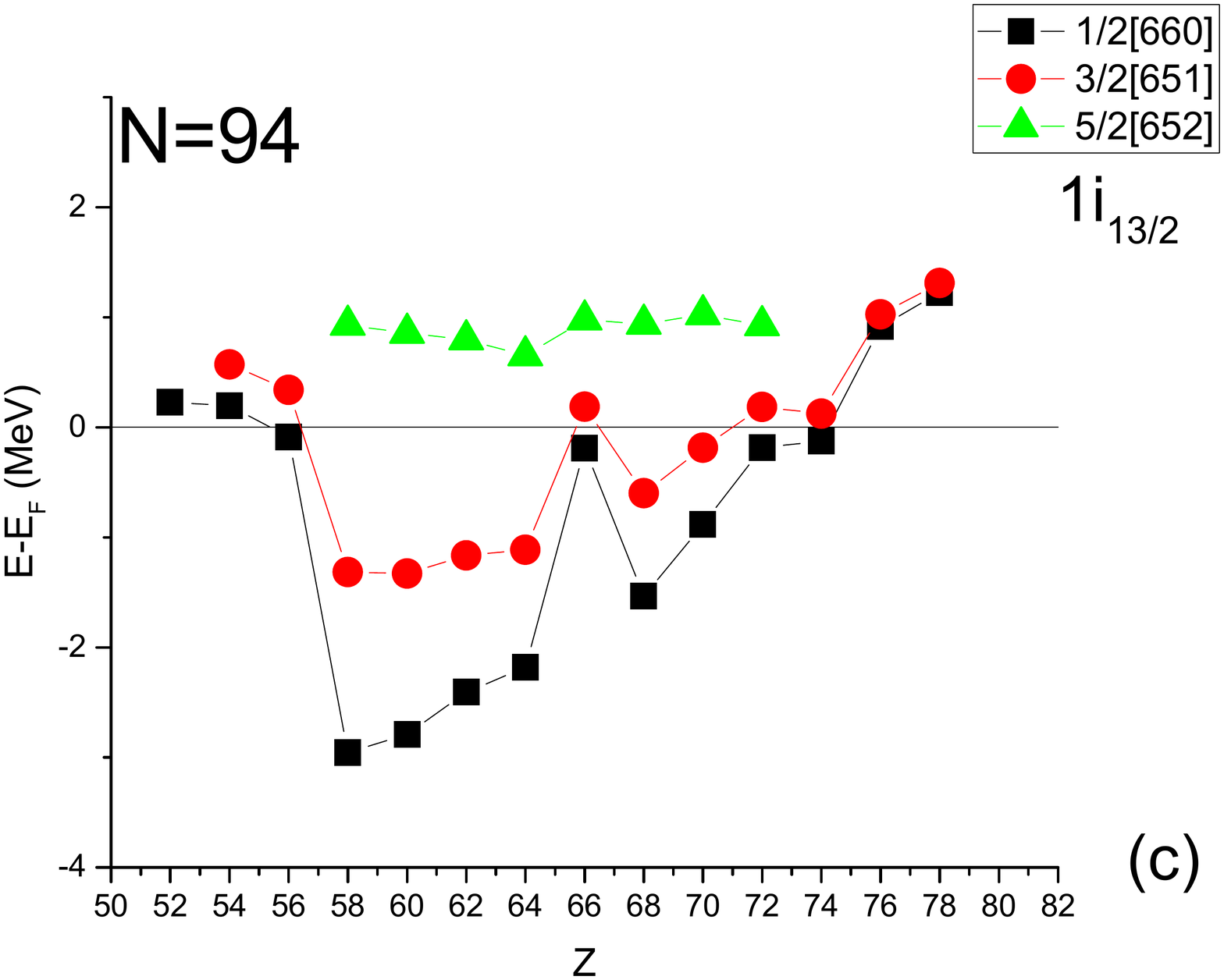}\hspace{5mm}
 \includegraphics[width=75mm]{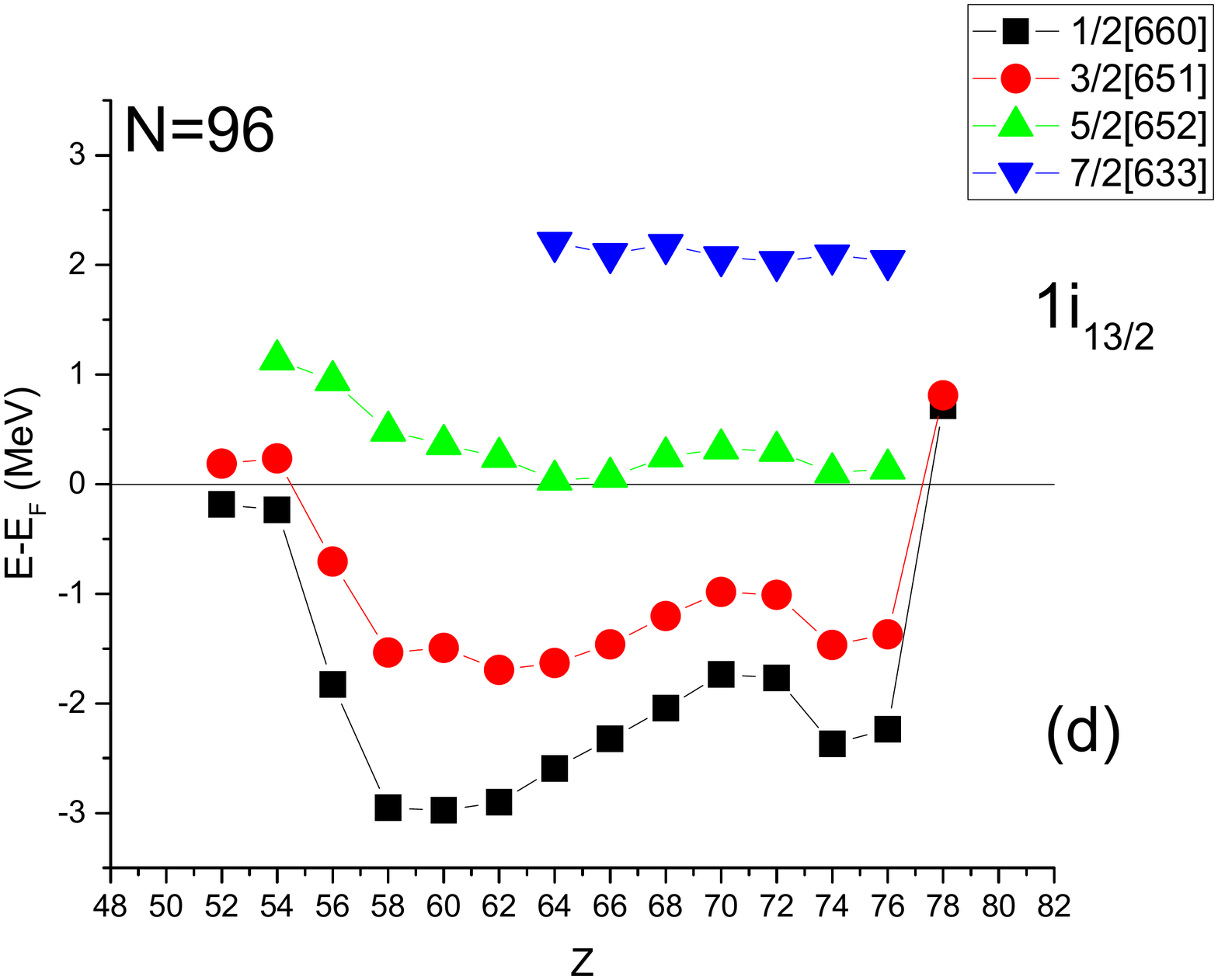}\hspace{5mm}}

\caption{Energies (in MeV) of single-particle neutron  orbitals relative to the Fermi energy obtained by CDFT for $N=90$ isotones. 2p-2h neutron excitations are seen for $Z=60$-64. The orbital 1/2[660] of $1i_{13/2}$ (normally lying above $N=112$) shown in this figure is occupied, while the orbital 5/2[523] of $2f_{7/2}$ (normally lying below $N=112$) shown in the next figure is vacant. Panels (a), (b) adapted from Ref.~\cite{PLB}. See Section IV for further discussion. 
} 

\end{figure*}


\begin{figure*}[htb]

{\includegraphics[width=75mm]{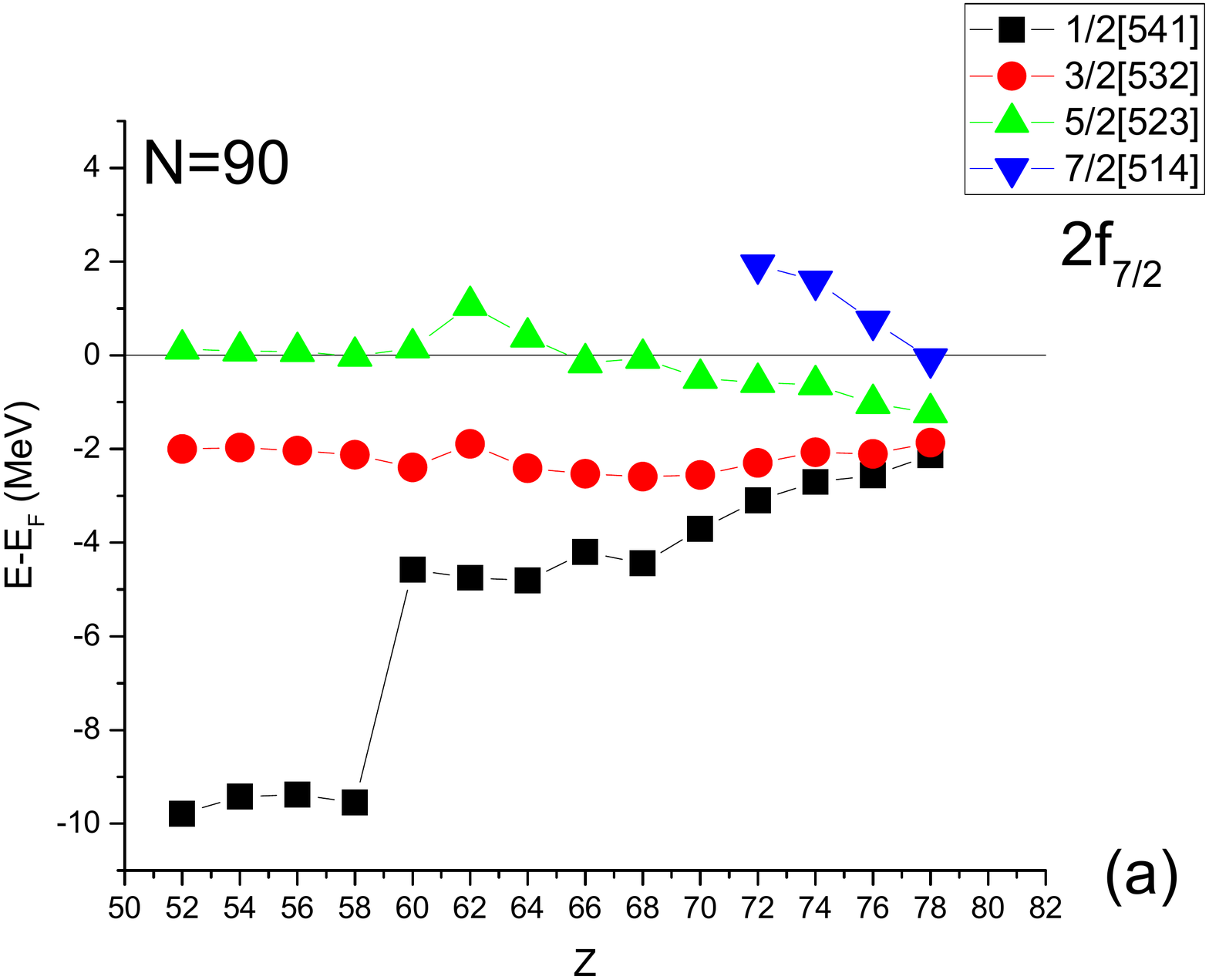}\hspace{5mm}
\includegraphics[width=75mm]{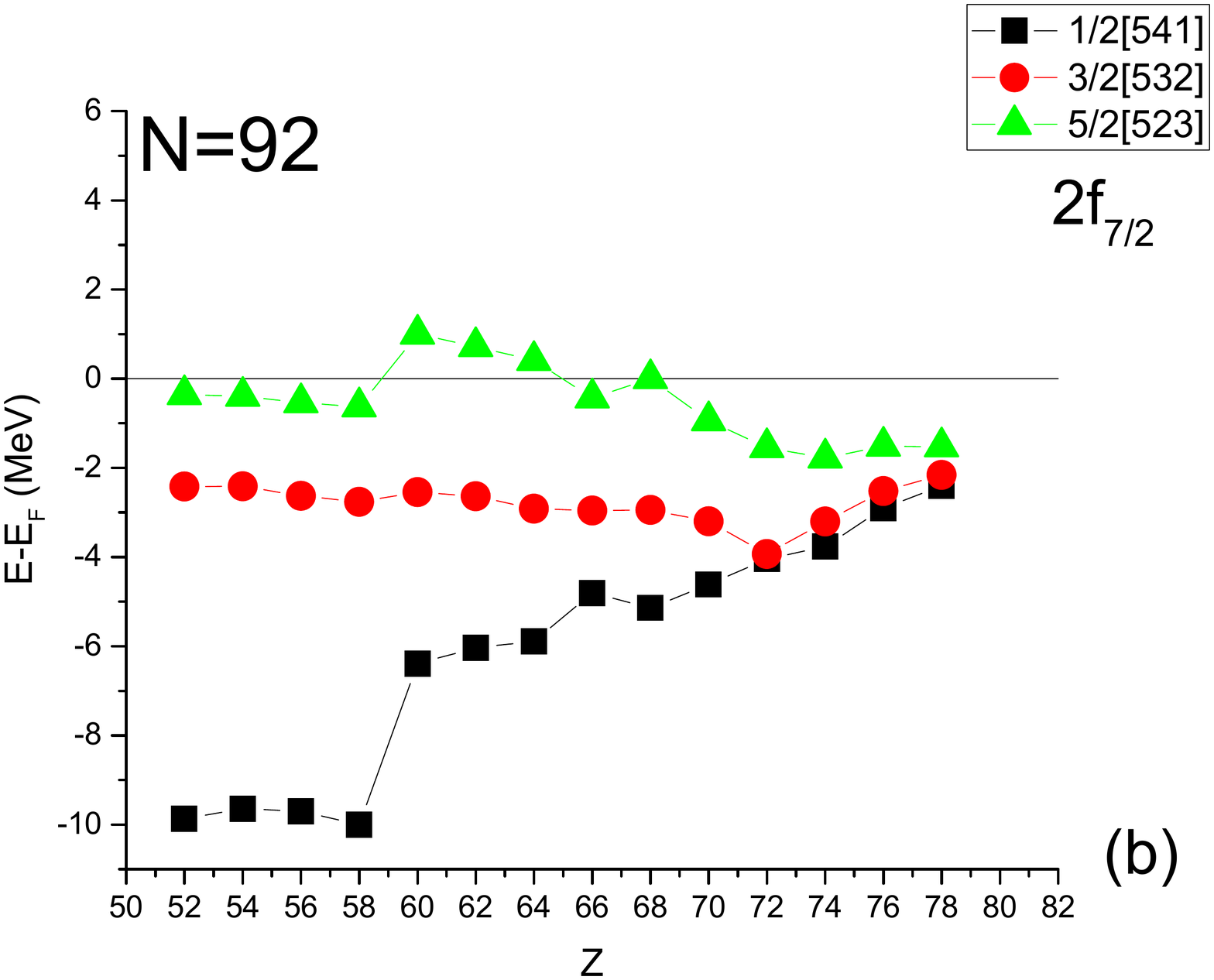}\hspace{5mm}}
{\includegraphics[width=75mm]{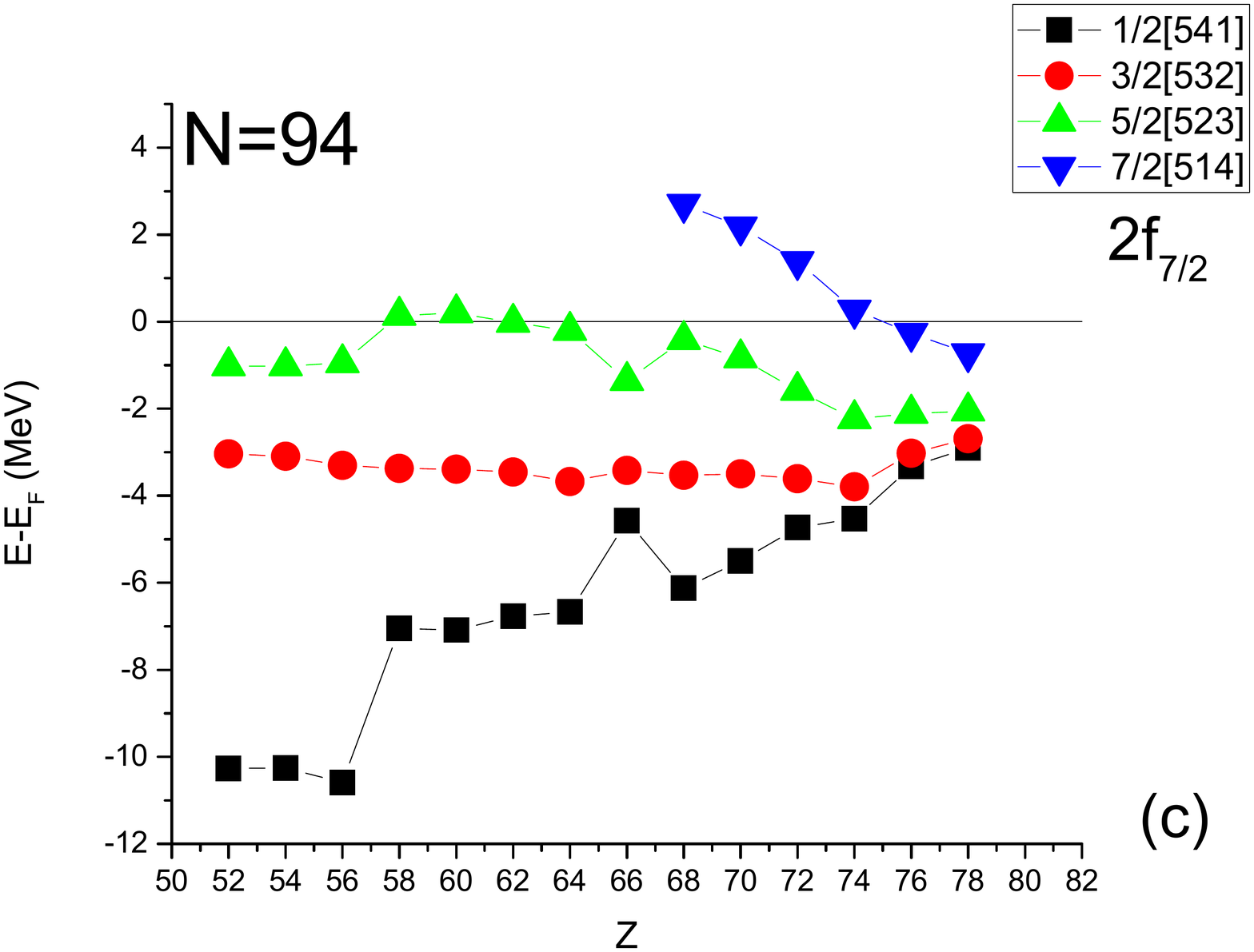}\hspace{5mm}
 \includegraphics[width=75mm]{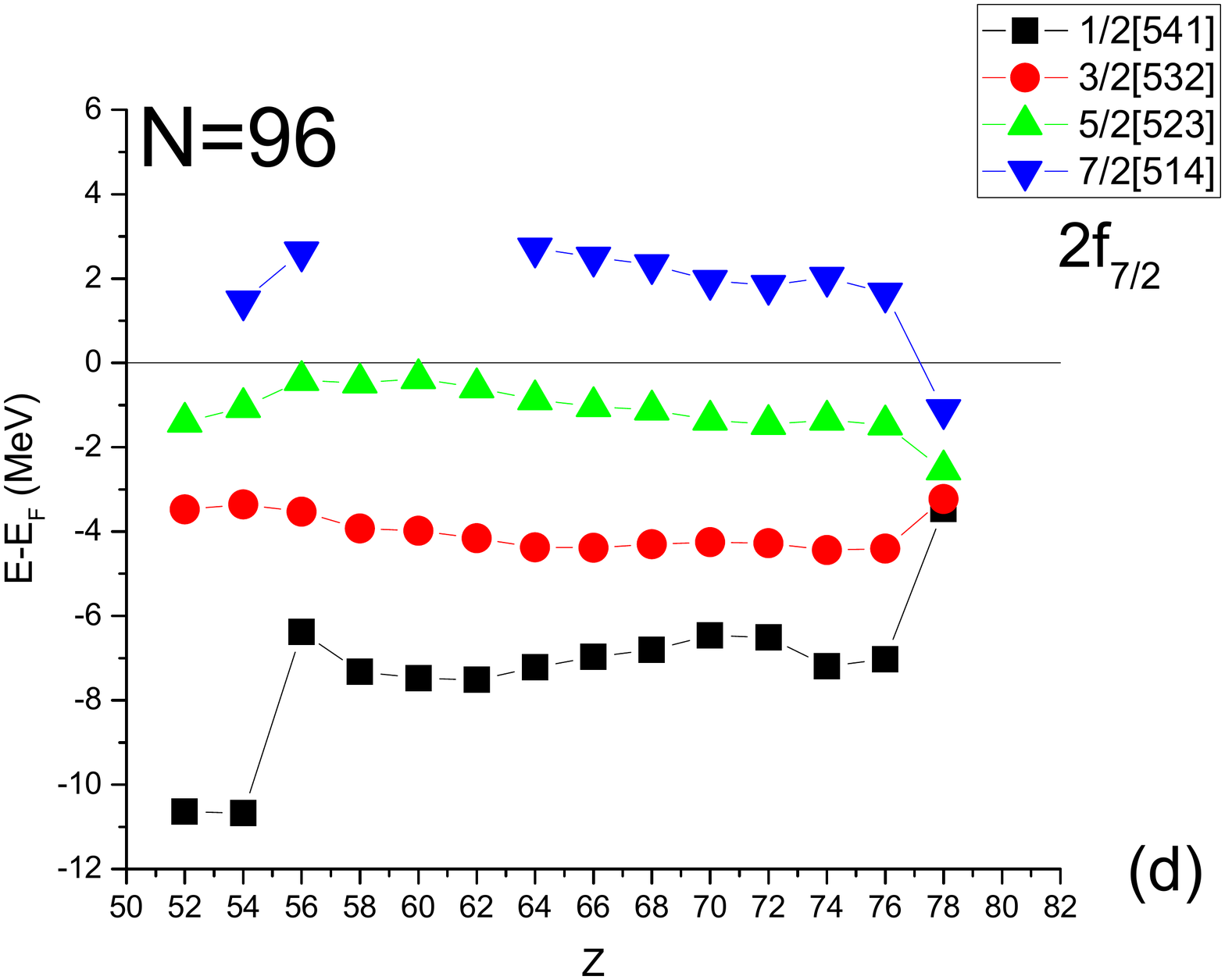}\hspace{5mm}}

\caption{Energies (in MeV) of single-particle neutron orbitals relative to the Fermi energy obtained by CDFT for $N=90$ isotones. 2p-2h neutron excitations are seen for $Z=60$-64. The orbital 5/2[523] of 2f7/2 (normally lying below $N=112$) shown in this figure is vacant, while the orbital 1/2[660] of $1i_{13/2}$ (normally lying above $N=112$) shown in the previous figure is occupied. Panels (a), (b) adapted from Ref.~\cite{PLB}. See Section IV for further discussion. 
} 

\end{figure*}


\begin{figure*}[htb]

{\includegraphics[width=75mm]{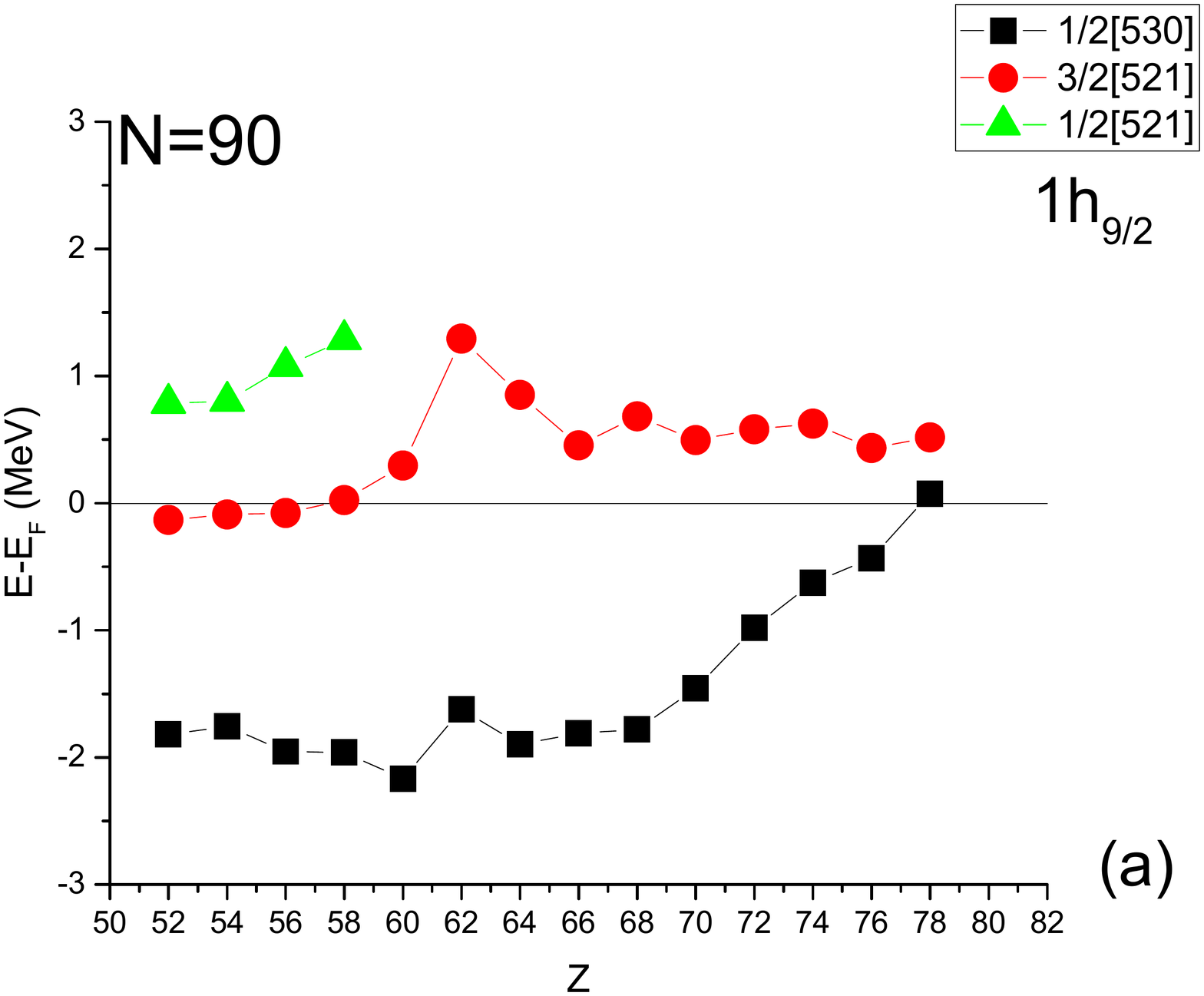}\hspace{5mm}
\includegraphics[width=75mm]{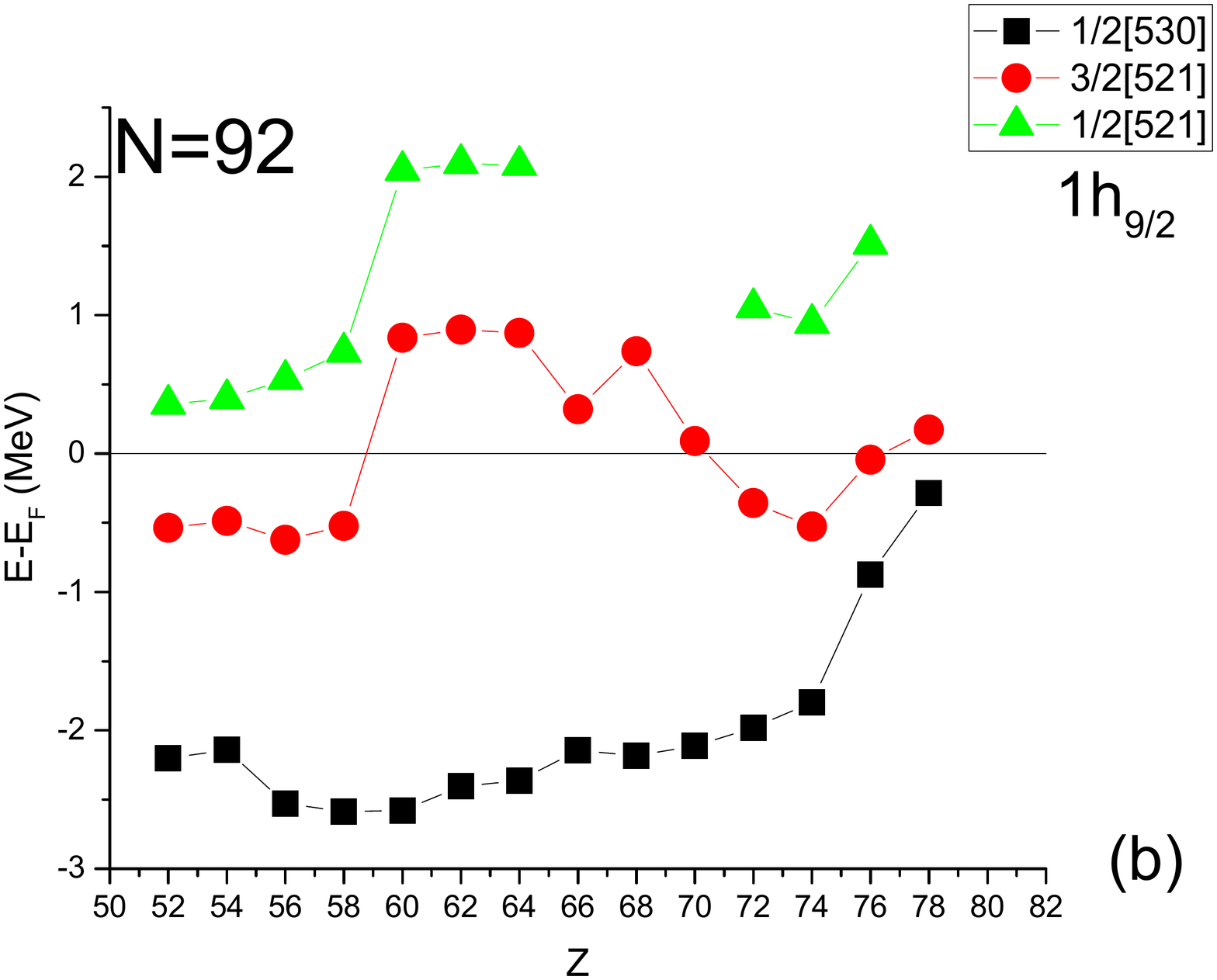}\hspace{5mm}}
{\includegraphics[width=75mm]{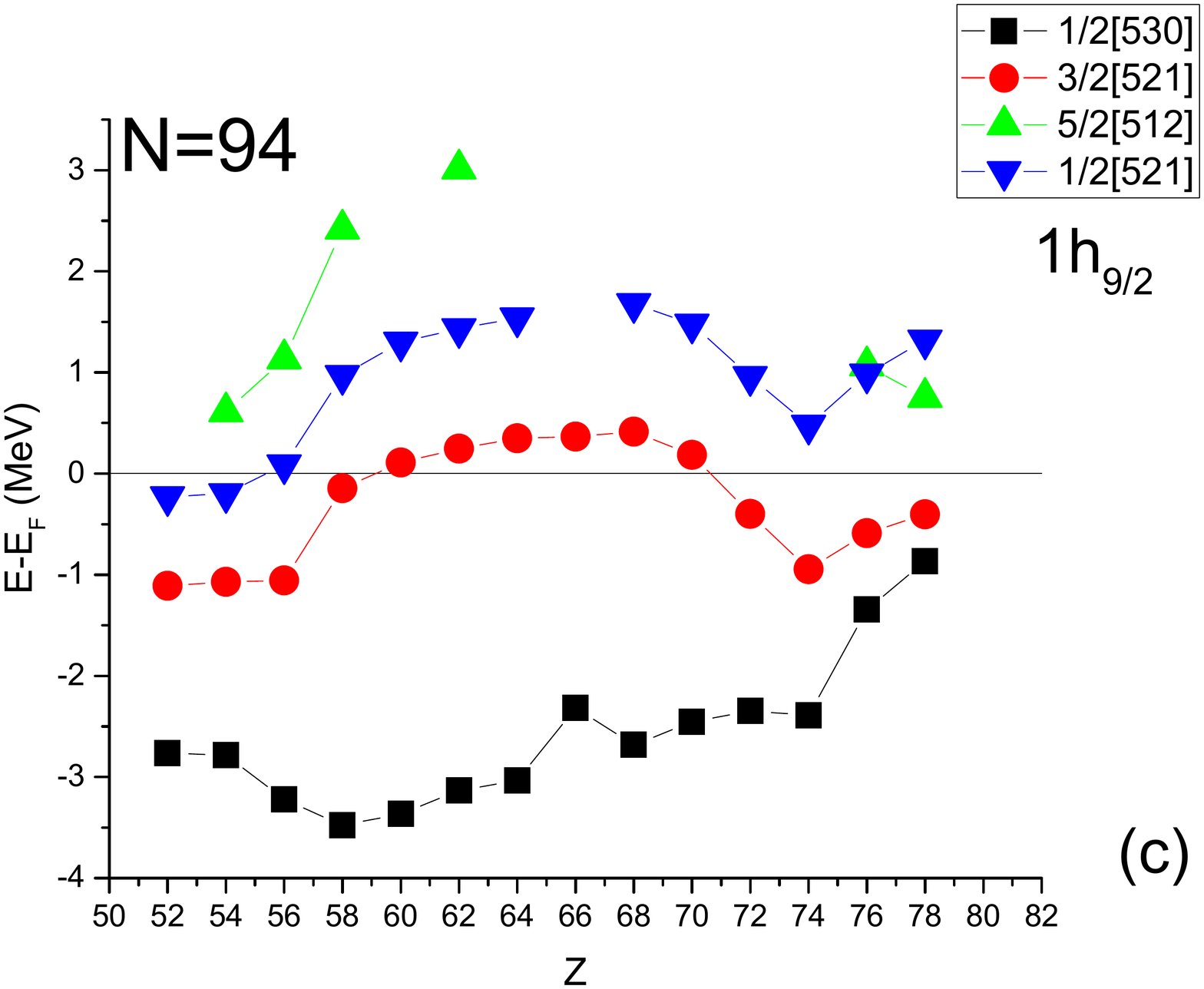}\hspace{5mm}
 \includegraphics[width=75mm]{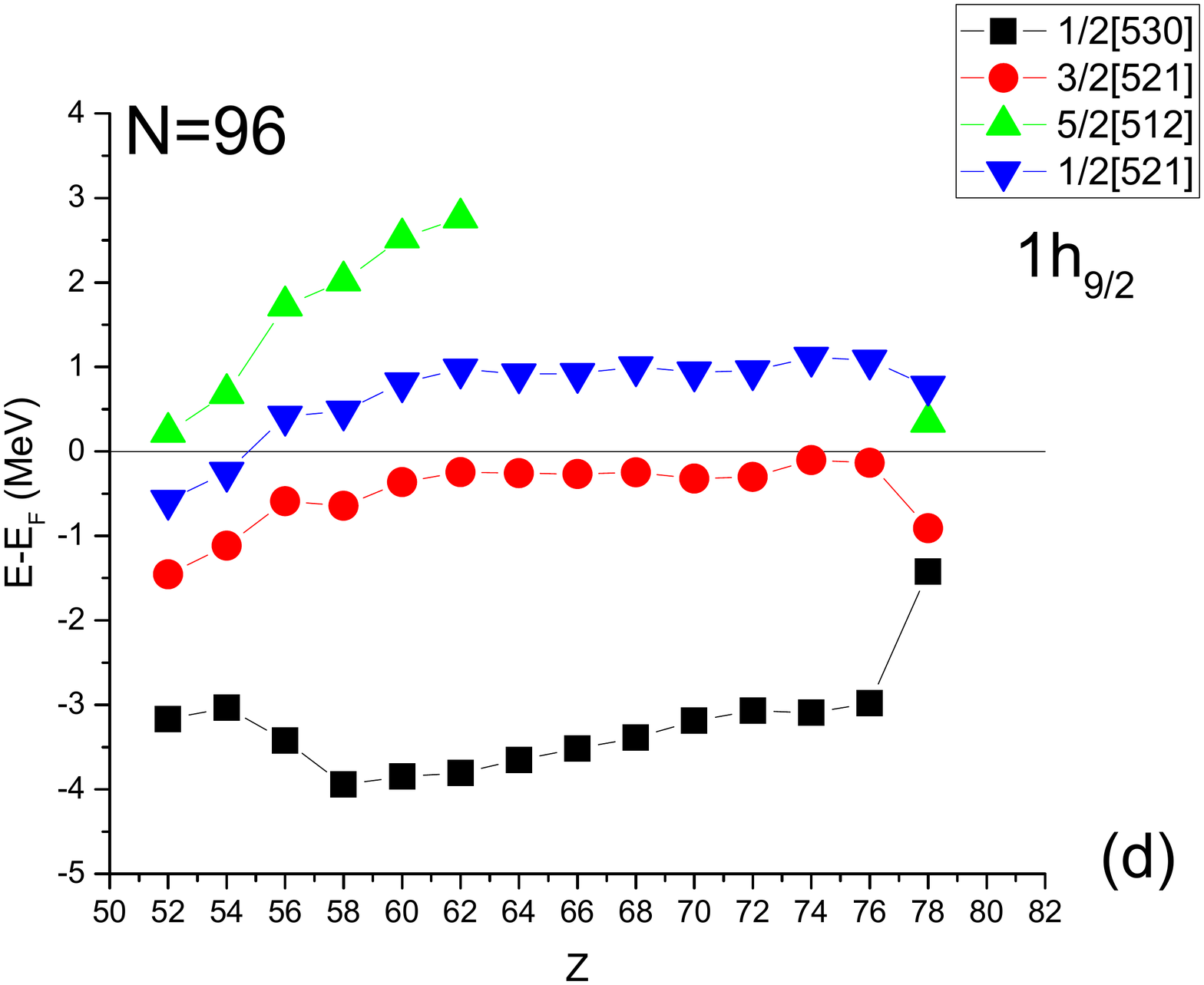}\hspace{5mm}}

\caption{Energies (in MeV) of single-particle neutron orbitals relative to the Fermi energy obtained by CDFT for $N=90$ isotones. The orbital 3/2[521] of $1h_{9/2}$ is gradually falling with increasing $N$. For $N=90$ it is empty everywhere above $Z=58$, for $N=92$, 94 it is still empty at $Z=60$-70, while at $N=94$ it has already sunk below the Fermi energy. See Section IV for further discussion. 
} 

\end{figure*}



\begin{figure*}[htb]

{\includegraphics[width=75mm]{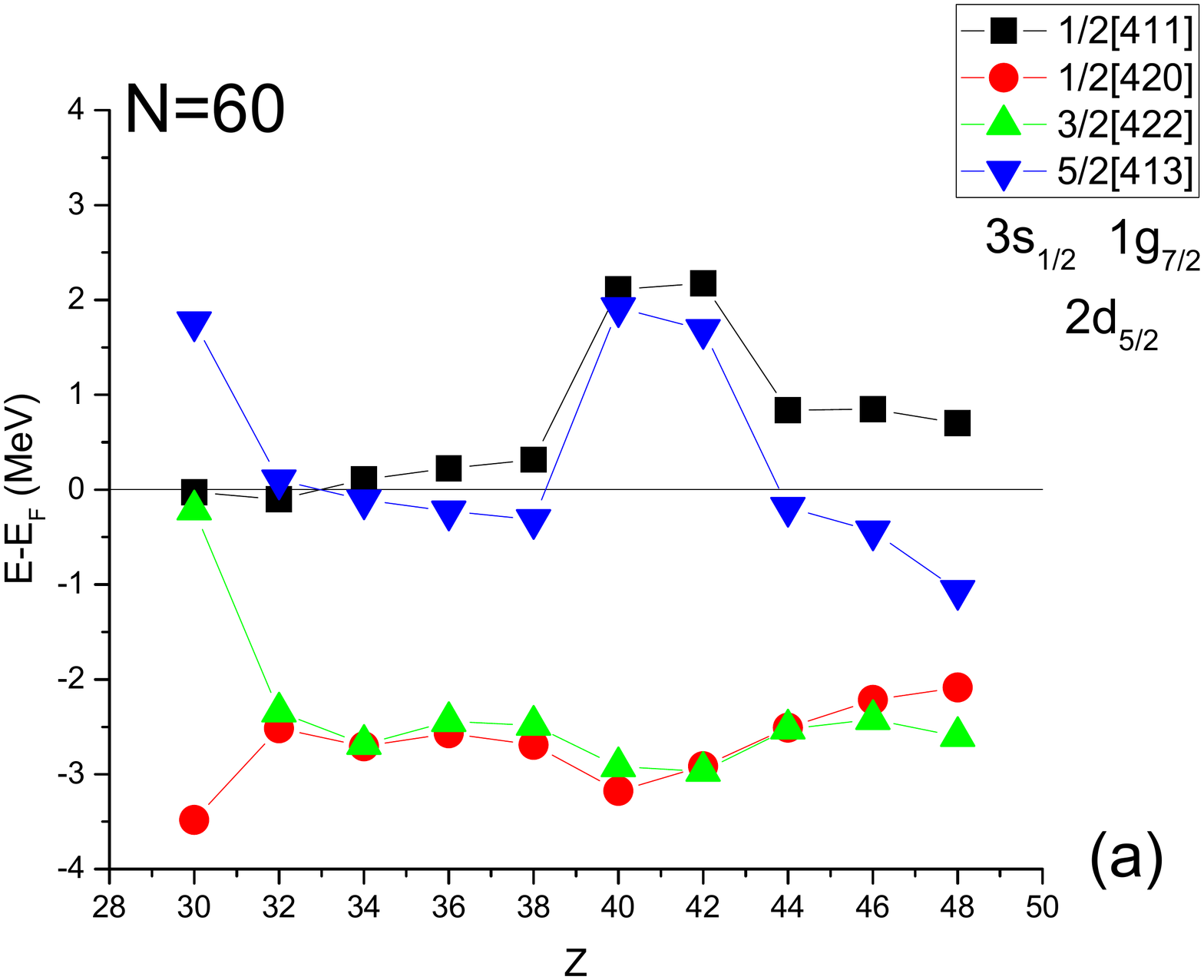}\hspace{5mm}
\includegraphics[width=75mm]{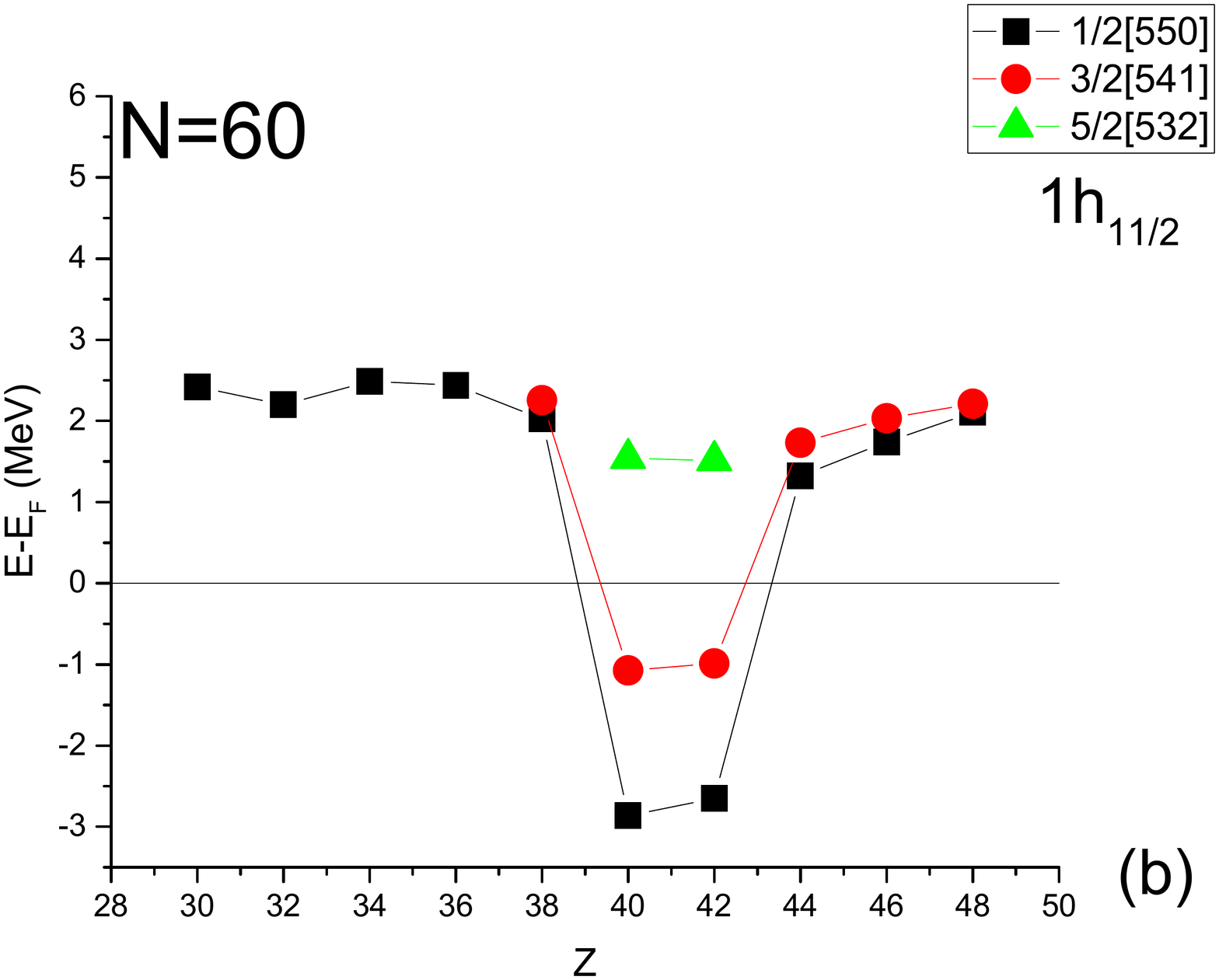}}

\caption{Energies (in MeV) of single-particle neutron orbitals relative to the Fermi energy obtained by CDFT for$N=60$ isotones. 4p-4h proton excitations are seen for $Z=40$-42. The orbitals 1/2[411] of $3s_{1/2}$ and 5/2[413] of $2d_{5/2}$ (a) (normally lying below $N=70$) are vacant for $Z=40$,42, while the orbitals 1/2[550], 3/2[541] of $1h_{11/2}$ (b) (normally lying above $N=70$) are occupied. Adapted from Ref.~\cite{PLB}. See Section IV for further discussion. 
} 

\end{figure*}


\begin{figure*}[htb]

{\includegraphics[width=75mm]{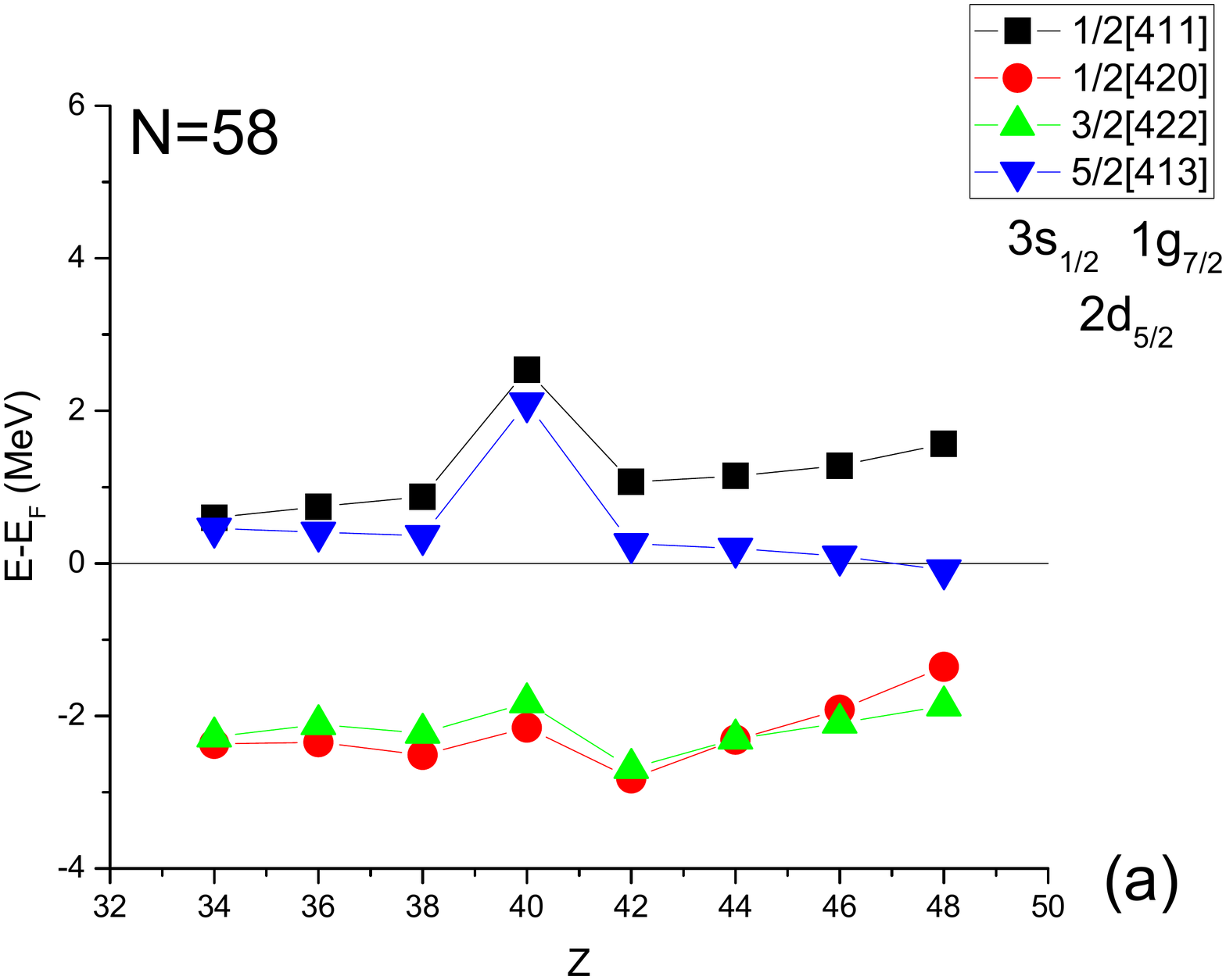}\hspace{5mm}
\includegraphics[width=75mm]{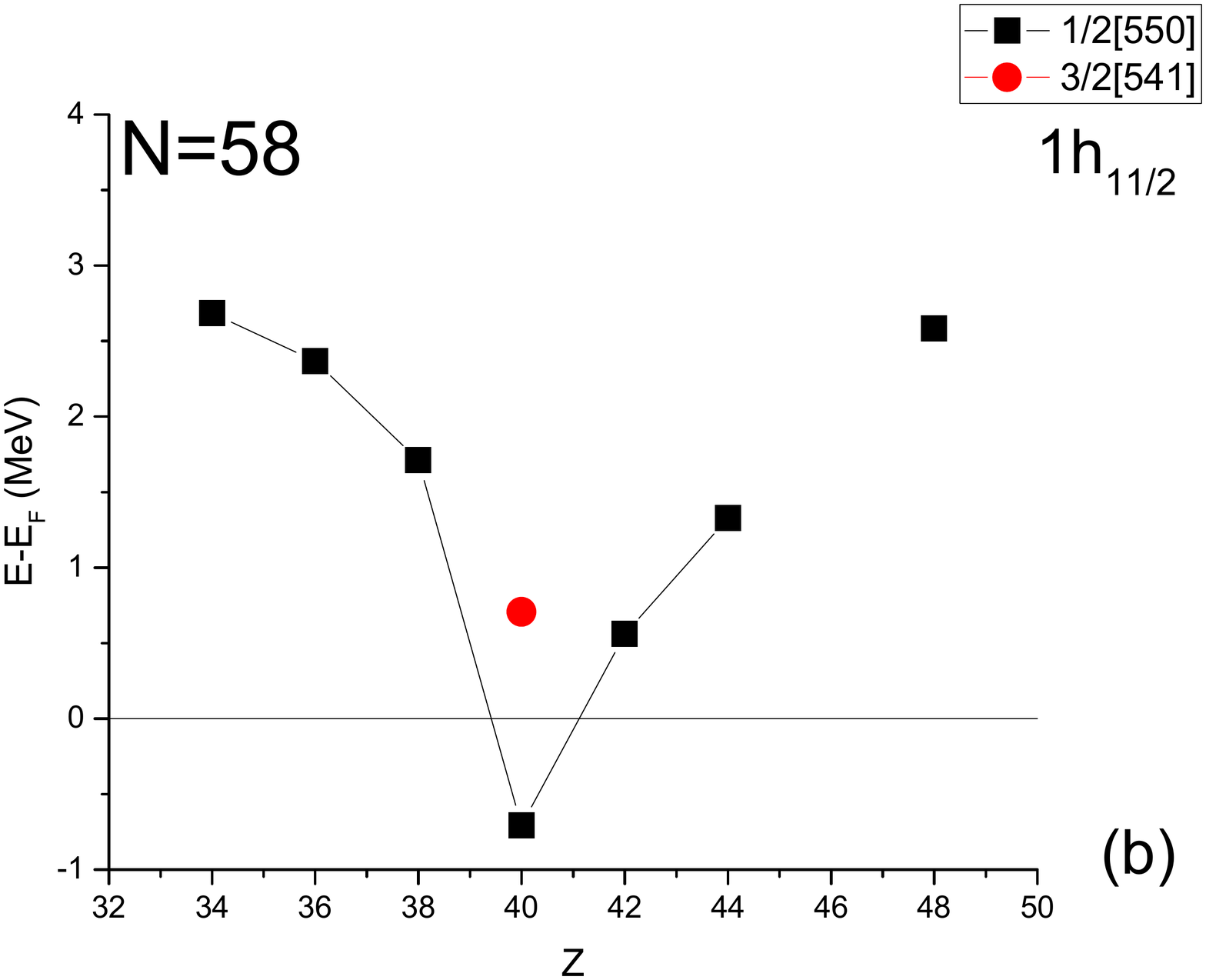}}

\caption{Energies (in MeV) of single-particle neutron orbitals relative to the Fermi energy obtained by CDFT for $N=58$ isotones. 4p-4h proton excitations are seen for $Z=40$-42. The orbitals 1/2[411] of $3s_{1/2}$ and 5/2[413] of $2d_{5/2}$ (a) (normally lying below $N=70$) are vacant for $Z=40$, while the orbital 1/2[550] of $1h_{11/2}$ (b) (normally lying above $N=70$) is occupied. See Section IV for further discussion. 
} 

\end{figure*}



\begin{figure*}[htb]

{\includegraphics[width=75mm]{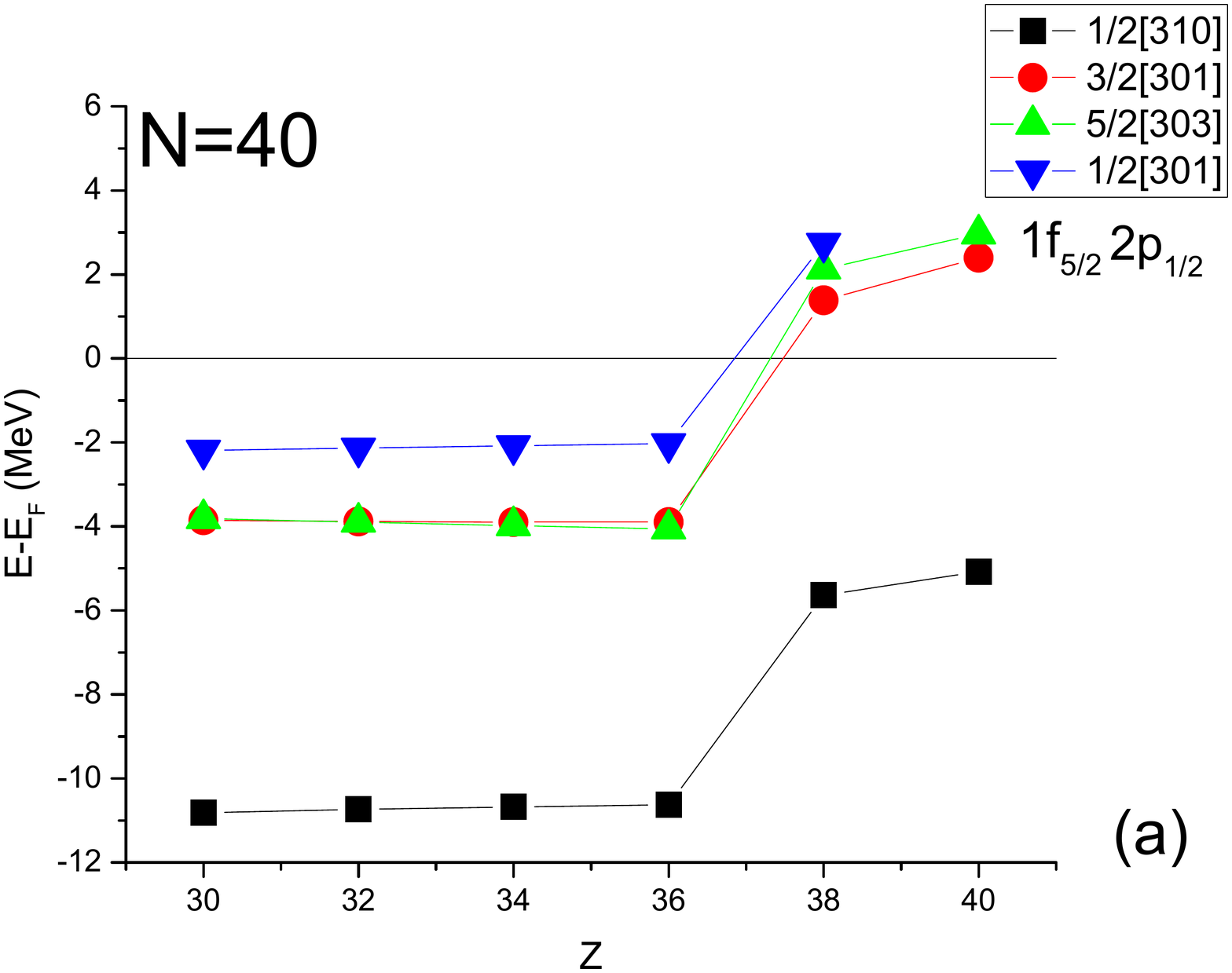}\hspace{5mm}
\includegraphics[width=75mm]{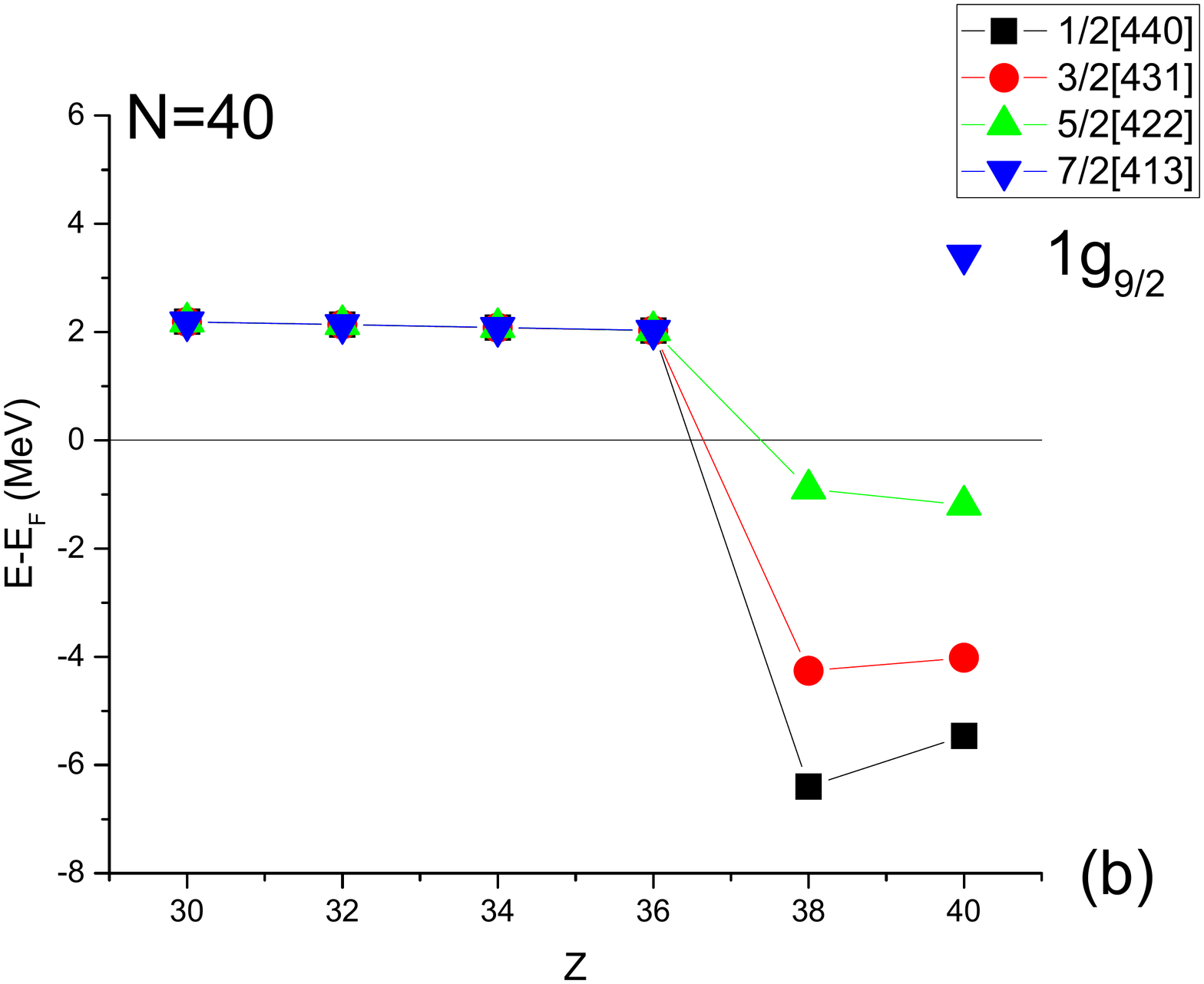}}

\caption{Energies (in MeV) of single-particle neutron orbitals relative to the Fermi energy obtained by CDFT for $N=40$ isotones. 6p-6h proton excitations are seen for $Z=40$. The orbitals 5/2[303], 3/2[301] of $1f_{5/2}$ and 1/2[301] of $2p_{1/2}$ (a) (normally lying below $N=40$) are vacant for $Z=40$, while the orbitals 1/2[440], 3/2[431], 5/2[422] of $1g_{9/2}$ (b) (normally lying above $N=40$) are occupied. See Section IV for further discussion. 
} 

\end{figure*}


\begin{figure*}[htb]

{\includegraphics[width=75mm]{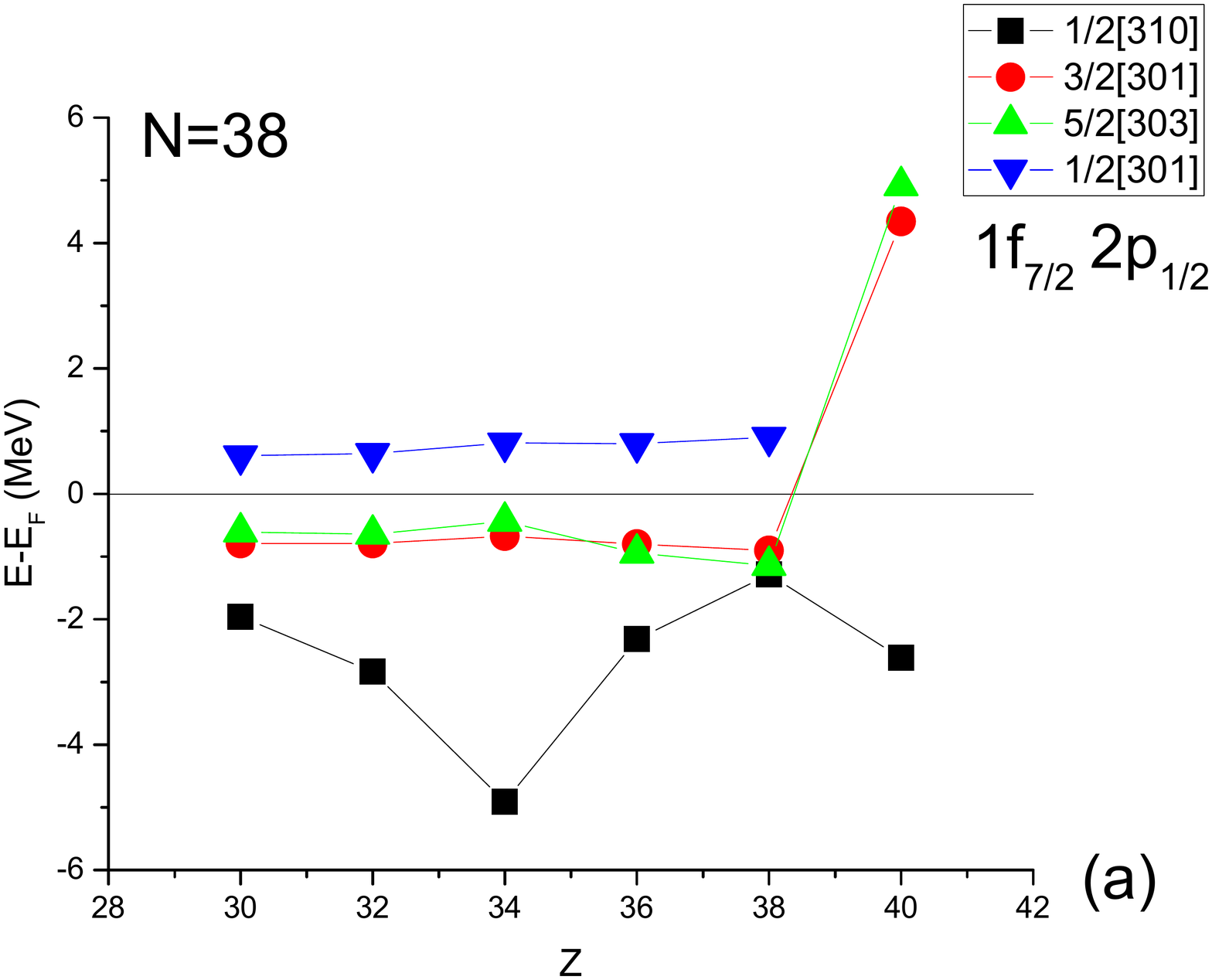}\hspace{5mm}
\includegraphics[width=75mm]{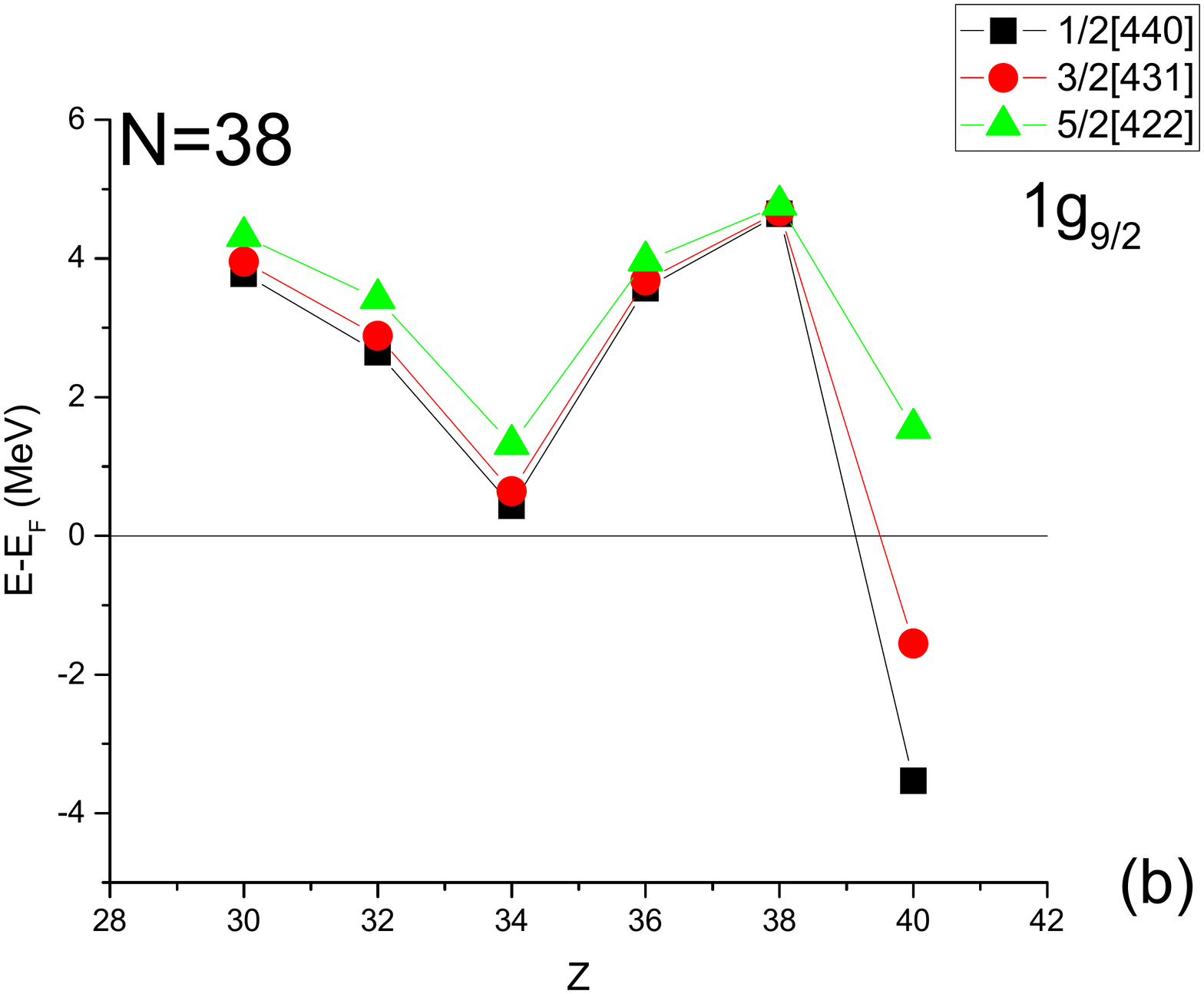}}

\caption{Energies (in MeV) of single-particle neutron orbitals relative to the Fermi energy obtained by CDFT for $N=38$ isotones. 4p-4h proton excitations are seen for $Z=40$. The orbitals 5/2[303], 3/2[301] of $1f_{5/2}$ (a) (normally lying below $N=40$) are vacant for $Z=40$, while the orbitals 1/2[440], 3/2[431] of $1g_{9/2}$ (b) (normally lying above $N=40$) are occupied. Adapted from Ref.~\cite{PLB}. See Section IV for further discussion. 
} 

\end{figure*}


\begin{figure*}[htb]

\includegraphics[width=150mm]{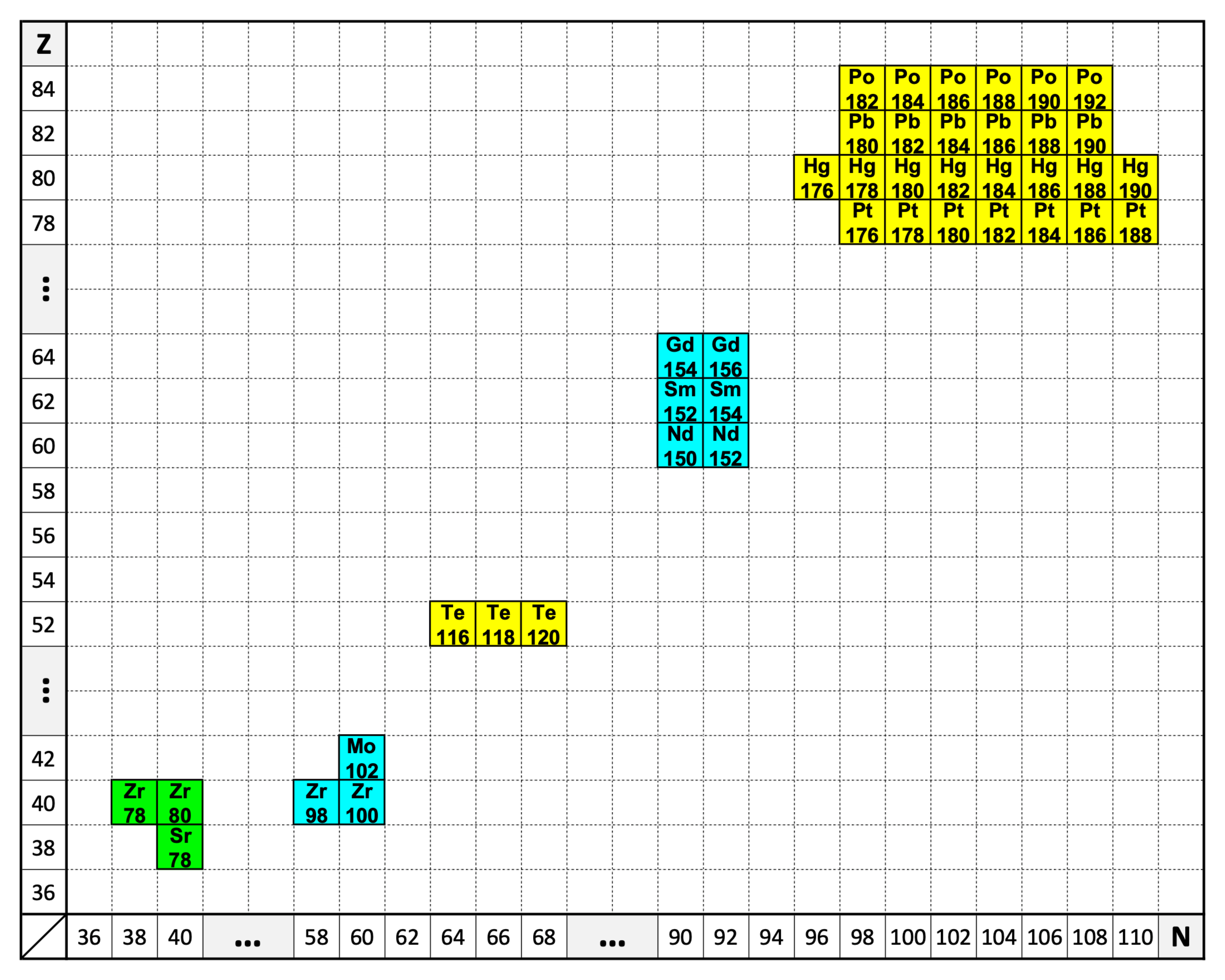}

\caption{Islands of shape coexistence (SC) found in the present study. Islands corresponding to neutron-induced SC are shown in yellow, islands due to proton-induced SC are exhibited in cyan, while islands due to both mechanisms are shown in green. See Sections III and IV for further discussion. }

\end{figure*}

\end{document}